\documentclass[11pt,a4paper]{article}
\usepackage{amsmath}
\usepackage{amsfonts}
\usepackage{amssymb}
\usepackage{hyperref}
\usepackage{inputenc}
\usepackage{t1enc}
\usepackage{times}
\usepackage{graphicx}

\newtheorem{axiom}{Theorem}[section]
\newtheorem{proposition}{Proposition}[section]

\newtheorem{Appendix}{APPENDIX}[section]
\newtheorem{definition}{Definition}[section]
\newtheorem{remarks}{Remarks}[section]
\newtheorem{remark}{Remark}[section]

\begin{document}

\bibliographystyle{unsrt}

\def\boxit#1#2{\setbox1=\hbox{\kern#1{#2}\kern#1}%
\dimen1=\ht1 \advance\dimen1 by #1 \dimen2=\dp1 \advance\dimen2 by
#1
\setbox1=\hbox{\vrule height\dimen1 depth\dimen2\box1\vrule}%
\setbox1=\vbox{\hrule\box1\hrule}%
\advance\dimen1 by .4pt \ht1=\dimen1 \advance\dimen2 by .4pt
\dp1=\dimen2 \box1\relax}

\def\build#1_#2^#3{\mathrel{\mathop{\kern 0pt#1}\limits_{#2}^{#3}}}

\def\K{\Bbb K}
\def\C{\Bbb C}
\def\R{\Bbb R}
\def\N{\Bbb N}
\def\ecart{\noalign{\medskip}}
\font\tenmsb=msbm10 \font\sevenmsb=msbm7 \font\fivemsb=msbm5
\newfam\msbfam
\textfont\msbfam=\tenmsb \scriptfont\msbfam=\sevenmsb
\scriptscriptfont\msbfam=\fivemsb
\def\Bbb#1{{\fam\msbfam\relax#1}}
\vfill\eject
{\title{  Local contractivity   of the $\Phi_4^4$
mapping}
\author{
 by\ \ Marietta Manolessou\\
EISTI\, \, -\, \, Department of Mathematics}
\maketitle

\begin{abstract}
We show  the existence and uniqueness of a  solution to a  $\Phi_4^4$  non linear renormalized system of equations of motion in Euclidean space. This  system represents  a non trivial model which describes the dynamics of the $\Phi_4^4$ Green's functions in the Axiomatic Quantum Field Theory (AQFT) framework.
 
 The main  argument is the local contractivity of the so called \emph{``new mapping''}  in the neighborhood  of a  particular ``tree type'' sequence of Green's functions. This neighborhood (and the $\Phi_4^4$ non trivial solution) belongs to a particular subset of the appropriate Banach space characterized by signs, splitting (analogous to that of the $\Phi_0^4$ solution), axiomatic analyticity properties and ``good'' asymptotic behavior with respect to the four-dimensional euclidean external momenta.
  \end{abstract}
\vfill\eject

\tableofcontents
\listoffigures

\vfill\eject

\section{ Introduction}
\pagenumbering{arabic}

\subsection{A new non perturbative method}

Several years ago we initiated a program for
 the construction of a nontrivial
 $\Phi^4_4$  model consistent with the
general principles of a Wightman Quantum Field
 Theory
($Q.F.T.$) \cite{(Q.F.T.)}. In references \cite{MM1} we 
introduced
 a non perturbative method for the
 construction of a nontrivial solution of
 the system
 of the $\Phi^4$ equations of motion
for the Green's functions, in the Euclidean space
 of zero, one and two dimensions.
 In references \cite{MM2}
we tried to apply an extension of this method to the case
 of a four (and a fortiori of a three)-dimensional
Euclidean momentum space through  the proof of a global contractivity principle inside an appropriate Banach space.
However that method was  rather complicated and also could profit from more precise norm definitions.

Using a Banach space ``analogous'' to that of zero dimensions and the local contractivity of the mapping (renormalized equations of motion in four -dimensional problem) in a neighbourhoud of the zero dimensional solution, we will be able to present an  easily convincing proof of the $\Phi^4_4$ nontriviality.

This method is different in approach from the work
 done in the Constructive $Q.F.T.$ framework of
 Glimm-Jaffe and others \cite{G.J.} \cite{G.J.S.}, and the methods
 of Symanzik who created the basis for a pure
 Euclidean approach to $Q.F.T.$ \cite{Sym}.

    In the $Q.F.T.$ language the interaction of
 four scalar fields $\Phi(x)$ is represented by a Lagrangian of ${\cal L}_{I} \sim \Lambda \Phi^4$ type, for example
 \cite{Zim1}:
 \begin{equation}{\cal L} \sim \Lambda \Phi^4 -  \partial_\mu\Phi\partial^\mu\Phi - m^2\Phi^2
 \end{equation}
 It is
 mathematically described
 in the four-dimensional Minkowski space
with coordinates:
\begin{equation}   x=\lbrace    \vec x\in
\Bbb R ^3, x_0\in \Bbb R;
 ||x||=\sqrt{ [x_0^2 -\vec x ^2 ]} \rbrace,
\label{(1.2)}
\end{equation}
by the following two-fold set of dynamical equations:
    \begin{itemize}
\item[i.] a nonlinear differential equation (the equation of
motion) resulting from the corresponding Lagrangian by application
of
 the variational principle:
    \begin{equation}    - (\boxit{1pt}{${ \ }^{ }$} + m^2 )
\Phi(x) =\frac {\Lambda}{\rho +\gamma} \lbrack\,:\Phi^{3}(x): - a\Phi(x)\, \rbrack
\label{(1.3)}
    \end{equation}
 \item[ii.] the ``conditions of quantization''
 of the field $\Phi(x)$ expressed by the equal time commutation relations:
            \begin{equation}[\Phi(x), \Phi(y)]  =
 [\dot {\Phi}(x), \dot {\Phi}(y)] = 0,\  \ \ \mbox{at}\ \   x_0=y_0
\label{(1.4)}
\end{equation}
    \begin{equation}    [\Phi(x),\dot {\Phi}(y)]  =
 i{\gamma\over\rho +\gamma}\delta^3(\vec x - \vec y),   \ \ \mbox{at} \ \  x_0=y_0
\label{(1.5)}
\end{equation}
\end{itemize}
Here $m>0$  and $\Lambda>0$ are the physical mass
 and coupling constant of the interaction model,
and $a$, $\rho$, $\gamma$, are physically
 well defined quantities associated to this model,
the so-called renormalization constants.
 For the precise definition of the latter
 and of the normal
product  $:\Phi^{3}(x) :$ we refer the reader
 to references \cite{Zim1}.

From these equations one can formally derive an equivalent
infinite system of nonlinear integral
 equations of motion
for the Green's functions (the {\it ``vacuum expectation
values''}) of the theory,
 analogous but not identical, to the Dyson
-Schwinger equations \cite{Dys}\cite{Sch}. This dynamic system
has been established in $4$ dimensions
  by using the Renormalized Normal Product
of \cite{MM3}.
\begin{figure}[h]
\begin{center}
\hspace*{-5mm}
 \includegraphics[width=13cm]{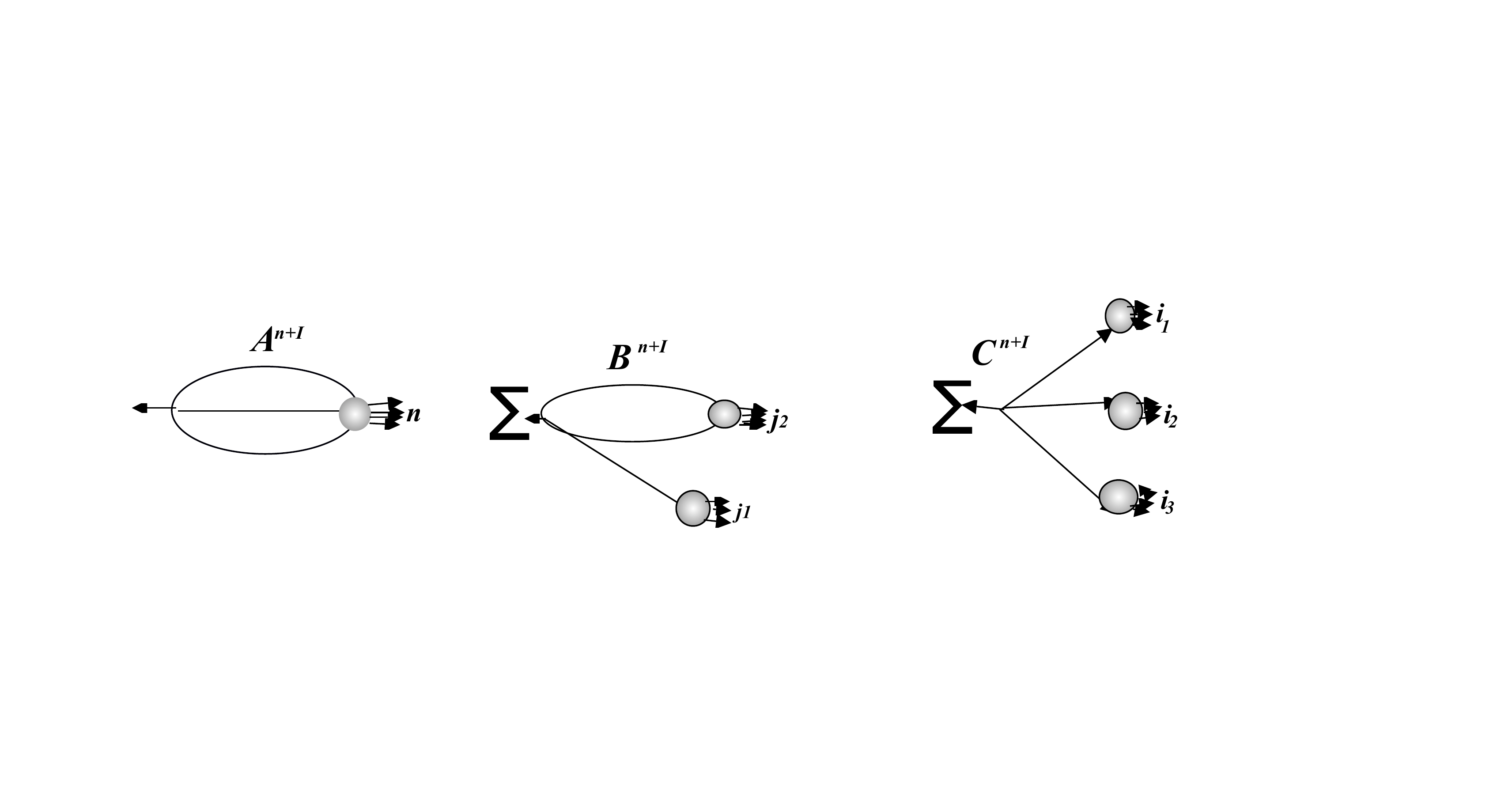}
\end{center} 
 \caption{\small\textrm{Graphical representation of the global terms of the $\Phi^4$ equations of motion}}
\label{fig.1.2}
\end{figure}
\subsection{The dynamic system of the Green's functions}

\begin{definition}[The $\Phi^4_4$ operations]\label{def.1.1}\

The   infinite system of equations for the Schwinger
 functions (i.e. the connected, completely
 amputated with respect to the free propagator
Green's functions, $ H =\{H^{n+1}\}_{n =2k+1, k \in\Bbb N}$),
 in the Euclidean $4$-dimensional
 momentum space, ${\cal E}^{4n}_{(q)}$
derived from the system
\ref{(1.2)}, \ref{(1.3)}, \ref{(1.4)}, \ref{(1.5)}  has  
the following form:
    \begin{equation}\begin{array}{l}
    H^2 (q,\Lambda)= - \displaystyle{{\Lambda\over{\gamma+\rho}}}
\lbrace\lbrack N_3H^4\rbrack
 -a
H^2 (q,\Lambda)\Delta_F(q) \rbrace
+\displaystyle{{(q^2+m^2)\gamma\over{\gamma+\rho}}}  \ \\    
\mbox{and}\ \  \forall\,  n \geq3, \, (q,\Lambda) \in
{\cal E}^{4n}_{(q)}\times \R^+ \\
 H^{n+1}(q,\Lambda) =  \displaystyle{{1\over{\gamma+\rho}}} \lbrace\,
 \lbrack A^{n+1}+ B^{n+1}
 + C^{n+1}\rbrack(q,\Lambda) + \Lambda
a H^{n+1}(q,\Lambda) \rbrace\\
    \mbox{with:}\\
A^{n+1}(q,\Lambda) = - \Lambda\lbrack
N^{(n+2)}_3H^{n+3}\rbrack(q,\Lambda);\\
\ \ \\
B^{n+1}(q,\Lambda) = - 3\Lambda\displaystyle{\sum_{\varpi_n(J)}}
\lbrack N^{(j_2)}_2H^{j_{2}+2}
 N^{(j_1)}_1H^{j_{1}+1}\rbrack(q,\Lambda); \\
 \ \ \\
 C^{n+1}(q,\Lambda) = - 6\Lambda\displaystyle{\sum_{\varpi_n(I)}\prod_{l=1,2,3}}
\lbrack N^{(i_l)}_1H^{i_{l}+1}\rbrack (q_{i_{l}},\Lambda);\\
    	a(\Lambda)=[N_{3}H^{4}]_{q^{2}+m^{2}=0};	\ \ \rho(\Lambda)=-\Lambda\left[
  	\displaystyle{\frac{\partial^2}{\partial q^{2}}[N_{3}^{(3)}H^{4}]}
  	\right]_{q^{2}+m^{2}=0};\\
  		\ \ \\
\mbox{and}\ \ \ \gamma(\Lambda)  =
		\left[\displaystyle{\frac{-6\Lambda \prod_{l=1,2,3}H^{2}(q_l)\Delta_{F}(q_l)}{H^{4}(q)}}\right]_{q=0} 
   \end{array} 
   \label{1.6}
   \end{equation}  
    \label{def.1.1}
\end{definition}
\begin{remarks}\
\begin{enumerate}
\item{}\emph{Here the notations:}
  $$\lbrack N^{(n+2)}_3H^{n+3}\rbrack,\ \ \ \lbrack N^{(j_2)}_2H^{j_{2}+2}
 N^{(j_1)}_1H^{j_{1}+1}\rbrack \ \mbox{\emph{and}} \displaystyle\prod_{l=1,2,3}
\lbrack N^{(i_l)}_1H^{i_{l}+1} \rbrack$$
\emph{represent the so called ``$\Phi^4_4$
operations'' that we introduce in the
 \emph{ Renormalized
 \,  G-Convolution\ \ Product} $(R.G.C.P) $  context of the references}
 \cite{Br.MM},
\cite{MM4}, \cite{MDu}.
  \emph{Briefly, the two
 loop  $\Phi^4_4$ - operation is defined by:}
 \begin{equation}
\lbrack N^{(n+2)}_3H^{n+3}\rbrack   = \int R^{(3)}_G \lbrack\
 H^{n+3}\prod_{i=1,2,3}
\Delta_F(l_i)\ \rbrack d^{4}k_1d^{4}k_2
\label{1.7}   
\end{equation}
\emph{with $ R^{(3)}_G$ being the corresponding renormalization operator
 for the two loops graph with bubble vertex the $H^{n+3}$ Green's function.}

\emph{The analogous expression for the one loop  $\Phi^4_4$ -
 operation is the following:}
\begin{equation}
[N^{(j_2)}_2H^{j_{2}+2}][ N^{(j_1)}_1H^{j_{1}+1}] =
H^{j_{1}+1}\Delta_F\int R^{(2)}_G  
[H^{j_{2}+2}\prod_{i=1,2}\Delta_F(l_i)]d^{4}k 
\label{1.8}
\end{equation}
\emph{with $R^{(2)}_G$, the corresponding renormalization  operator for
the
 two loops graph.}

\emph{The notation $\Delta_F$ indicates the free propagator, and the
$\Phi^4_4$
  operation
$N^{(j_1)}_1$ is exactly the multiplication (\emph{``trivial
convolution''})
 by  the corresponding
 free propagator }$\displaystyle\Delta_F ={ 1\over (\|q\|^2+m^2)}$.
 \emph{Here $\|q\|^2$
 means the
 Euclidean norm of the vector $q\in{\cal E}^{4n}_{(q)}$.}
 \item{} \emph{In equations \ref{1.6} the notation $q$ in the arguments of the two-point and four-point Green's functions 
 is used indifferently for $q\in{\cal E}^{12}_{(q)}$ or $q\in{\cal E}^{4}_{(q)}$.}
 \item{}
\emph{ In the previous notations
 and in all that follows, $\Bbb N$ always means the set
 of non negative integers and $n$ will always be
 an odd positive
 integer.}

\end{enumerate}
\end{remarks}

\subsection{The ``primary $\Phi$-Iteration''  }
 \subsubsection{The fixed point method}

The method is based on the proof of the existence and uniqueness
of the solution of the above infinite system of dynamical
equations
 of motion verified by the
Schwinger functions, following a fixed point theorem argument. 

    The information concerning the special features
 of the dynamics of four interacting fields
 has been obtained through an iteration at fixed coupling constant and \emph{at zero external momenta} of these
 integral equations of motion in the two dimensional
 case, taking the free solution as starting point.
 
 This is what was called the \emph{``$\Phi$-Iteration''}
 in \cite{MM1}.
 The exploration of the detailed  organization of the different structural global terms
 of the $H^{n+1}_{\nu}$ functions at every order $\nu$ 
of what we call   now the 
\emph{``primary $\Phi$-Iteration''},  has brought forth particular properties as:
 \begin{itemize}
 \item{(a)} \emph{alternating signs}  and \emph{splitting}
 (or \emph{factorization})
 properties at zero external momenta:
\begin{equation}    H^{n+1}(q=0,\Lambda)=
 - n(n-1)\delta_n(\Lambda)
H^{n-1}(q=0,\Lambda) [H^2(q=0,\Lambda)]^2   
\label{1.9}  
\end{equation}
with $\{\delta_n\}$ a bounded increasing sequence of continuous
functions of $\Lambda$ and uniformly convergent to some finite
positive constant $\delta_{\infty}$.
\item{(b)} \emph{ bounds at zero external momenta}  which in turn yield  global bounds
 of the general  form:
                    \begin{equation}
|H^{n+1}(q,\Lambda)| \leq  n\ !\   \  K^n \ \mbox{with $K$  a finite positive constant independent of $n$.}
\label{1.10}
\end{equation}
\end{itemize}
  These features formed a self-consistent
 system of conditions conserved by the
\emph{``$primary \ \Phi$-Iteration''}.
 In particular they implied
 precise \emph{ ``norms''}
 of the sequences of the Green's functions $H^{n+1}$:
\begin{equation}
\| H \| = \sup_{q, n, \Lambda}\{ M_n^{-1}(q, \Lambda)
 \ \ | H^{n+1}(q,\Lambda)| \}
\label{1.11}
\end{equation}


 These norms in turn were conserved and
automatically ensured
 the convergence of this
  \emph{``primary $\Phi$-Iteration''} to the solution.
 So,  in references \cite{MM1} and \cite{MM2} we thought about obtaining  an answer
to the problem by first
 defining a Banach space ${\cal B}_{initial}$
 using the norms provided by the \emph{``primitive $\Phi$-Iteration''}.
 and seeked a fixed point of
 the equations of motion inside a characteristic
 subset $\Phi\subset {\cal B}_{initial}$
  which exactly \emph{imitated}
 the fine structure of the $\Phi$-Iteration.\\

    Now, taking into account  the divergence of the perturbative series of
 a $\Phi^4_2$ model, a result
 that A. Jaffe \cite{Jaf} established several years ago, another (two-part)
 question immediately arises.
Does the two-dimensional  \emph{``$\Phi$-Iteration''} generate the
 perturbative series exactly?
If yes, then is there any contradiction between the divergence of
the
 perturbative expansion and the
 convergence of the  $\Phi$-Iteration? The answer is
that the $\Phi$-Iteration
 has nothing to do with the perturbation series. The method is not a
 reconstruction of perturbation
 theory. More precisely, all terms of the $N^{th}$ order
 of perturbation
 series are included together
 with subsequent ones at every order $N$  of the $\Phi$-Iteration
 (cf.\cite{MM1}a,b,). In other
 words
 the series like that of perturbation can be divergent, but
 a series of polynomials of the
 terms of the former may still be convergent.
 The difference between the
 two approaches comes from the
 different way in which the polynomials in $\Lambda$ are arranged
 and summed up in each of the two
 approximations. So automatically there is no contradiction if the
 \emph{``$\Phi$-Iteration''} converges
 to a nontrivial solution despite the divergence of
 perturbation theory.

                The reasons that motivated us for a study in smaller dimensions and
 not directly in four, were
 the absence of the difficulties due to the renormalization in two
 dimensions, and the pure
 combinatorial character of the problem in zero dimensions \cite{MM6}.

 Another useful aspect  of the zero dimensional case is the fact that
 it provides a direct way to
 numerically test the validity of the method \cite{GASM}, \cite{MMST}.
 
\subsubsection{The new mapping ${\cal M}^*$ and the local contractivity}
These ``conserved
norms''  \ref{1.11} lead to the convergence of the 
\emph{
``$Primary \  \Phi$\ -Iteration''}.
So, by
 introducing an appropriate Banach space ${\cal B}_{initial}$ defined by
 these norms and a characteristic subset $\Phi\subset {\cal B}_{initial}$
 which exactly
imitates the fine structure of the  ``Primary $\Phi$-Iteration''}
 one expects
to establish
 by a fixed point theorem the existence of a unique nontrivial solution
 inside this subset.

    Unfortunately this is not the case \cite{Vor}. The  global terms $A^{n+1}$,
 $B^{n+1}$,
and $C^{n+1}$,
 (tree terms) (with alternating signs)
 have identical  asymptotic behavior with respect to $n$, but not what we schould expect of the corresponding $H^{n+1}$.
 More precisely,
at every fixed value of the external momenta (precisely we proved them at zero external momenta) , we obtain:

\begin{equation} A^{n+1}\build\sim_{n\rightarrow\infty}^{}
(-\ \delta_{\infty})^{n-1\over2}
  n !\  n^2 
  \label{1.12}
  \end{equation}
    \begin{equation}B^{n+1}\build\sim_{n\rightarrow\infty}^{}
 -  (-\ \delta_{\infty}^{ })^{n-1\over2}  n !
  \ n^2
  \label{1.13}
\end{equation}

    \begin{equation}C^{n+1}\build\sim_{n\rightarrow\infty}^{}
(-\ \delta_{\infty}^{\Lambda})^{n-1\over2} n !\, n^2
\label{1.14}
\end{equation}

 As far as the behavior with respect to the external four momenta
 is concerned they follow
 the behavior of the norm functions  \ref{1.11} (i.e. the
corresponding structure of the Banach space).

But,  the above $n^2$ dependence of the global terms prevents the mapping 
 \begin{equation}{\cal M}:\, \ {\cal B}_{initial}\stackrel{\cal M}
 \longrightarrow {\cal B}_{initial} 
 \label{1.15} 
\end{equation}
from being \emph{contractive} in ${\cal B}_{initial}$, despite  the convergence of  the \emph{ ``$\Phi$-Iteration''},
(thanks  to the alternating signs of the global
 terms).
  
 This is the reason that motivated us to define  a new
 mapping ${\cal M}^*$ (equivalent to the initial  mapping) given
 by the following equations, and which is $\emph{contractive}$:
            \begin{equation}
            H^{n+1'}(q,\Lambda) = {\delta_n^{'}(q,\Lambda)
 C^{n+1'}(q, \Lambda) \over 3\Lambda
 n (n-1)}
 \label{1.16}
 \end{equation}
    with:
    \begin{equation}
    \delta_n^{'}(q,\Lambda)=\frac{3\Lambda n(n-1)}
 {(\gamma+\rho)+D_n(H)-\Lambda\alpha}
 \label{1.17}
 \end{equation}
and
\begin{equation}
 D_n(H)={|B^{n+1}| - |A^{n+1}| \over |H^{n+1}|}
  \label{1.18}
\end{equation}
One can intuitively understand the contractivity of the new
mapping ${\cal M}^*$  by looking at the
 behavior with respect to $n$  of the function $D_n$
 (at fixed external momenta). Precisely,
    \begin{equation}
    D_n(H ) \build\sim_{n\rightarrow\infty}^{}\displaystyle{\frac{ n^2 n\ !
(\delta_{\infty})^{n-3\over2}}{  n\ !
(\delta_{\infty})^{n-1\over2}}} \build\sim_{n\rightarrow
\infty}^{}\displaystyle{\frac{n^2}{\delta_{\infty}}}
\label{1.19}
\end{equation}
Consequently:
\begin{equation}
\delta_n^{'}(q,\Lambda)
\build\sim_{n\rightarrow\infty}^ {}\displaystyle{\frac{\delta_{\infty}  n
(n-1)}{ n^2}}\build\sim_{n\rightarrow\infty}^{}
\delta_{\infty}
\label{1.20}
\end{equation}
By this last argument one can show not only the conservation of
the norms 
 but also the contractivity  of  the ``new mapping'' ${\cal M}^*$
to a fixed point
 inside a characteristic subset $\Phi\subset {\cal B}_{initial}$,
 under a sufficient condition of the following type
imposed on the renormalized coupling constant:
\begin{equation}0\leq\Lambda\leq \Lambda_0
\label{1.21}
\end{equation}
 In an equivalent way, this result implies the existence and uniqueness
of a nontrivial solution (even in four dimensions), of the
system. Under the condition \ref{1.21},
 this solution
lies in a neighborhood of a precise point-sequence of the
appropriate subset $\Phi\subset {\cal B}$, the so-called
\emph{fundamental sequence $\{H_{T0}\}$}.

Consequently, the construction of this non perturbative solution
can be obtained by the iteration
 of the mapping ${\cal M}^*$ inside $\Phi\subset {\cal B}$
 starting from the corresponding, to every dimension,
\emph{fundamental sequence $\{H_{T0}\}$}.
 So, this solution verifies automatically, the\emph{``alternating signs''} and
\emph{``splitting''} properties  at every value of the external
momenta, analyticity properties
 together with the
physical conditions imposed on $H^2$- and $H^4$ - Green's
functions for the definition of the renormalization parameters.

 The essential  results of the method in 4 dimensions are the following:
\begin{enumerate}
 \item{}  The proof of the
 \emph{``alternating signs''} and \emph{``splitting''}
 properties \emph{at every value of the external momenta} (and not only at zero external momenta as we had originally
 established in the
 \emph{ ``primary $ \Phi$-Iteration''}).
 \item{} The proof of specific asymptotic increase properties with  respect to the four dimensional external momenta of the renormalized Green's functions, in particular the powers of $q^2 \log(q^2)^k$  which dominate the $H^2$-point function's behaviour.
 \item{} The proof of A.Q.F.T. properties in complex Minkowski space.
 \item{} The proof of the existence and uniqueness of solution of the system   \ref{1.6}   obtained as a \emph{limit} of a (new)  renormalized
 iteration procedure (the so called  ``$\Phi^4_4$- iteration") or in other words as a fixed point of a locally contractive mapping in the neighborhood of a precise \emph{``tree type'' } sequence.
\end{enumerate}

\subsection{Planning of the paper}
The paper is organized as follows:
\begin{enumerate}
\item{}
In section $2$ we introduce the general vector space ${\cal B} $, the \emph{tree type sequences}, and the Renormalized $\Phi$ 
 Convolutions (R.$\Phi$.C.) - Green's functions using as building block one of the tree type sequences the  \emph``fundamental tree type sequence'' $H_{T0}$. (cf.def. \ref{def.2.5} )\\
  Then we define the ``renormalized subspace'' ${\cal B}_{R}\subset {\cal B}$ which (being provided  with the appropriate norm) is a Banach  space.
  
 \item{}  In section $3$ 
 \begin{itemize}
 \item{} We introduce a  particular subset $\Phi_{R}\subset {\cal B}_{R}$ 
  characterized by the detailed bounds, signs, splitting properties, consistent definitions of the renormalization parameters $a,\ \rho, \mbox{and} \ \gamma$, together with the analyticity properties of  general n-point functions (in the A.Q.F.T. framework).  The non triviality of $\Phi_{R}\subset {\cal B}_{R}$ is established by proving that $H_{T0}\in\Phi_{R}$. 
  \item{} We define the Generalized Renormalized $\Phi$- 
 Convolutions (G.R.$\Phi$.C.) and the \emph{new mapping ${\cal M}^*$} on ${\cal B}_{R}$. 
 \item{}
 By successive application of ${\cal M}^*$ on   $H_{T0}$  a \emph{$\Phi_4^4 $- iteration} is defined which preserves   the good properties of  $H_{T0}$ and automatically establishes the stability of a neighbourhood of $\Phi_R$ under ${\cal M}^*$.
 \end{itemize}
  \item{}
   Section  $4$ contains  the construction of
 the solution: the local contractivity of the mapping ${\cal M}^*$ inside a precise closed ball in 
 ${\cal B}_{R}, \ \ {\cal S}(H_{T0},  r_0)\subset \Phi_R$.
 \item{}
In the Appendices we present the necessary proofs
 of the theorems stated in sections $3$ and $4$.
\end{enumerate}

\section{The vector space ${\cal B}$ - the``splitting sequences'' \\
the ``tree type sequences'' and the Banach space ${\cal B}_R$}
 
 The fundamental difference between two and four (or three)
 dimensions is the divergence in the latter case of a finite number of
 $\Phi-Convolutions$
 for every  fixed integer $n =2k +1,\ \  k\in {\N}$. Therefore, it has been necessary to introduce the precise definition of the renormalization operations and choose a more specific space of Green's functions sequences in the  A.Q.F.T.  framework.
  
	To this purpose, we introduce the basic elements of this space,
 the so-called {\emph tree type sequences} $H_T$, and from a particular choice of one of them, we
 recursively define the Renormalized $\Phi-Convolutions$ $(R.F.C)$.
 For the detailed definitions and statements concerning the recursive
 procedure of the renormalization, we refer  the reader  to the references   \cite{Br.MM}
\cite{MM4}\cite{MDu}.

 Apart  from a brief reminder  of certain crucial properties, we apply the results 
 of these papers without any detailed description.

\subsection{The space ${\cal B}$ - the``splitting sequences'' }

\begin{definition}{The space ${\cal B}$} \ \

 We define the vector space ${\cal B}$
of the sequences $H=\{ H^{n+1}\} _{{n =2k+1;}\atop {k\in\N}}$
as follows:

 For every $n$ the function $H^{n+1}$ belongs to the space
 $C^{\infty}({\cal E}^{4n}_{(q)}\times {\R}^{+})$
 of continuously  differentiable real numerical functions of the set 
 $(q,\Lambda)$ of $4n+1$ real independent variables and verifies the following 
  properties. There exists a finite positive constant $C_n$, 
 such that the following  bounds hold:
   
 \begin{equation}\begin{array}{l}
   \forall\  (q,\Lambda) \in ({\cal E}^{4}_{(q)}\times {\R}^{+})\\
 	\vert H^{2}(q,\Lambda)\vert\leq C_1
 	[  (\Vert q\Vert^{2}+m^2)^{(1+\pi^2/18)}]
 \end{array}
 \label{2.22}
 \end{equation}
 \emph{and} $\forall\, n=2k+1\,\quad k\in\N^{*},\ (q,\Lambda)\in ({\cal E}^{4n}_{(q)}\times {\R}^{+})$
 \begin{equation}
 	\vert H^{n+1}(q,\Lambda)\vert\leq n!C_n
 	[  (\Vert q\Vert^{2}+m^2)^{(1+\pi^2/18)}]^n
 	\label{2.23}
 \end{equation}
 Here the notation    $\Vert q\Vert=\displaystyle{\sqrt{\sum_{i=1}^{n}q_{i}^2}}$ means the Euclidean norm of the vector 
$q\in{\cal E}^{4n}_{(q)}$. In the following we  often use the 
equivalent notation: $\Vert q\Vert^{2}=q^{2}$.  
\label{def.2.1}
\end{definition}

\begin{definition}
  The "splitting sequences" $\delta\in\mathcal{B}$  \label{def.2.2}\  
  
 We introduce the particular class $\mathcal{D}\subset \mathcal{B}$ of the so-called ``splitting sequences'' 
$\delta=\{\delta_{n}(q,\Lambda)\}_{n=2k+1\ k\in\N}\in  \mathcal{B}$.

There exists a  finite positive constant $C_{0}$ such that 
the corresponding bounds (\ref{2.23}) take the following form: 
\begin{equation}
\begin{array}{l}
	\forall\, n=2k+1\quad k\in\N^{*}  \\
	 0\ < \delta_{n}(q,\Lambda)\leq C_{0}
	\quad \ õ\forall
	(q,\Lambda)\in{\cal E}^{4n}_{(q)}\times \R^{+}.
\end{array}	
	\label{2.24}
\end{equation}
\label{def.2.2}
\end{definition}

\subsection{The "tree type sequences" $H_{T}$}

\begin{definition}\ \

We define the following class of sequences $H_{T}$ associated with a given splitting sequence 
$\delta\in\mathcal{D}$:

\begin{equation}
\begin{array}{l}
\ \ \ \forall \ (q, \Lambda)\ \in ({\cal E}^{4n}\times\R^{+*})\\
\ \ \\
	H_{T}^{2}(q,\Lambda)= (q^2+m^2) ( b_{0}(\Lambda)+ b_1(\Lambda)(q^2+m^2)^{(1+\pi^2/18)}\\
	 \mbox{(here $b_0$ and $b_1$ are in general continuous bounded positive   functions of $\Lambda$)} \\
	H_{T}^{4}(q,\Lambda)=\ - \delta_{3}(q,\Lambda) 
	\displaystyle{\prod_{j=1,2,3}}H_{T}^{2}
	(q_{{j}},\Lambda)\Delta_{F}(q_{{j}})\\
\mbox{and}\quad
	\forall\, n\geq 5\\
	H_{T}^{n+1}(q,\Lambda)=-\displaystyle{\frac{\delta_{n}(q,\Lambda)}{n(n-1)}
	\sum_{\varpi_{n}(I)}\prod_{j=1,2,3}H_{T}^{i_{j}+1}
	(q_{i_{j}},\Lambda)}\Delta_{F}(q_{i_{j}})
		\end{array}
		\label{2.25}
\end{equation}
\label{def.2.3}
\end{definition}
	
	\begin{figure}[h]
\begin{center}
\hspace*{-15mm}
 \includegraphics[width=16cm]{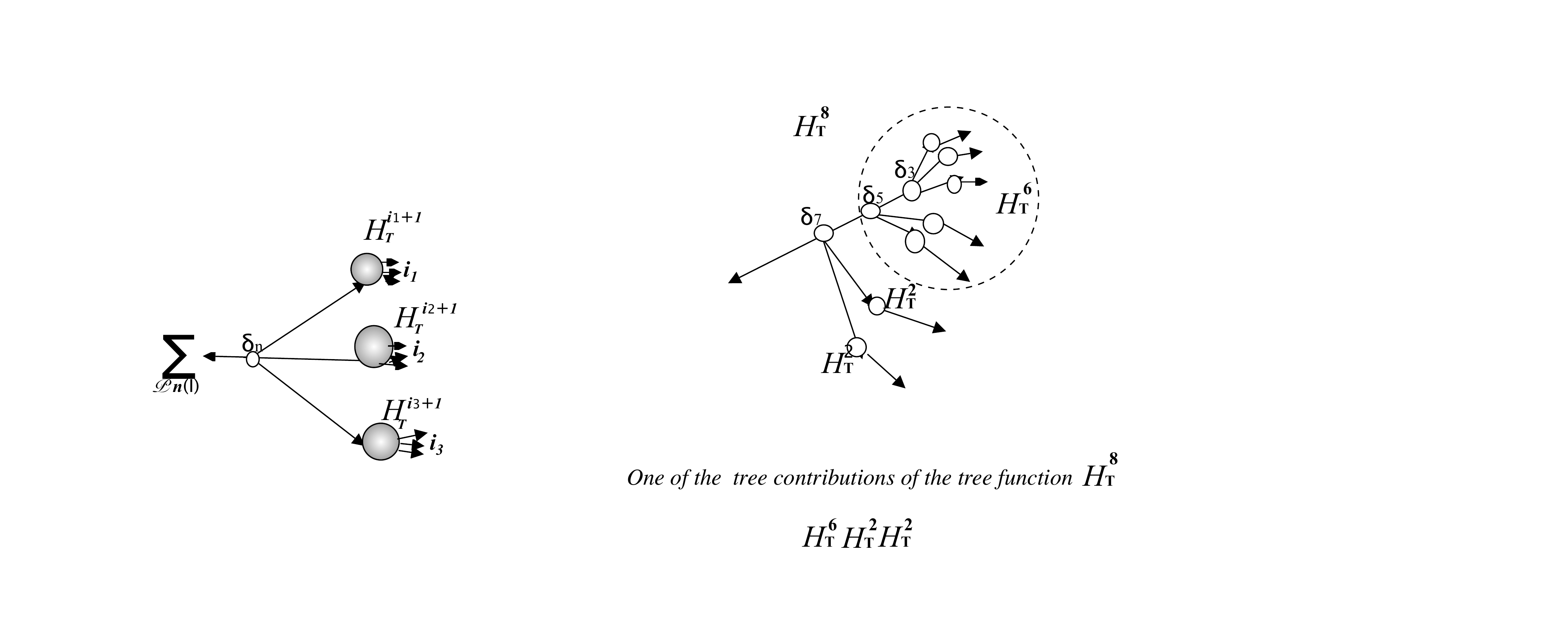}
\end{center} 
 \caption{\small\textrm{Graphical representation of a general tree function $H_{T}^{n+1}$, and a zoom on the bubble vertices of one of the tree  terms of the $H_{T}^{8}$ function.}}
\label{fig.Treegraph}
\end{figure}  
	  We call these sequences $H_{T}$ (resp. every $H_{T}^{n+1}$) the 
"tree type sequences" (res. the tree functions $H_{T}^{n+1}$). For every $n\geq 5$ the graphical representation of every $H_{T}^{n+1}$ 
is a finite sum of ``tree graphs'', with a   four point bubble vertex 
associated  with the corresponding 
$\delta_{n}(q,\Lambda)$ connected by three simple free propagators $\Delta_{F}(q_{i_{j}})$ to three ``bubble vertices''. These bubbles represent each one of the three tree functions   $H_{T}^{i_{j}+1}$ of the corresponding partition in the sum
 (cf.fig.\ref{fig.Treegraph}).
 
 \subsubsection{Particular splitting sequences-Reminders }\
 In  \cite{MM6} we introduced the following particular splitting sequences for the zero dimensional problem. 
\begin{definition}\label{def.2.4}
\begin{enumerate}
\item{} \textbf{The upper and lower bounds - splitting sequences}
\begin{equation}\begin{array}{l}
 \ \ \ \ \    \delta_{3, max}(\Lambda) =\displaystyle{\frac{6\Lambda}{ \gamma_0+\rho_0+\Lambda|a_0| +6d_0}};\qquad
 \delta_{3, min}(\Lambda) = \displaystyle{{6\Lambda\over 1+9\Lambda(1+6\Lambda^2)}}\\
\mbox{and}   \ \  \forall   n \geq 5 \\
 \delta_{n, max}(\Lambda) = \displaystyle{\frac{3\Lambda  n(n-1)}{ \gamma_0+\rho_0+\Lambda|a_0| +n(n-1)d_0}}\\
 \mbox{with:}\\
  \gamma_0=1,\ \quad  a_0 = -\delta_{3,min}[N_{3}\tilde ]_{q^2+m^2=0};\quad 
	 \rho_{0}= \Lambda\delta_{3,min}\lbrack \displaystyle{\frac{\partial}{\partial q^{2}}[N_{3}\tilde ]\rbrack_{q^2+m^2=0}}\\
 \mbox{and}\\ 
\delta_{n, min}(\Lambda ) =\displaystyle{{3\Lambda\ n(n-1)\over \gamma_{max}+\rho_{max}+\Lambda|a_{max}| +3\Lambda  n\ (n-1)} 
 }\\
\mbox{with}\\
\gamma_{max}=1+9\Lambda (1+6\Lambda^2),\quad \rho_{max}=6\Lambda^2 \displaystyle{\frac{\partial}{\partial q^{2}}}[N_{3}\tilde ]_{q^2+m^2=0}\ \ \ \mbox{and} \\
|a_{max}|=  6\Lambda[N_{3}\tilde ]_{q^2+m^2=0}\\
\end{array}
\label{2.26}
\end{equation}

\begin{remarks}\

\begin{itemize}
\item{} \
Notice that  for the constant $d_0$ appearing in the definition of $\delta_{n, max}$ we put $d_0=3\Lambda 10^{-2}$ that is precisely the value we determined and used  in  \cite{MM6} for the zero dimensional case. 
\item{} For the $\delta_{n,min}$'s  in 4-dimensions we need to use the maximal values of the renormalization constants obtained directly by the definitions \ref{1.6}.
 \item{}Concerning the $\delta_{n,max}$'s in  4-dimensions we need to use the minimal values of the renormalization constants $\gamma_0$, $\rho_0$, and $a_0$ (cf. def.\ref{def.2.5}).
\end{itemize}
\end{remarks}
\item{}
 \textbf{The solution of the zero dimensional mapping ${\cal M}^*_0$}.
 
\emph{In \cite{MM6} we proved the existence and uniqueness
of the splitting sequence $\{\delta_{n0}\}$-solution of the zero dimensional mapping  ${\mathcal M}^*_{0}$ defined as 
 follows:} 
 ${\cal M^*}:\, \Phi_0 \stackrel{\cal M^*}
\longrightarrow
 {\cal B}_0 $:
\begin{equation}\begin{array}{l}
H^{2'} (\Lambda) = 1-\Lambda H^4 (\Lambda)\\
H^{4'} (\Lambda) = - \delta_3'(\Lambda)[H^{2'}]^3\quad
\hbox{with}\quad \displaystyle{\delta_3^{'}(\Lambda)= \frac{6\Lambda}{1+\ D_3}}\\
\mbox{and}\quad D_3=6\Lambda
H^2\left(3/2
 - \frac{|H^6|}{6|H^4||H^2|}\right);\\
\mbox{moreover:}\\
\qquad
\forall\  n\geq 5 \qquad
    H^{n+1'} (\Lambda) = \displaystyle{ \delta_n^{'}(\Lambda)  C^{n+1'}\over \displaystyle
 3\Lambda n(n-1)}\\
\mbox{Here:} \qquad \delta_n^{'}(\Lambda)=\displaystyle{\frac{3\Lambda n(n-1)}
 {1+D_n(H)}}\\
 \mbox{ with:}\quad
D_n(H)=\displaystyle{{|B^{n+1}| - |A^{n+1}| \over |H^{n+1}|}}
\end{array} 
\label{2.27}
\end{equation}
\end{enumerate}
 \label{def. 2.4}
 \end{definition}

\subsubsection{The ``fundamental tree type sequence'' $H_{T0}$}

 For further purposes, we introduce the particular tree 
type sequence $H_{T0}$   that 
we shall call ``fundamental'' defined as follows:
 \begin{definition}\label{def.2.5} 
 \begin{equation}\begin{array}{l}
	 \gamma_0=1\\	
a_0 = -\delta_{3,min}[N_{3}\tilde ]_{(q^2+m^2) =0};\\	
\ \ \\ 
	 \rho_{0}= \Lambda\delta_{3,min}\lbrack \displaystyle{\frac{\partial}{\partial q^{2}}[N_{3}\tilde ]\rbrack_{(q^2+m^2) =0}} \\
	 \ \ \\
\mbox{(Reminder:	$\quad\delta_{3,min}= \displaystyle{\frac{6\Lambda}{1+9\Lambda(1+6\Lambda^2)}}$)}\\	
 \forall(q,\Lambda)\in(\mathcal{E}_{(q)}^{4}\times\R^{+*}) \\
	H_{T0}^{2} = (q^2+m^2)(1+\delta_{10}(q,\Lambda)\Delta_F )\\
	\mbox{with}\quad
	\delta_{10}(q,\Lambda)\Delta_F  =\displaystyle{\frac{-\rho_{0}+ \Lambda \delta_{3,min}([N_{3}\tilde] - [N_{3}\tilde ]_{(q^2+m^2) =0})\Delta_F } {1+\rho_{0}}}\\
 \forall(q,\Lambda)\in(\mathcal{E}_{(q)}^{12}\times\R^{+*})\quad\ 
    H_{T0}^{4} = -  \delta_{3,min}(\Lambda)\displaystyle{\prod_{l=1,2,3}H^{2}_{T0}(q_l)\Delta_{F}(q_{l})}\\
 \ \ \\   
 \mbox{and for every $n\geq 5$ and 
$\forall\ (q,\Lambda)\in(\mathcal{E}_{(q)}^{4n}\times\R^{+*})$}\\
H^{n+1}_{T0}(q,\Lambda) = \displaystyle{{\delta_{n, min}(\Lambda)
 C^{n+1 }_{T0}(q,\Lambda) \over 3\Lambda
 n (n-1)}};\\
\mbox{where}:\\
C^{n+1}_{T0}(\Lambda) = - 6\Lambda\displaystyle{\sum_{\varpi_n(I)}
 {n\ !\over i_{1}!i_{2}!i_{3}!\
 \sigma_{sym}(I)}
\prod_{l=1,2,3}H^{i_{l}+1}_{T0}\Delta_{F}(q_{i_{l}})}\\
 \mbox{and\  $\{\delta_{n,min}\}_{n\geq 3}$\ is the splitting sequence 
of definition \ref{def. 2.4}.}
\end{array}
\label{2.28}
\end{equation}
\end{definition}
 

\vfill\eject
\subsection{The   Renormalized $\Phi$-Convolutions }\ \

\begin{figure}[h]
\begin{center}
\hspace{10mm}
\includegraphics[width=14cm]{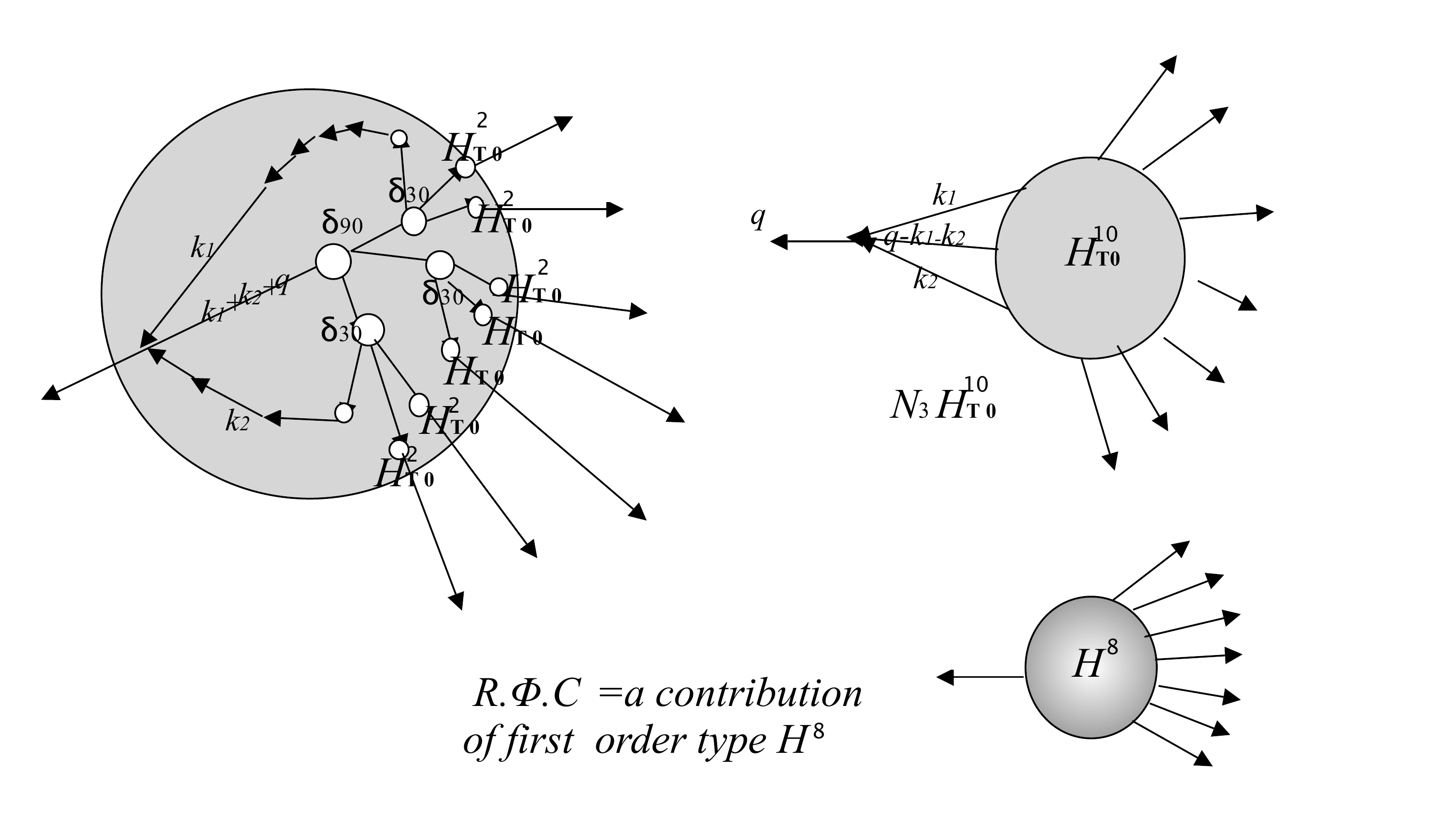}
\end{center} 
 \caption{\small\textrm{Graphical representation of a ``first order R. $\Phi$. C.'' - $H_{\nu=1}^8$-point function, coming from applying once  $\lbrack N^{(9)}_3\rbrack-\Phi_4^4$ operation, on the ``zero order'' tree contribution of the tree function $H_{T0}^{10}$ .}}
\label{fig.firstorderH8}
\end{figure}

\begin{figure}[h]
\begin{center}
\includegraphics[width=14cm]{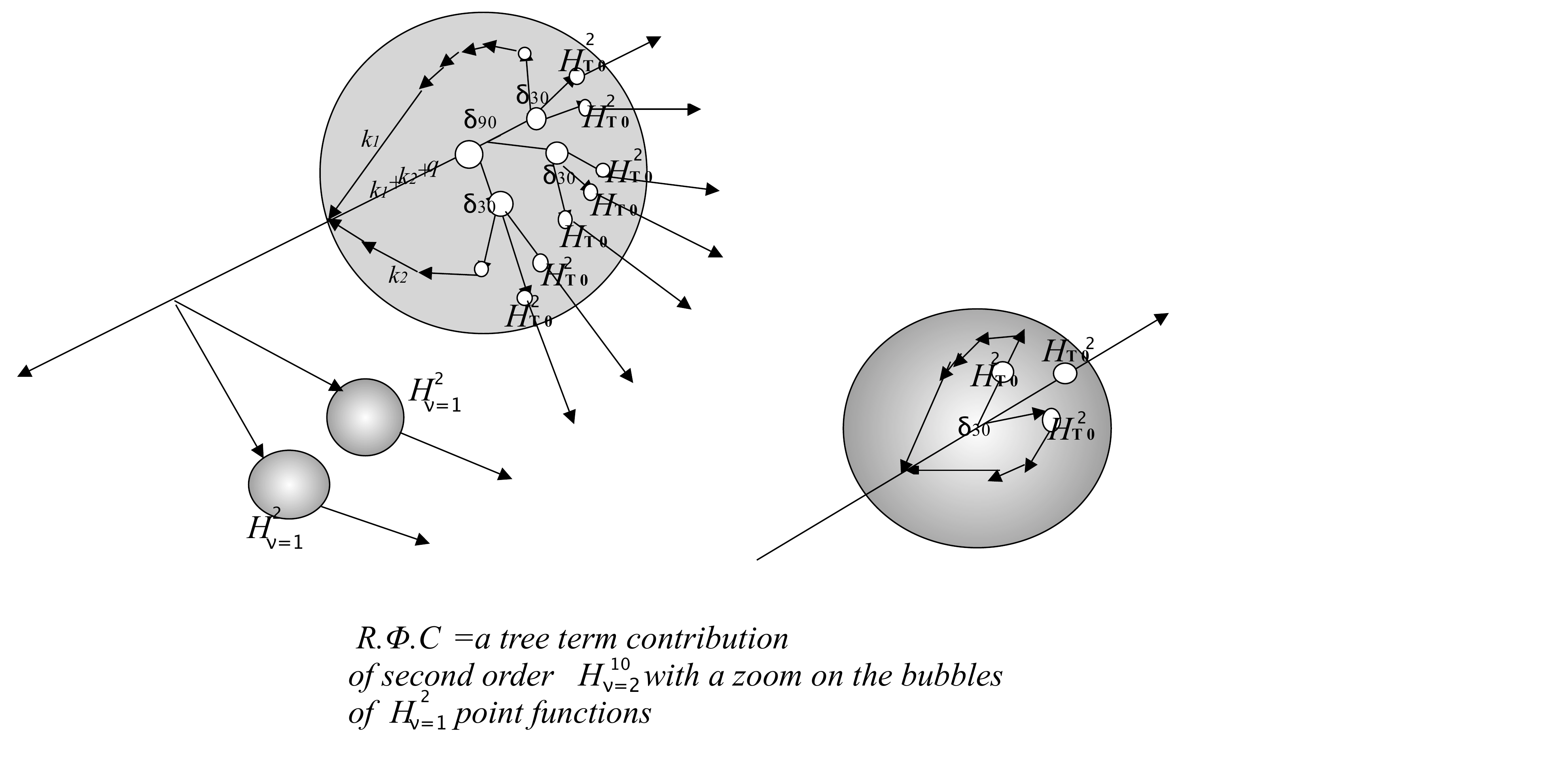}
\end{center} 
 \caption{\small\textrm{ Graphical representation of a tree type $\Phi^4$ operation of order $\nu=2 $  for the first term $C^{10}_{\nu=2}$ of  $H_{\nu=2}^10$-point function,   by using the bubble-vertex of figure  \ref{fig.firstorderH8}
 of $H_{\nu=1}^8$-point function and the bubble-vertices $H_{\nu=1}^2$-point functions (coming from a  $\Phi_4^4$ on $H_{T0}^4$-point function)}}
\label{fig.H102treecontrib.pdf}
\end{figure}

\begin{figure}[h]
\begin{center}
\includegraphics[width=14cm]{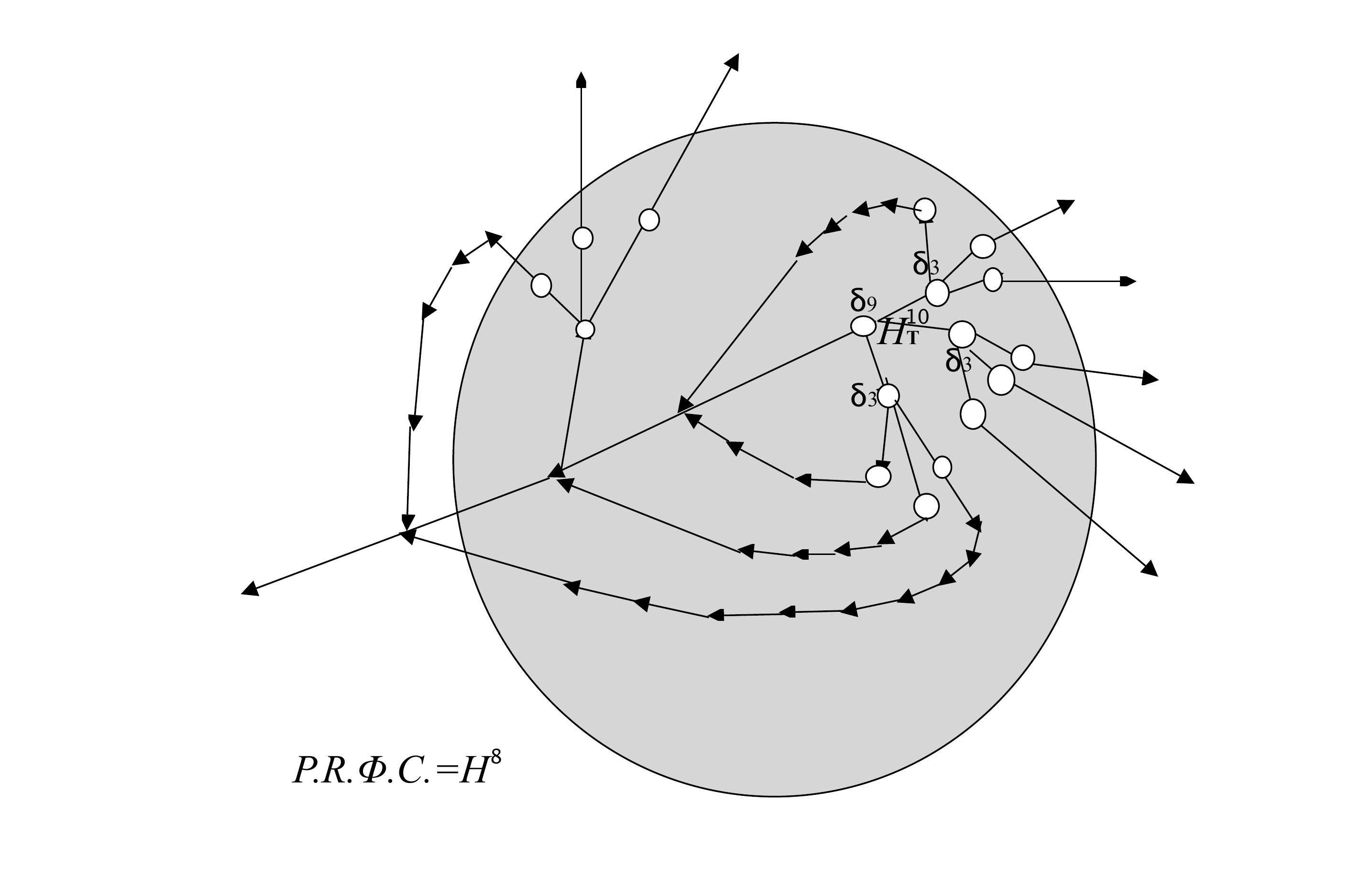}
\end{center} 
 \caption{\small\textrm{ Graphical representation of $H^8$-type  point bubble-function (third order $H_{\nu=3}^8$-point function  R. $\Phi$. C.), obtained after three successive applications of $\Phi_4^4$ operations on the ``zero order'' tree contribution of the tree function $H_{T0}^{10}$. It could constitute a bubble vertex of a tree contribution $H^8H^2H^2$ of third order R. $\Phi$. C. $H_{T0}^{10}$. Here $L=5$ and $\mathcal{L}=12$ so that 20-24=-4 is the asymptotic Weinberg  indicatrix of an $H^8$ point function}}
\label{fig.RFC8-point}
\end{figure}

\begin{definition}\ \label{def.2.6}

Using the previously defined fundamental tree type sequence $H_{T0}$ we recurrently construct 
the infinite family of the so-called  Renormalized $\Phi$-convolutions as follows: 

 We successively apply in an arbitrary way the   $\Phi^4_4$-operations (defining the mapping  $\mathcal {M}$ by  definition \ref{def.1.1});  (cf. example fig.\ref{fig.firstorderH8} and  fig \ref{fig.RFC8-point}).

At a certain order $\nu$ of this iteration we consider the corresponding result from an arbitrary $ H_{T0}^{ n+1}$ tree function. Graphically it is the sum of tree  type functions with``bubbles'', the corresponding images coming from successive applications of the $\Phi^4_4$ operations on the vertices of every tree contribution of the original $H_{T0}^{ n+1}$ tree function. 

We denote by $\Phi_{n}^{\bar n(\nu)}(q,\Lambda)$  such a bubble (with $\bar n(\nu)+1$ external lines), and call it the \textbf{Renormalized $\Phi$-Convolution  (R.$\Phi$.C) associated with the $ H_{T0}^{ n+1}$ tree-type function}.
 
 Every  $\Phi_{n}^{\bar n(\nu)}(q,\Lambda)$, depends on the set 
$q=(q_{1},\cdots, q_{\bar n(\nu)}) (q\in\mathcal{E}_{(q)}^{4\bar n(\nu)}$) of the (remaining after the integrations)  external independent 
momenta. It constitutes a candidate bubble vertex for   new tree type sequence in $\cal B$
 (cf. figures\  \ref{fig.H102treecontrib.pdf}  and\ \ref{fig.RFC8-point}).

 Using the general prescription of renormalization of \cite{Br.MM} we introduce the 
renormalization operator at every step of the above recursive 
construction. More precisely suppose that $\Phi_{n}^{\bar n(\nu)}(q,\Lambda)$ has been already 
well defined and we want to construct the newly composed convolution 
$[N_{3}^{(\bar n(\nu)+2)}\Phi_{n}^{\bar n(\nu)}(q,\Lambda)]$. We define: 
\begin{equation}
	[N_{3}^{(\bar n(\nu)+2)}\Phi_{n}^{\bar n(\nu)}(q,\Lambda)]\equiv
	\int R_{G}^{(3)}[\Phi_{n}^{\bar n(\nu)}(q,\Lambda)
	\prod_{k=1,2,3}\Delta_{F}(\ell_{k})]\ d^{4L}k
\label{2.29}
\end{equation}
\end{definition}

Here $G$ is the total graph representing the  R.$\Phi$.C.
$[N_{3}^{(n'=\bar n(\nu)+2)}\Phi_{n'}^{\bar n}]$ and $R_{G}^{(3)}$ the renormalization operator. 
 For the precise momentum assignment (following \cite{Br.MM}) we consider the 
product vector space defined by:
$\mathcal{E}_{(q,k)}^{4N}\equiv \mathcal{E}_{(q)}^{4n}\otimes E_{(k)}^{4L}$ 
with $N=n+L$. We associate these notations with the set of external 
independent $(q)$ (resp. internal 
$(k)$ or integration) variables of the given   R.$\Phi$.C. 
$[N_{3}^{(n')}\Phi_{n'}^{\bar n}]$. The integer $L$ indicates the number of independent 
loops of $G$ (i.e. the integration variables $k\in E_{(k)}^{4L}$ of the
 R.$\Phi$.C.). We also use the notation $\mathcal{L}$ for the set of all 
internal lines of $G$. The non renormalized integrand $I_{G}(q,k)$, is simply 
the product of the vertex functions (bubble vertices) and free propagators (simple 
internal lines) involved in the initial R.$\Phi$.C, $\Phi_{n'}^{\bar n}$ 
and the product of free propagators associated with $N_{3}^{(n')}$ so,
\begin{equation}
	I_{G}(q,k)=\prod_{v_{j}}H_{T}^{n_{j}+1}
	\prod_{i\in\mathcal{L}}\Delta_{F}(\ell_{i})
	\prod_{k=1,2,3}\Delta_{F}(\ell_{k}).
	\label{2.30}
\end{equation}
 The argument $\ell_{i}$, of every free propagator means the total momentum 
carried by the corresponding internal line $i\in\mathcal{L}$ associated with 
the linear application $\lambda_{i}$: 
\begin{equation}
	(q,k)\stackrel{\lambda_{i}}{\leftrightarrow}\ell_{i}(q,k)
	\label{2.31}  
	\end{equation}
 The precise form of the function $\ell_{i}(q,k)$ is given by the conditions of 
energy momentum conservation imposed on the momentum 
assignement at every vertex of $G$. A definition analogous to that of  
 \ref{2.30}  holds for the non 
renormalized integrand associated with every subgraph $\gamma$ of $G$. 

 Following \cite{Br.MM}, the abbreviated notation for the renormalized integrand 
means precisely:
\begin{equation}
	R_{G}^{(3)}\left[
	\Phi_{n'}^{\bar n}\prod_{k}\Delta_{F}(\ell_{k})\right]=
	\sum_{U_{\mathcal{F}(G)}}(1-t^{d(G)})Y_{G}^{(3)}(U_{\mathcal{F}})
	\label{2.32} 
\end{equation}

 The sum $\displaystyle{\sum_{U_{\mathcal{F}(G)}}}$ extends over all complete forests 
$U_{\mathcal{F}}$ of $G$ (with respect to a nested set $\mathcal{S}$ 
of subspaces $S\subset E_{(k)}^{4L}$ containing nontrivial renormalization 
parts $\gamma$ subgraphs of $G$. The functions 
$Y_{G}^{(3)}(U_{\mathcal{F}})$ (and the corresponding,
$Y_{\gamma}^{(3)}(U_{\mathcal{F}})$
for every $\gamma\in U_{\mathcal{F}}$) are also recursively defined in the
ref.\cite{Br.MM}. Notice that the degree of Taylor operators $d(G)$ (resp. 
$d(\gamma)$) coincides with the superficial degree of divergence of the graph 
$G$ (resp. of $\gamma$) and below (cf. proposition 2.1 ) we shall give precise 
upper bounds for these numbers in terms of the asymptotic indices of the tree 
functions. In an analogous way we define the renormalized $\Phi^{4}$ operation 
$[N_{2}^{(n'-1)}\Phi_{n'}^{\bar n}(q,\Lambda)]$. 

 For the convergence proof of the  R.$\Phi$.C's we shall use Weinberg's 
criterion of convergence applied to a certain class of Weinberg 
functions.

 A class of Weinberg functions (cf.\cite{Wein}, \cite{Br.MM})
$f\in C^\infty(\mathcal{E}_{(q)}^{4n})$
is denoted by $\mathcal{A}_{4n}^{(\alpha,\beta)}$ and 
is characterized by two bounded real valued functions $\alpha(S),\ \beta(S)$, 
on the set of all linear subspaces $S\subset \mathcal{E}_{(q)}^{4n}$, which are 
called the "asymptotic indicatrices".

	 In \cite{Br.MM} particular classes of Weinberg functions have been introduced 
(the classes of admissible Weinberg functions) and they have played a 
fundamental role for the convergence proof and good asymptotic behavior 
of the renormalized convolutions by an extension of the\  B.P.H.Z \cite{Bog.Par} 
renormalization procedure. 

The study in \cite{Br.MM} concerned the most general 
convolutions (the so-called $G$-convolutions) in a  space of arbitrar dimensions
space. The corresponding graphs were defined by bubble vertices $v$ (resp. complete 
internal lines) associated with general $H^{n_{v}}$ point functions (resp. with 
general $H^{2}$-point functions) satisfying all appropriate A.Q.F.T. properties. 

Under the assumption that the general $H^{n_{v}},\ H^{2}$ - point 
functions belong to the classes of symbols of pseudodifferential operators 
$\sum_{4(n_{v}-1)}^{\mu_{v}},\ \sum_{4}^{\mu_{i}}$ 
it has been proven (cf. theorem 4.1 of \cite{Br.MM}) that for every such $G$-convolution, 
the corresponding renormalized integrand belongs to the class of Weinberg 
functions with the appropriate asymptotic indices. So the Weinberg's 
criterion of convergence is verified. 

	 Moreover in \cite{MM4} (resp. in \cite{Du}) the asymptotic behavior with respect to 
the powers of external momenta (resp. with respect to powers of logarithms 
also) has been established for the $G$-convolutions, by using the same 
prescription of renormalization. We shall use here the  
definition of projection of a subspace $S$ of $\mathcal{E}_{(q,k)}^{4N}$ onto 
$E_{(k)}^{4L}$ established in \cite{Br.MM} and \cite{Du} concerning the classes of symbols and the 
classes of admissible Weinberg functions which have been denoted by 
$\mathcal{A}_{4N}^{(\alpha,\beta,\sigma,\omega)}$.

 We do not give here the corresponding precise definitions; we simply 
notice that $(\sigma,\omega)$ is a couple of sets of such subspaces in 
$E_{(k)}^{4L}$ and $E_{(q,k)}^{4N}$ respectively, where under differentiation 
the corresponding asymptotic indicatrices $\alpha, \beta$ decrease. We also recall 
the notation $\pi(S)$ for the canonical projection of a subspace $S$ of 
$E_{(q,k)}^{4N}$ onto $E_{(k)}^{4L}$. 

 Before giving the main theorem for the convergence of every  R.$\Phi$.C we show an 
auxiliary statement. Let us start with some useful notations and definitions. 

 \begin{definition} \
  We denote by $q_{T}^j$ the set 
of independent momenta of the tree function $H_{T}^{n_{j}+1}$ associated with 
the vertex $v_{j}$ of a given graph $G$. We define the following linear mapping 
\begin{equation}
	\lambda_{\nu_{j}}:\qquad(q,k)\stackrel{\lambda_{\nu_{j}}}{\leftrightarrow}
	q_{T}^j(q,k)
	\label{2.33} 
\end{equation}
\end{definition}

 We notice that, as previously, the function $q_{T}^j(q,k)$ is 
precisely defined by the momentum assignment, following the prescription of \cite{Br.MM}, 
when all constraints of energy momentum conservation at each vertex 
$\nu_{j}$ of $G$ are taken into account. We then state the following: 

\begin{proposition}\ \

 Given a  R.$\Phi$.C and the associated $G$-graph the following properties 
hold:
\begin{description}
\item{a)}	Every vertex function $H^{n_{j}+1}$ associated with the bubble vertex 
$v_{j}$ of $G$, belongs to the class 
$\mathcal{A}_{4N}^{(\alpha_{v_{j}}\beta_{v_{j}},\sigma_{v_{j}},\omega_{v_{j}})}$
of admissible Weinberg functions. The sets $\sigma_{v_{j}},\ \omega_{v_{j}}$  
are defined by:
\begin{equation}
	\sigma_{v_{j}}=\{
	S\subset E_{(k)}^{4L}:\quad S\not\subset 
	\mathcal{K}er\lambda_{v_{j}}\}\ ;
	\label{2.34} 
\end{equation}
\begin{equation}
	\omega_{v_{j}}=\{
	S\subset E_{(k)}^{4N}:\quad \pi(S)\in\sigma_{v_{j}}\}.
\label{2.35} 
\end{equation}
For every $S \subset\mathcal{E}_{(q,k)}^{4N}$ the corresponding asymptotic 
indicatrices are given by:
\begin{equation}
	\alpha_{v_{j}}(S)=\left\{
	\begin{array}{rl}
		-(n_{j}   -3),  & \mbox{if } \ S\not\subset\mathcal{K}er\ \lambda_{v_{j}}  \\
		0 & \mbox{if }\ S\subset\mathcal{K}er\ \lambda_{v_{j}} 
	\end{array}
	\right\}
	\label{2.36} 
\end{equation}
\begin{equation}
	\beta_{v_{j}}(S)=\left\{
	\begin{array}{l}
		2\nu_{(n_{j})}  \  \mbox{if} \ S\not\subset\mathcal{K}er\ \lambda_{v_{j}}  \\
		0 \ \   \mbox{if }\  S\subset\mathcal{K}er\ \lambda_{v_{j}}\\
		\mbox{ and $\nu_{(n_{j})}\in\N$    arbitrary large  depending on  $n_j $}
	\end{array}
	\right\}
	\label{2.37} 
\end{equation}
\item{b)}
The non renormalized integrand $I_{G}(q,k)$ associated with $G$ (cf. equation \ref{2.30}  of definition \ref{def.2.6}) 
belongs to the class of admissible  Weinberg  functions 

$\mathcal{A}_{4N}^{(\alpha_{G},\beta_{G},\sigma_{G},\omega_{G})}$
characterized by the following sets and indices:
\begin{equation}
	\sigma_{G}=\displaystyle{\bigcap_{i\in\mathcal{L}}\sigma_{i}\bigcap_{v_j}\sigma_{v_{j}}} 
		\label{2.38}
\end{equation}
\begin{equation}
	\omega_{G}=\{
	S\subset\mathcal{S}_{(q,k)}^{4N}:\ 
	S\not\subset\mathcal{K}er\ \lambda_{i}\quad 
	\forall\;i\in\mathcal{L},\ \pi(S)\in\sigma_{G}
	\}
\label{2.39}
\end{equation}
 and for every $S\subset\mathcal{E}_{(q,k)}^{4N}$~: 
\begin{equation}
	\alpha_{G}(S)=\sum_{v_{j}}\alpha_{v_{j}}(S)+
	\sum_{i\in\mathcal{L}}\mu_{i}(S)
	\label{2.40}
\end{equation}
 Here  $\forall\; i\in\mathcal{L}$,
\begin{equation}
	\mu_{i}(S)=\{
	-2 \mbox{ if } S\not\subset\mathcal{K}er\ \lambda_{i}\; \ \mbox{ and }
	0 \mbox{ if }S\subset\mathcal{K}er\ \lambda_{i}
	\}
		\label{2.41}
\end{equation}
\emph{and}
\begin{equation}
	\beta_{G}(S)=\sum_{v_{j}}\beta_{v_{j}}(S)
		\label{2.42}
\end{equation}

\item{c)} Analogous result holds for the non renormalized integrand associated with 
every subgraph $\gamma$ of $G$. 
\item{d)}
 The degree $d(G)$ (resp. $d(\gamma)$) of the Taylor operators 
associated with $G$ (resp. with $\gamma$) in formulas
(\ref{2.45}) is bounded as follows:
\begin{equation}
	d(G)\leq 2\ ;\mbox{ (resp. }d(\gamma)\leq 2).
		\label{2.43}
\end{equation}
\end{description}
\label{prop.2.3}
\end{proposition}

\textbf{Proof of proposition} \ref{prop.2.3}
 
The proof is obtained by application of the previous definitions (in 
particular  the definitions of the admissible classes of Weinberg ) and is a direct consequence of \cite{Br.MM} and  \cite{Du}. 
Notice that for every $n\geq3$ 
 the function $\delta_{n}(q,\Lambda)$ belongs to the class of Weinberg functions 
$\mathcal{A}_{4N}^{(0, 0,\sigma_{\nu j},\omega_{\nu j})}$.

We use the notation 
$\vert\mathcal{L}\vert$ for the total number of internal lines of
$G$ we have~:
\begin{equation}
	d(G)=-2\vert\mathcal{L}\vert+\max_{S}
	\sum_{\nu_{j}}\alpha_{v_{j}}(S)+4L\leq 2
	\      \quad \blacksquare 
		\label{2.44}
\end{equation} 
Taking into account the above results we notice that the conditions for the 
non renormalized integrand of $G$ established in \cite{Br.MM} are all verified. 
So we are allowed to apply directly the corresponding theorems of 
refs. \cite{Br.MM}, \cite{MM4} and \cite{Du} in order to obtain  the following result 
that we present without proof~:

\begin{axiom}\ \label{th.2.1}\

Every R.$\Phi$.C. $\Phi_{n}^{\bar n}(q,\Lambda),\ (q\in\mathcal{E}_{(q)}^{4n}$) with $n$ external independent  variables and $L$ integration 
variables $k\in E_{(k)}^{4L}$,
 (with $\Lambda$ a fixed real 
positive number) verifies the following properties:
\begin{itemize}
	\item  [(i)]
Defined as an integral of $k\in E_{(k)}^{4L}$, it is absolutely convergent and it 
belongs to the class $\mathcal{A}_{4n}^{(\alpha_{n},\beta_{n})}$
of Weinberg functions with the following precise asymptotic indicatrices:
\begin{equation}\begin{array}{l}
	\forall S \subset\mathcal{E}_{(q)}^{4n} :\\
	\alpha_{n}(S)=d(G)\quad ;\quad \beta_{n}(S)=\tilde\nu_{(n)}\\
	\mbox{with $\tilde\nu$ an arbitrary big natural number depending on n }
	\end{array}
	\label{2.45}
\end{equation}
	\item  [(ii)]
It satisfies Euclidean invariance and all linear axiomatic field 
theory properties of a general $n+1$-point function in complex Minkowski space.	
\end{itemize}
\end{axiom}

\vfill\eject
\subsection{ The Generalized Renormalized $\Phi$- 
 Convolution (G.R.$\Phi$.C.)} 
   
\begin{definition}\

A Generalized R.$\Phi$.C. (G.R.$\Phi$.C.) is defined as the image of different $\Phi_4^4$ operations on an arbitrary finite sum of R.$\Phi$.C's. 
The corresponding to a G.R.$\Phi$.C,  
$G_{\Phi}$ graph contains sums of disconnected graphs associated to each one of the connected components R.$\Phi$.C's.

	The renormalization operator 
$R_{G}$ corresponding to the operations $N_{3}^{(n)}$ and $N_{2}^{(n-1)}$ 
respectively is defined as a consistent extension of the scheme presented
previously for the R.$\Phi$.C's. 
 More precisely, the renormalized integrand (analog 
of formula \ref{2.32} ) corresponding to the convolution 
$[N_{3}^{(r)}\Phi_{(r)}^{(n)}]$ reads: 
\begin{equation}
	R_{G_{\Phi}}^{(3)}[\Phi_{(r)}^{(n)}(H)
	\prod_{k}\Delta_{F}(\ell_{k})]=
    \sum_{U_{\mathcal{F}}(G_{\Phi}(H))}
	(1-t^{d(G_{\Phi})})Y_{G_{\Phi}}^{(3)}(U_{\mathcal{F}}(H))
	\label{2.46}
\end{equation}
	 Here, the argument $H$ means that the summation is multiple because the fine 
structure of the corresponding $H\in\mathcal{B}$ must be taken into account and 
every $H^{n_{j}+1}$-function associated with the bubble vertex $v_{j}$ of 
$G_{\Phi}$, \emph{must be expanded in its terms  R.$\Phi$.C's} (cf. def.\ref{def.2.6}). 
\label{def.2.10}
\end{definition}
	 In other words, the total graph $G_{\Phi}(H)$ is, in fact, a sum of 
disconnected graphs $G^{*}$ coming from the expansion of all bubble vertices in 
their \emph{disconnected components} graphs $G$ associated with the 
R.$\Phi$.C's involved in the definition of the corresponding $H^{n_{j}+1}$'s. 
Therefore the sum in \ref{2.46} contains all possible nontrivial individual 
forests of every such component $G^{*}$. 
  Notice that now, the associated vertex functions are not in
 general tree functions. For simplicity we keep the same mode of notation 
 for the number $n$ of the initial function $H^{n+1}$ (superscript) and 
 respectively the number $r$ of external independent variables
 (subscript) as in the definition \ref{def.2.6} of the R.$\Phi$.C's.

\subsection{The Banach space $\mathcal{B}_{R}\subset \mathcal{B}$}
 
 \begin{definition} \label{def.2.11}

We say that a sequence  $H\in \mathcal{B}$ belongs to the linear subspace $\mathcal{B}_{R}\subset \mathcal{B}$,
if \  $\forall\, n=2k+1,\ k\in\N$ the corresponding $H^{n+1}$ function is a  (G.R.$\Phi$.C.) 
in the sense of definition \ref{def.2.10}.
\end{definition}

\vfill\eject
\begin{definition} [$\mathcal{B}_{R}\subset \mathcal{B}$  a Banach space]\label{def.2.10}\

 We introduce the 
following positive mapping on  $\mathcal{B}_{R}$   
$$\mathcal{N}:\ \ \mathcal{B}_{R}\to\R^{+}$$

\noindent $H\mapsto\Vert H\Vert$ \emph{Here}
\begin{equation}
	\begin{array}{l}
	H\mapsto\Vert H\Vert  \ \ \ \mbox{where}\ \ \ 
		\Vert H\Vert=\displaystyle{\sup_{\Lambda}}
		\left\{
		\displaystyle{\sup_{n, q}}\left\{\displaystyle{\frac {\vert H^{n+1}\vert}{ M_n}};\displaystyle{\frac{\vert \frac{\partial^{(0,1)}}{\partial q^{2}}\vert N_3H^{4}\vert}{\hat M_3^{(0,1)}}};\displaystyle{\frac{\vert N_2H^{n+1}\vert }{\hat M_{(n,2)}};\frac{|\gamma|}{{\cal  N}_{\gamma}}}\right\}\right\}
    \ \ \\	
    		\mbox{where:} \quad 
	\forall\ (q,\Lambda)\in\mathcal{E}_{(q)}^{4n}\times\R^{+*}\\
    M_{1}(q)= \gamma_{max} ( q^{2}+m^2)(1+6(q^2+m^2)^{\pi^2/54})\\
   {\cal  N}_{\gamma}= \gamma_{max}\displaystyle{\prod_{i=1}^{3}M_{1}(q_{i})\Delta_{F}(q_i)|_{q^{2}=0}}\\
    \gamma_{max}=1+9\Lambda (1+6\Lambda^2)\\
      M_3(q)= 6\Lambda\displaystyle{\prod_{i=1}^{3}M_{1}(q_{i})\Delta_{F}(q_i) }\\
         \hat M_3^{(0,1)}(\Lambda)= 6\Lambda\ \displaystyle{\sup_{ q}}\left[\vert \displaystyle{\frac{\partial}{\partial q^{2}}}[N_{3}\tilde I_{1,3}]\vert M_1; \ \vert [N_{3}\tilde I_{1,3}]\vert M_1 \right]\ \ \\
         \mbox{with} \quad\tilde I_{1,3}=\displaystyle{\prod_{i=1}^{2}M_{1}(k_{i}) [\Delta_{F}(k_i)]^2\Delta_F(k_1+k_2+q) } \\
         \mbox{(cf. also \ref{2.30} , \ref{2.32})     \ \ and figures\  
          \ref{fig.H102treecontrib.pdf} )}\\     
             \mbox{by analogy}\quad
         \hat M_{(3,2)}= 6\Lambda[N_2^{(3)} \tilde I_{1,2}]\displaystyle{
	\prod_{i=2}^{3}M_{1}(q_{i})}\Delta_{F}(q_i) \\
	\mbox{with} \quad\tilde I_{1,2}=M_{1}(k)[\Delta_{F}(k)]^2 \Delta_F(k+q) \\
\mbox{ and for every $n\geq 5$  }  \\ 
	\hat M_{(n,2)}=[N_2^{(n)}\tilde]_{q=0}n(n-1)\delta_{n, max}(\Lambda)M_{n-2} M_1(q_{{1}} )\Delta_F(q_{{1}} )\\ 
		 M_{n}= n(n-1)\delta_{n, max}(\Lambda)M_{n-2}(q_{(n-2)},\Lambda)\Delta_{F}(q_{(n-2)})\displaystyle{
	\prod_{i=2}^{3}M_{1}(q_{i})}\Delta_{F}(q_i) 
	\end{array}
	\label{2.47}
\end{equation}
\end{definition}
Now, one easily verifies that $\mathcal{N}$ defines a finite norm on 
	$\mathcal{B}_R$, and that $\mathcal{B}_R$ is a complete metric space with respect  to the induced distance (of uniform convergence) so the following is established:
\begin{proposition}\ \
	$\mathcal{B}_R$ is a 
 Banach space with respect to the distance associated with the norm $\mathcal{N}$ of definition \ref{def.2.10}.
\end{proposition}

We cconclude this section with a crucial result for the subsequent 
sections. It ensures the good convergence, asymptotic behavior, 
Euclidean and linear A.Q.F.T. (in complex Minkowski space) 
properties of the G.R.$\Phi$.C's - $H\in \mathcal{B}_{R}$. 
\vspace{2mm}

\begin{axiom}\ \ \label{th.2.2}
	 	
The system of equations  presented in.def.\ref{def.1.1} of the introduction  constitutes a well 
defined non linear mapping 
$\mathcal{M}\ 
:\mathcal{B}_{R}\stackrel{\mathcal{M}}{\rightarrow}\mathcal{B}_{R}$ in 
the following sense~:		
	
	\begin{itemize}
		\item  [a)]
    For every $n$, the good convergence of integrals, asymptotic behaviour, 
	symmetry and Euclidean invariance of $H^{n+1}$ (G.R.$\Phi$.C's),\  $\forall 
	q\in\mathcal{E}_{(q)}^{4n}$, are preserved by $\mathcal{M}(H)$.
		\item  [b)]
	For every $n$, the corresponding Green's function $H^{n+1'}(q)$ (the image under $\mathcal{M}$  of $H^{n+1}(q)$), verifies 
	the analyticity (primitive domain) and algebraic A.Q.F.T. (\cite{Br.Ru.Ar}) 
	properties in complex Minkowski space.
		\item  [c)]
	The G.R.$\Phi.C's -H^2$ functions, which depend on only one external variable 
	$q^{2}$, 
	$\Phi_{(r)}^{(n)}(H)$ satisfy and conserve under the action of 
	$\mathcal{M}$ the real analyticity character for every 
	$q\in\mathcal{E}_{(q)}^{4}$ and at $q^{2}+m^{2}=0$. The same property holds 
	for all order derivatives of $\Phi_{(r)}^{(n)}(H)$.
	\end{itemize}
		\end{axiom}	

\textbf{Proof of theorem \ref{th.2.2}}
\begin{itemize}
	\item[a)] The proof follows directly by 
application of the previous definitions, of  proposition \ref{prop.2.3}, and of theorem \ref{th.2.1}  (the latter 
being applied to every graph $G^{*}$ component of $G_{\Phi}(H)$). 
\item[b)] The verification of axiomatic field theory properties 
are obtained as a trivial application of a most general result of  
 \cite{MDu} concerning $\Phi^4$ type renormalized convolutions. 

The Euclidean invariance and 
symmetry of every R.$\Phi$.C,  $\Phi_{(r)}^{(n)}(H)$ can be verified in another more direct way. 
Following the recursive construction presented in definition \ref{def.2.6} for the
 R.$\Phi$.C's and by choosing an appropriate coordinate transformation 
(spherical coordinates in four dimensions) we can eliminate by integration all angular 
dependence. Then, the limit of the total multiple integration at 
$q^{2}+m^{2}=0$ is (at fixed $\Lambda$) a real finite positive number. The analogous 
results hold for every order derivatives, with respect to $q^{2}$ (at the 
point $q^{2}+m^{2}=0$)
The real analyticity property  comes from the 
fact that the primitive domain of analyticity of every $\Phi_{(r)}^{(n)}(H)$ contains the corresponding
Euclidean region. 
\item[c)]
The  proof is a direct consequence of the properties a) and b).
\end{itemize}

 \section{The subset $\Phi_{R}$ -The  mapping ${\cal M}^*$ - The  $\Phi_4^4 $ iteration}
 \subsection{The subset $\Phi_{R}\subset {\cal B}_{R}$}
 In this subsection we  describe the subset $\Phi_R\subset 
	\mathcal{B}_R$ which is 
characterized by the splitting and sign properties (tree-structure), together with the 
physical conditions implemented by the renormalization (which is 
associated with the four-or three-dimensional problem). The "splitting" or 
factorization properties are the analogs of the properties displayed by 
the $\Phi$-subset defined previously in the case of the zero-dimensional 
problem of \cite{MM6}. As it should become evident, apart from the 
renormalization constraints, the structure of $\Phi_R$ given here can entirely 
be applied to smaller dimensions $1\leq r\leq 3$, with non-zero external momenta.

	\begin{definition}{The subset $\Phi_R \subset{\cal B}_R $}\label{def.3.1}

We say that a sequence $H \in \mathcal{B}_R $ belongs to the 
subset $\Phi_R$, if the following properties are verified:
\begin{enumerate}
\item{} $\forall(q,\Lambda)\in(\mathcal{E}_{(q)}^{4}\times\R^{+*})$
\begin{equation}\begin{array}{l}
H^2 (q,\Lambda)= (q^2+m^2)(1+\delta_{1}(q,\Lambda)\Delta_F )\\
\mbox{with} \\
 \displaystyle{\lim_{(q^2+m^2)= 0}}\delta_{1}(q,\Lambda)\Delta_F(q)=0\ \ \ \mbox{or}  \displaystyle{\lim_{(q^2+m^2)= 0}} H^2 \Delta_F(q)=1\\
 \ \ \\
\mbox{and} \quad	 H^2_{min}(q)  \leq H^{2}(q,\Lambda) \leq H^{2}_ {(max)}(q,\Lambda) \\
\ \ \\
	 \mbox{with}\ \ \ H^{2}_ {(max)}(q,\Lambda)= \gamma_{max} (  ( q^{2}+m^2)+6\Lambda^2(q^2+m^2)^{\frac{\pi^2}{54}});\\
	   H^2_{min}(q)=q^2+m^2
	 \end{array}
	 \label{3.48}
  \end{equation}
\item{} 
For every  $n=2k+1, k\in\N^*$  the function $H^{n +1}$, belongs to the class 
$\mathcal{A}_{4n}^{(\alpha_{n}\beta_{n})}$
of Weinberg functions such that 
 $ \forall\  S\subset\mathcal{E}_{(q)}^{4n}$ the corresponding asymptotic 
indicatrices are given by:
\begin{equation}
	\alpha_{n}(S)=\left\{
	\begin{array}{l}
		-(n-3) \  \mbox{if } \ S\not\subset\mathcal{K}er\ \lambda_{n}  \\
		0 \ \  \mbox{if}\ S\subset\mathcal{K}er\ \lambda_{n}\\ 
	\beta_{n}(S)=n\beta_{1} \ \ \forall\  S\subset{\mathcal E}_{(q)}^{4n}\\
 \mbox{(with $\beta_{1}\in\N$  arbitrarily large)}
	\end{array}
	\right\}	
		\label{3.49}
\end{equation}
\item{} There is an increasing and bounded (with respect 
to $n$)  associated positive sequence  (cf. definition \ref{def.2.2}):
$\{\delta_{n}(q,\Lambda)\}_{n=2k+1,k\in\N^*}$, 
  of splitting functions $ \in  \mathcal{D}$ which belong to the 
class   $\mathcal{A}_{(n)}^{(0, 0)}$ of Weinberg functions for every $n\geq 3 $ such
that $H$ is a \textbf{tree type sequence} in the sense of definition \ref{def.2.3}. More precisely:
\begin{description} 	
 	\item{i)}\ \
$\forall(q,\Lambda)\in(\mathcal{E}_{(q)}^{12}\times\R^{+*}) $
\begin{equation}\begin{array}{l}
	 H^4 (q\Lambda) = -\delta_{3}(q,\Lambda)
	 \prod_{\ell=1,2,3}H^{2}(q_{\ell},\Lambda)\Delta_{F}(q_{\ell})\\ 
	  \mbox{ with}\  \delta_{3}(q,\Lambda)\build\sim_{q\rightarrow\infty}^{}\Lambda\\ 
	\mbox{  For every finite fixed }  \ \tilde q\in\mathcal{E}_{(q)}^{12}\ \ \ 
	 \displaystyle{\lim_{\Lambda\to 0} 
 	\frac{\delta_{3}(\tilde q,\Lambda)}{\Lambda}}=6\\
	\ \ \\
  \mbox{and}\ \ \forall\Lambda\in\R^{+*}\quad
	\delta_{3,min}(\Lambda) \leq \delta_{3}(\tilde q, \Lambda)\leq
	\delta_{3,max}(\Lambda)  
	\end{array}
	\label{3.50}
\end{equation}
 	\item {ii)}
	For every $n\geq 5$ and  \ $\forall 
 	(q,\Lambda)\in(\mathcal{E}_{(q)}^{4n}\times\R^{+*})$~:
\begin{equation}
	\begin{array}{l}
 	H^{n+1}(q,\Lambda)=\displaystyle{\frac{\delta_{n}(q,\Lambda)C^{n+1}(q,\Lambda)}
 	{3\Lambda n(n-1)}}\\
 \mbox{ with}\ \    \delta_{n}(q,\Lambda)\build\sim_{q\rightarrow\infty}^{}\Lambda\\
 \ \ \\
	\mbox{ For every finite fixed } \tilde q\in\mathcal{E}_{(q)}^{4n}\\
	\ \displaystyle{\lim_{\Lambda\to 0} 
 	\frac{\delta_{n}(\tilde q,\Lambda)}{\Lambda}\sim 3n(n-1)}\\
	\ \ \\
\mbox{and}\ \ \forall\Lambda\in\R^{+*},\quad
	\delta_{n,min}(\Lambda) \leq \delta_{n}(\tilde q, \Lambda)\leq
	\delta_{n,max} (\Lambda) 
	\end{array}
	\label{3.51}
\end{equation}
 Here  $\{\delta_{n,min}\}$, (but not $\{\delta_{n,max}\}$) (cf. equations \ref{2.26}) are the splitting sequences lower  bounds of the  \textbf{solution of the zero dimensional problem}  previously presented in definition \ref{def. 2.4}).
	\item {iii)}	
Moreover there is a finite number  $\delta_{\infty}\in\ \R^+$ a uniform bound 
independent of $H$ such that : 
\begin{equation}
	\displaystyle{\lim_{n\to \infty}\delta_{n}(\tilde q, \Lambda)}\leq\delta_{\infty}\ \  \ \forall\ \Lambda\in\R^{+*}
	\label{3.52}
\end{equation}
 \end{description}
\item {} 
The \emph{renormalization functions} $a,\rho$ and $\gamma$, appearing in the 
    definition of $\mathcal{M}$ are well defined real analytic functions of 
    $q^2$ and $\Lambda$, and yield at the limits $(q^{2}+m^{2})=0$\  and \ $q=0$  the physical conditions of renormalization  required by the two-point and four
    point functions: 
    \begin{equation}\begin{array}{l}  
    	a (q, \Lambda)= [N_{3}^{(3)}H^4(q, \Lambda)]  \ \ \mbox{and} \ \  \tilde a(\Lambda)=\displaystyle{\lim_{(q^2+m^2)= 0}}[N_{3}^{(3)}H^4(q, \Lambda)]\\ 
\mbox{with:}	\ a_{min}(\Lambda)\leq \tilde a(\Lambda) \leq  a_{max}(\Lambda)
	\end{array}
	\label{3.53}
\end{equation}
\begin{equation}\begin{array}{l}
   	\rho(q, \Lambda)= \left[
  	\displaystyle{\frac{\partial }{\partial q^{2}}}[N_{3}^{(3)}H^4(q\Lambda)]	\right], \ \ \mbox{and} \ \  \tilde \rho (\Lambda)=\displaystyle{\lim_{(q^2+m^2)= 0}} \rho(q, \Lambda) \\
	\mbox{with:}\ \ \  \rho_{min}(\Lambda)\leq \tilde \rho (\Lambda)\leq\rho_{max}(\Lambda)
	\end{array}
	\label{3.54}
\end{equation} 
\begin{equation}\begin{array}{l}
 \gamma(q,\Lambda)=\left[\displaystyle{\frac{-6\Lambda \prod_{l=1,2,3}H^{2}(q_l)\Delta_{F}(q_l)}{H^{4}(q)}}\right]\\  \ \\
\mbox{and}\ \tilde\gamma(\Lambda)  =
		\left[ \gamma(q,\Lambda)\right]_{q=0} \quad
	\mbox{with} \ \ \gamma_{min}(\Lambda)\leq \tilde\gamma(\Lambda)\leq\gamma_{max}(\Lambda)\\ 
	\end{array}  
	\label{3.55}
\end{equation}
\end{enumerate}
\end{definition}

\vspace{3mm}

\begin{remarks} \

\begin{enumerate}
\item{}
\emph{We first remark that the "splitting" or factorization properties ii) and iii) are
 general formulae which simply define the functions 
$\delta_{n}(q,\Lambda)$ and they can 
formally be written for every sequence $H$ of 
$\mathcal{B}_{R}$. \\
The particular character of the subset $\Phi_R$ comes 
from the fact that the splitting sequence  
$\{\delta_{n}\}\in\mathcal{D}$, is such that $\forall n=2k+1, k\geq 1$ the corresponding  splitting function $\delta_{n}(q,\Lambda)$
 belongs to the class $\mathcal{A}_{(n)}^{(0, 0)}$ of Weinberg functions and
verifies the limit and asymptotic properties of definition \ref{def.3.1}}. 
\item{}
\emph{We point out that  the symbol $\build\sim_{q\rightarrow\infty}^{}$ is  used  as
 an abbreviated notation of the fact that both sides  of the appropriate relations belong to the same 
class of Weinberg functions, or to put it differently, they have an 
asymptotically equivalent behavior.}
\end{enumerate}
\label{Rem.3.1}
  \end{remarks}


\subsection{The non triviality of the subset $\Phi_R$} 

\ \begin{axiom}\ \label{th.3.1}

The subset $\Phi_R$ is a nontrivial subset of $\mathcal{B}_{R}$

\end{axiom}

\noindent \textbf{Proof of theorem \ref{th.3.1}}\

  We consider the fundamental  
 sequence $H_{T0}$ (cf. definition \ref{def.2.5}) and verify successively all properties of $\Phi_R$. For details cf. the reader in Appendix \ref{APP.6.1}

 \vspace{3mm}

\subsection{The new mapping ${\cal M}^*$ on $\Phi_R  $ and equivalence with $\mathcal{M}$}

\begin{proposition}\label{prop.3.1}

Let  $H  \ \in \  \Phi_R$. The following mapping, \begin{equation}\begin{array}{l}
\mathcal{M}^{*}:\Phi_R\stackrel{\cal M^{*}}
 \longrightarrow {\mathcal{B}_R}\\
 \qquad \ H\ \ \mapsto \mathcal{M}^{*}(H)\\
  \end{array}
 \label{3.56}
  \end{equation}
defined by equations \ref{3.57}$\dots$\ref{3.60}, is equivalent to the mapping ${\cal M}$ (cf. equations  \ref{1.6} of the introduction).
  \begin{itemize}   
  \item [i)]
  \begin{equation}   \begin{array}{l}
		a'(q,\Lambda)  =\left[N_{3}^{(3)}H^{4}(q,\Lambda)\right];  \  \rho'(q,\Lambda)=-\Lambda
		\left[\displaystyle{\frac{\partial}{\partial q^{2}}[N_{3}^{(3)}H^{4}(q,\Lambda)]} \right]; \\
		\ \  \gamma'(q,\Lambda)  =
		\left[\displaystyle{\frac{-6\Lambda \prod_{l=1,2,3}H^{2}(q_l)\Delta_{F}(q_l)}{H^{4}(q)}}\right]\\
\mbox{Moreover we define:}\\
 \tilde a'(\Lambda)=a'(q,\Lambda)|_{(q^2+m^2=0)}; \ \ 	
\tilde \rho'(\Lambda)=\rho'(q,\Lambda)|_{(q^2+m^2=0)}\\  \mbox{and},\ \tilde \gamma'(\Lambda)=\gamma'(q,\Lambda)|_{q=0}
		\end{array} 
	\label{3.57}	 
\end{equation}
\item[ii)]
\begin{equation}\begin{array}{l}
	 H^{2'} (q,\Lambda)= (q^2+m^2)(1+\delta'_1(q,\Lambda)\Delta_F(q^2))\\
		\mbox{with:}\  \delta'_1(q,\Lambda)\Delta_F(q^2)=\displaystyle{\frac{- \tilde\rho-\Lambda\lbrace\lbrack
		  N^{(3)}_3H^4\rbrack -\tilde a H^2 (q,\Lambda)\Delta_F(q^2)\rbrace\Delta_F(q^2)}{ (\tilde\gamma+\tilde\rho)}}
		 \end{array} 
	\label{3.58}	 
\end{equation}
\		\item  [iii)]
		\begin{equation}\begin{array}{l}
			H^{4'}(q,\Lambda)=-\delta'_{3}(q,\Lambda)
			\prod_{\ell=1, 2, 3}H^{2'}(q_{i},\Lambda)\Delta_{F}(q_{i})\\
			 \mbox{with:}	\\ 	
			\delta_{3}'(q,\Lambda)=\displaystyle{\frac{6\Lambda}{(\tilde\gamma+\tilde\rho)+D_{3}(H)-
			\Lambda\tilde a }}\\
			\ \ \\
			\mbox{and} \quad D_{3}(H)= \displaystyle{\frac{ |B^{4}|- |A^{4}| }{\ |H^{4}|}}\end{array} 
			\label{3.59} 
		\end{equation}
		\item  [iv)]
	\emph{for every $n\geq 5$ :}
	\begin{equation}\begin{array}{l}		
			H^{{n+1}'}(q,\Lambda)=\displaystyle{\frac{\delta'_{n}(q,\Lambda)C^{{n+1}'}(q,\Lambda)}{3\Lambda n(n-1)}}\\			
\emph{with:}\ \ \ \delta'_{n}(q,\Lambda)=\displaystyle{\frac{3\Lambda n(n-1)}
				{(\tilde\gamma+\tilde\rho)+D_{n}(H)-\Lambda \tilde a}}
				\end{array}
				\label{3.60}
			\end{equation}
			and  $C^{{n+1}'}$ is obtained  recursively, in the usual way, from the sum of all the partitions of the products\ \  
			$$\prod_{l=1,2,3} \lbrack N^{(i_l)}_1H^{i_{l}+1}\rbrack (q_{i_{l}},\Lambda)\   \forall  \ i_l\leq n-2$$

 Notice that  in the denominators of equ. \ref{3.59}, we defined the 
function $D_{n}(H)$ by: 
\begin{equation}
	D_{n}(H)=\frac{ |B^{n+1}|- |A^{n+1|} }{ |H^{n+1}|}
	\label{3.61} 
\end{equation}
where, in view of the 
hypothesis $H\in \Phi_R$ (sign properties) we used the absolute values.
\end{itemize}
\end{proposition}

\emph{Proof of proposition \ref{prop.3.1}}

Taking into account the infinite system of equations \ref{1.6} of the 
introduction 
and the splitting or factorization properties ii) and iii) 
 in $\Phi_R$  (cf. also remarks \ref{Rem.3.1}), we write:
 \begin{itemize} 
 \item [i)]  
  \begin{equation}\begin{array}{l}
    H^{2' }=(q^2+m^2)(1+\delta'_1(q,\Lambda)\Delta_F(q^2)\\
    = - \displaystyle{{\Lambda\over{(\tilde\gamma+\tilde\rho)}}}
\lbrace\lbrack N^{(3)}_3H^4\rbrack
 -\tilde a
H^2  \Delta_F(q) \rbrace
+\displaystyle{\frac{ (q^2+m^2)\tilde\gamma}{(\tilde\gamma+\tilde\rho)}}\\
\mbox{or} \\
  (q^2+m^2) (\tilde\gamma+\tilde\rho) + (\tilde\gamma+\tilde\rho)\delta'_1 =-\Lambda \lbrace\lbrack N^{(3)}_3H^4\rbrack
 - \tilde a
H^2  \Delta_F(q)\rbrace +(q^2+m^2)\tilde\gamma\\
 \mbox{or}\\
 \delta_1'\Delta_F=\displaystyle{\frac{- \tilde\rho-\Lambda\lbrace\lbrack
		  N^{(3)}_3H^4\rbrack -\tilde a H^2  \Delta_F \rbrace \Delta_F }{(\tilde\gamma+\tilde\rho)}}  \qquad \blacksquare
		 \end{array} 
		\label{3.62} 
\end{equation} 
\item [ii)] and iii) In an analogous way:
\begin{equation}
\begin{array}{l}
\forall\,  n \geq3, \, (q,\Lambda) \in
{\cal E}^{4n}_{(q)}\times \R^+ \\
 H^{n+1}(q,\Lambda) =  \displaystyle{{1\over{(\tilde\gamma+\tilde\rho)}}} \lbrace\,
 \lbrack A^{n+1}+ B^{n+1}
 + C^{n+1}\rbrack(q,\Lambda) + \Lambda
\tilde a  H^{n+1}(q,\Lambda)  \rbrace\\
\mbox{or by using the splitting property in $\Phi_R$:} \ \  \displaystyle{H^{{n+1}}=\frac{\delta_{n}(q,\Lambda)C^{{n+1}}}
			{3\Lambda n(n-1)}}\\
			\ \ \\
 H^{n+1} =  \displaystyle{{1\over{(\tilde\gamma+\tilde\rho)}}} 
 \lbrack A^{n+1}+ B^{n+1}+ \Lambda
\tilde a H^{n+1} \rbrack 
 + \displaystyle{ \frac{H^{n+1}3\Lambda n (n-1)}{\delta_n (\tilde\gamma+\tilde\rho)}}\\
 \mbox{or}\\
 \delta_n\lbrace (\tilde\gamma+\tilde\rho) H^{n+1}-\lbrack A^{n+1}+ B^{n+1}+ \Lambda
\tilde a  H^{n+1} \rbrack \rbrace  =  H^{n+1}3\Lambda n (n-1)\\
\mbox{and finally}\\
\delta'_{n}(q,\Lambda)=\displaystyle{\frac{3\Lambda n(n-1)}
				{(\tilde\gamma+\tilde\rho)+D_{n}(H)-\Lambda \tilde a}}  \ \qquad  \qquad\blacksquare
 \end{array}
 	\label{3.63} 
\end{equation}
Notice that as far as the renormalization parameters $a$, $\rho$ and $\gamma$ are concerned
 the corresponding equations of the mapping ${\cal M}^* $ are  the same as in \ref{1.6}.	
\end{itemize}

 \subsection {The  $\Phi_4^4 $- iteration} 
 \begin{definition}\ \label{def.3.2}
 
 By successive application of the mapping $\mathcal{M}^{*}$ to the fundamental sequence $H_{T0}$ we construct a sequence  of G.R.$\Phi$ C's:
\begin{equation}\Phi_{\nu}(H_{T0})=\mathcal{M}^{*}(\Phi_{\nu-1}(H_{T0}))
	\label{3.64} 
	\end{equation}
 the so called   $\Phi_4^4 $-iteration. 
 \end{definition}
 The following theorem   shows recurrently that this sequence is a subset of $\Phi_R$ and automatically constitutes a neighbourhood of the fundamental sequence $H_{T0}$. Then in the next section we show, by a contractivity argument, the convergence of the $\Phi_4^4 $- iteration to the unique non trivial solution inside a precise closed ball $S_r(H_{T0})\subset \ \Phi_R\subset {\cal B}_R$. 

\vspace{3mm} 

\begin{axiom}\label{th.3.2}\ {The ``stability''}

 Every order  $\Phi_{\nu}(H_{T0})$ of the $\Phi_4^4 $- iteration belongs to $\Phi_R$.
\end{axiom}

\begin{remarks}\
\begin{itemize}
\item[a)]

 The zero order of the $\Phi_4^4 $- iteration being the sequence $H_{T0}$ we establish the recurrence starting from the order $\nu=1$. The arguments of the proof of order $0 \to 1$ being similar  we only  present  them  for the transition $\nu-1 \to  \nu$ order of the $\Phi_4^4 $-iteration. In order to simplify the notations we often omit the arguments $(q, \Lambda)$ and $(q^2)$.
 \item[b)]
For the proof of the stability   we use the following auxiliary statements  verified  when $\Phi_{(\nu)}(H_{T0})$ belongs to $\Phi_R$. For the proof of them we refer the reader to Appendix \ref{Ap.6.2}.
\end{itemize}
 \end{remarks}
 
  \subsubsection{The signs and bounds }
 
\begin{proposition} 

Let $\Phi_{(\nu )}(H_{T0})\in \Phi_R$ then \ \ $\forall \Lambda\leq 0.05$:

\begin{itemize}
\item[i)] 
$ \forall   q\in \mathcal{E}_{(q)}^{4}$
\begin{equation}\begin{array}{l}
 H^2_{(\nu)}(q,\Lambda)>0 \\
 \ \ \\
 H^2_{(\nu)}(q,\Lambda)\Delta_F\leq  1+6\Lambda^2(log(q^2+m^2))^{\beta_{1,(\nu)}}\\
\mbox {where $\beta_{1,(0)}=1$ and recurrently $\forall  \nu\geq 1$ :}\  \beta_{1,(\nu)}= \beta_{1,(\nu-1)}3+1\\
 \mbox{and for  sufficienly large $q\in \mathcal{E}_{(q)}^{4}$},\ \ \  \mbox{with}\ \  H^2_{min}=q^2+m^2\\
 \hspace{2cm}H^2_{min}< H^2_{(\nu)}\leq  H^2_{(\nu,max)}\\
 \mbox{and:}\ \  H^2_{(\nu,max)}=\gamma_{max}((q^2+m^2)+6\Lambda^2(q^2+m^2)^{1/3\sum_{k=1}^{\nu} 1/k^2)}\\
 \ \ \\
\mbox {and} \displaystyle\lim_{\nu\to \ \infty}H^2_{(\nu,max)}\equiv H^2_{(max)}=\gamma_{max}[(q^2+m^2)+6\Lambda^2(q^2+m^2)^{\pi^2/54} ]
\end{array}
\label{3.65} 
\end{equation}
\item[ii)] The 
global term  ``$(`\Phi_4^{4}$ operation'')
\begin{equation} C_{(\nu )}^{n+1}(q,\Lambda) = - 6\Lambda\displaystyle{\sum_{\varpi_n(I)}\prod_{l=1,2,3}}
\lbrack N^{(i_l)}_1H_{(\nu)}^{i_{l}+1}\rbrack (q_{i_{l}},\Lambda)
\label{3.66} 
\end{equation} given by definition
 \ref{def.1.1}  verifies the following properties:
\begin{itemize}
\item[a.] The ``good sign'' property:
\begin{equation}\forall \ n=2k+1\  (k\geq 1)\ \  C_{(\nu)}^{n+1}=(-1)^{\frac{n-1}{2}}|C_{(\nu)}^{n+1}| 
\label{3.67}
\end{equation}
\item[b.] It  is a R.$\Phi$.C. in the sense  of definition \ref{def.2.6} consequently it verifies Euclidean invariance and linear axiomaric quantum field theory properties.
\item[c.]
For every $n=2k+1, k\geq 1$  the function $ C_{(\nu)}^{n+1}(q, \Lambda)$ , belongs to the class 
$\mathcal{A}_{4n}^{(\alpha_{n}\beta_{(n,\nu)})}$
of Weinberg functions such that 
 $ \forall\  S\subset\mathcal{E}_{(q)}^{4n}$ the corresponding asymptotic 
indicatrices are given by:
\begin{equation}
	\alpha_{n}(S)=\left\{
	\begin{array}{rl}
		-(n -3),  & \mbox{if } \ S\not\subset\mathcal{K}er\ \lambda_{n}  \\
		0 & \mbox{if }\ S\subset\mathcal{K}er\ \lambda_{n} 
	\end{array}
	\right\}
	\label{3.68}
\end{equation}
\begin{equation}
	\beta_{(n,\nu)}= \beta_{(1,\nu)}n\ \ \ \forall\  S\subset\mathcal{E}_{(q)}^{4n}
		\label{3.69}
\end{equation}
\item[d)] For every $n=2k+1, k\geq1$  
\begin{equation} \begin{array} {l}      \vert  C^{n+1}_{min}(q, \Lambda)\vert\leq	\vert  C_{(\nu)}^{n+1}(q,\Lambda)\vert\leq \vert  C^{n+1}_{(\nu, max)}(q, \Lambda)\\
\mbox{with}: \\
C^{n+1}_{(max)}= 3\Lambda  n(n-1) \mathcal{T}_{n} |H^{n-1}_{max}|\displaystyle{\prod_{l=2,3}  H^2_{(max)}(q_{{l}},\Lambda)\Delta_F(q_{{l}} )}\\
\end{array}
\label{3.70}
\end{equation}
Notice that in the last formula we take  into account  the result
of ref. \cite[c]{MM1} about the number $\mathcal{T}_{n}$
 of different partitions inside the tree terms.
\end{itemize}
\item[iii)]
\begin{equation}\begin{array}{l}
 \forall \ n=2k+1\  (k\geq 1)\ \  H_{(\nu)}^{n+1}=(-1)^{\frac{n-1}{2}}|H_{(\nu)}^{n+1}| 
\end{array}
\label{3.71}
\end{equation}
\item[iv)]
\begin{equation}
 \forall \ n=2k+1\  (k\geq 1) \quad 	\vert  H^{n+1}_{min} \vert\leq	\vert	H_{(\nu)}^{n+1} \vert\leq \vert H^{n+1}_{(\nu,max)}\vert
\label{3.72}
\end{equation}
\end{itemize}
Here $H^{n+1}_{(\nu,max)}$ is recurrently defined as follows: 
\begin{equation}\begin{array}{l}
H^{4}_{(\nu,max)}=-\delta_{3,max}\displaystyle{\prod_{l=1,2,3}}
\lbrack  H^2_{(\nu,max)}\Delta_F(q_{{l}},\Lambda)\rbrack\\
\mbox{and } \displaystyle\lim_{\nu\to \ \infty}H^4_{(\nu,max)}\leq H^4_{(max)}\equiv 6\Lambda\displaystyle{\prod_{l=1,2,3}  H^2_{(max)}(q_{{l}},\Lambda)\Delta_F(q_{{l}} )} \ \\
\mbox{Then recurrently}\ \ \  \forall n=2k+1 Ê\ k\geq 2\\
\ \ \\
|H^{n+1}_{(\nu,max)}|=\delta_{n,max}\mathcal{T}_{n} |H^{n-1}_{(\nu,max)}|\displaystyle{\prod_{l=1,2}  H^2_{(\nu,max)}(q_{{l}},\Lambda)\Delta_F(q_{{l}} )}\\
\mbox{and } \displaystyle\lim_{\nu\to \ \infty}H^{n+1}_{(\nu,max)}=| H^{n+1}_{(max)}|\equiv \delta_{n,max} \mathcal{T}_{n} |H^{n-1}_{max}|\displaystyle{\prod_{l=2,3}  H^2_{(max)}(q_{{l}},\Lambda)\Delta_F(q_{{l}} )}\\
\mbox{and by analogy:}\ \ |H^{n+1}_{(min)}|= \delta_{n,min} \mathcal{T}_{n} |H^{n-1}_{min}|\displaystyle{\prod_{l=2,3}  H^2_{(min)}(q_{{l}},\Lambda)\Delta_F(q_{{l}} )}
\end{array}
\label{3.73}
\end{equation}
\label{prop.3.2} 
\end{proposition}

 \subsubsection{The properties of the global terms $   B_{(\nu )}^{n+1},\   A_{(\nu )}^{n+1}$}

\vspace{1mm}

\begin{proposition} \label{prop.3.3}\ 

Let $\Phi_{(\nu)}(H_{T0})\in \Phi_R$.  Under  the condition $0\leq \Lambda\leq 0.05$, the global term 
 $B_{(\nu )}^{n+1}(H)$  given by definition
 \ref{def.1.1} precisely:
\begin{equation}\begin{array}{l}
B_{(\nu)}^{n+1}(q,\Lambda) = - 3\Lambda\displaystyle{\sum_{\varpi_n(J)}}
\lbrack N^{(j_2)}_2H_{(\nu )}^{j_{2}+2} 
 N^{(j_1)}_1H_{(\nu )}^{j_{1}+1}\rbrack(q,\Lambda)\\
 =\displaystyle{\sum_{\varpi_n(J)}}H_{(\nu )}^{j_{1}+1}\Delta_F\int R^{(2)}_G  
[H_{(\nu )}^{j_{2}+2}\prod_{i=1,2}\Delta_F(l_i)]d^{4}k \end{array}
\label{3.74}
\end{equation}
 verifies the following properties:
\begin{itemize}
\item[i)] \textbf{the ``opposite sign'' property:}
\begin{equation}\forall \ n=2k+1, \  (k\geq 1)\ \  B_{(\nu )}^{n+1} =(-1)^{\frac{n+1}{2}}|B_{(\nu )}^{n+1} 
|\label{3.75}
\end{equation}
\item[ii)] It  is a R.$\Phi$.C. in the sense  of definition \ref{def.2.6} consequently it verifies Euclidean invariance and linear axiomaric field theory properties  as follows from theorem \ref{th.2.2}.
\item[iii)]
For every $n=2k+1, k\geq 1$  the function $B_{(\nu)}^{n+1}(q, \Lambda)$ , belongs to the class 
$\mathcal{A}_{4n}^{(\alpha_{n}\beta_{(n,\nu)})}$
of Weinberg functions such that 
$ \forall\  S\subset\mathcal{E}_{(q)}^{4n}$ the corresponding asymptotic 
indicatrices are given by:
\begin{equation}
	\alpha_{n}(S)=\left\{
	\begin{array}{rl}
		-(n -3),  & \mbox{if } \ S\not\subset\mathcal{K}er\ \lambda_{n}  \\
		0 & \mbox{if }\ S\subset\mathcal{K}er\ \lambda_{n} 
	\end{array}
	\right\}
	\label{3.76}
\end{equation}
\begin{equation}
	\beta_{(n,\nu)}= \beta_{(1,\nu)}n\ \ \ \forall\  S\subset\mathcal{E}_{(q)}^{4n}
		\label{3.77}
\end{equation}
\item[iv)] 
$\exists$ a splitting - sequence $\delta_{\nu}^{B}=\{\delta^{B}_{n, \nu}(q,\Lambda)\}_{n} \in  \mathcal{D}$ such that  for every  $n\geq  3$ the following properties are verified:
 \begin{itemize}
\item{a)} 
\begin{equation}
	\begin{array}{l} 
\delta^{B}_{n, \nu }(q,\Lambda)\build\sim_{q\rightarrow\infty}^{}\Lambda\\
\ \ \\
	\mbox{and\  $\forall$ \  fixed \  $\tilde q $\  and   \ $\Lambda \leq 0.05$}\
	\ \ \\
 	B_{\nu}^{n+1}(\tilde q,\Lambda)=\ -\delta^{B}_{n, \nu }(\tilde q,\Lambda) n(n-1)H_{\nu}^{n+1}(\tilde q,\Lambda)
  	\end{array}
	\label{3.78}
 \end{equation} 
 \item[b)]
For all $n=2k+1, k\geq 1$  the function $ -\delta^{B}_{n, \nu }$ belongs to the same class of Weinberg as the corresponding  splitting function $\delta_n$ precisely:
\begin{equation} \delta^{B}_{n, \nu}\in \ \mathcal{A}_{4n}^{(\tilde \alpha_{n}=0,  \tilde\beta_{n}=0)} \quad \forall\  S\subset\mathcal{E}_{(q)}^{4n}
\label{3.79}
\end{equation}
\end{itemize}

\item[v)] $\forall   \ \ \mbox{fixed }(\tilde q, \tilde\Lambda)\in(\mathcal{E}_{(q)}^{4n}\times]0, 0.05])$, the sequence:
\begin{equation}\left\{\tilde\delta^{B}_{n}\right\}= \displaystyle{\frac{|B^{n+1}_{min}|}{ n(n-1)|H^{n+1}_{max}|}}
\label{3.80}
\end{equation}
increases with increasing $n$.
\item[vi)] $\forall   \ \ \mbox{fixed }(\tilde q, \tilde\Lambda)\in(\mathcal{E}_{(q)}^{4n}\times]0, 0.05])$, 
\begin{equation}\begin{array}{l}
 |B^{n+1}|\leq |B^{n+1}_{max}| \quad \ \ \mbox{with,}\\
 \ \ \\
 |B^{n+1}_{max}| = \frac{3\Lambda n(n-1}{2}\delta_{n,max}\mathcal{T}_{n} \lbrack N_2 |H^{n-1}_{( max)}|\displaystyle{\prod_{l=1,2}  H^2_{( max)}(q_{{l}},\Lambda)\Delta_F(q_{{l}} )}
\rbrack 
\end{array}
\label{3.81}
 \end{equation}
\end{itemize} 
\end{proposition}
\vspace{3mm}

\begin{proposition}\label{prop.3.4}\

Let   $\Phi_{(\nu)}(H_{T0})\in \Phi_R$. Under  the condition $0\leq \Lambda\leq 0.05$,
the global term   $(``\Phi_4^{4}$ operation'')
\begin{equation} A^{n+1}_{(\nu)}(H)=
-\Lambda\lbrack N^{(n+2)}_3H_{(\nu)}^{n+3}\rbrack   = -\Lambda \int R^{(3)}_G \lbrack\
 H_{(\nu)}^{n+3}\prod_{i=1,2,3}
\Delta_F(l_i)\ \rbrack d^{4}k_1d^{4}k_2
\label{3.82}
\end{equation} given by definition
 \ref{def.1.1} verifies the following properties:
\begin{itemize}
\item[i)] the ``good sign'' property:
\begin{equation}\forall \ n=2k+1\  (k\geq 1)\ \  A^{n+1}=(-1)^{\frac{n-1}{2}}|A^{n+1}| 
\label{3.83}
\end{equation}
\item[ii)] It  is a R.$\Phi$.C. in the sense  of definition \ref{def.2.6} consequently it verifies Euclidean invariance and linear axiomaric field theory properties  as follows from theorem \ref{th.2.2}
\item[iii)]
For every $n=2k+1, k\geq 1$  the function $A^{n+1}(q, \Lambda)$ , belongs to the class 
$\mathcal{A}_{4n}^{(\alpha_{n}\beta_{n})}$
of Weinberg functions such that 
 $ \forall\  S\subset\mathcal{E}_{(q)}^{4n}$ the corresponding asymptotic 
indicatrices are given by:
\begin{equation}
	\alpha_{n}(S)=\left\{
	\begin{array}{rl}
		-(n-3),  & \mbox{if } \ S\not\subset\mathcal{K}er\ \lambda_{n}  \\
		0 & \mbox{if }\ S\subset\mathcal{K}er\ \lambda_{n} 
	\end{array}
	\right\}
	\label{3.84}
\end{equation}
\begin{equation}
	\beta_{n(S)}=\left\{
	\begin{array}{rl}
		2(n+1)   & \mbox{if }S\not\subset\mathcal{K}er\ \lambda_{n}  \\
		0 & \mbox{if } S\subset\mathcal{K}er\ \lambda_{n}
	\end{array}
	\right\}
	\label{3.85}
\end{equation}
\item[iv)] $\exists$ a  splitting - sequence $\delta^{A}=\{\delta^{A}_{n}(q,\Lambda)\}_{n} \in  \mathcal{D}$ such that  for every  $n\geq  3$  the following properties are verified: 
\begin{itemize}
\item[a)] \begin{equation}
	\begin{array}{l}
\delta^{A}_{n}(q,\Lambda)\build\sim_{q\rightarrow\infty}^{}\Lambda^2\\	
	\mbox{and $\forall$  \ fixed\  $\tilde q$ \  and \  $ \Lambda \leq 0.05$}\\
	\ \ \\
 	|A^{n+1}_{\nu-1}|(\tilde q,\Lambda)\leq\delta^{A}_{n,\nu-1}(\tilde q,\Lambda) n(n-1)|H^{n+1}_{max}|(\tilde q,\Lambda)
 		\end{array}
	\label{3.86}
 \end{equation} 
 \item[b)] 
For all $n=2k+1, k\geq 1$  the function $\delta^{A}_{n}$ belongs to the same class of Weinberg as the corresponding  splitting function $\delta_n$ precisely:
$$\delta^{A}_{n} \in \ \mathcal{A}_{4n}^{(\tilde \alpha_{n}=0,  \tilde\beta_{n}=0)} \quad \forall\  S\subset\mathcal{E}_{(q)}^{4n}$$

\end{itemize}

\item[v)] $\forall   \ \ \mbox{fixed }(\tilde q, \tilde\Lambda)\in(\mathcal{E}_{(q)}^{4n}\times]0, 0.05])$ the sequence
\begin{equation}\left\{\tilde\delta^{A}_{n}\right\}_{n=2k+1, k\geq 1}= \displaystyle{\frac{|A^{n+1}_{max}|}{ n(n-1|H^{n+1}_{max}|}}
\label{3.87}
\end{equation} 
decreases with increasing $n$.
\end{itemize}

\end{proposition}

\begin{proposition}\label{prop.3.5} \

Let $H_{\nu}\!\in\!\Phi_R$ then, for every $n\geq 3$ and $\forall   \ \ \mbox{fixed }(\tilde q, \tilde\Lambda)\in(\mathcal{E}_{(q)}^{4n}\times]0, 0.05])$  there exist positive  continuous functions of 
$(\tilde q, \tilde\Lambda), \ D_{n,min}(\tilde q, \tilde\Lambda),\ D_{n,max}(\tilde q, \tilde\Lambda)$,
independent of $H$, such that the function $D_{n,\nu}(H)$ defined as follows :

\begin{equation}
\begin{array}{l}
	D_{n,\nu}(H)=
	\displaystyle{\frac{3\Lambda\vert
	\sum_{\varpi_{n}(J)}
	[N_{2}H^{j_{2}+2}_{\nu}][N_{1}H^{j_{1}+1}_{\nu}]\vert}
	{\vert H^{n+1}_{\nu}\vert}-
	\frac{\vert\Lambda[N_{3}^{(n+2)}H^{n+3}_{\nu}]\vert}
	{\vert H^{n+1}_{\nu}\vert}}\\
	\mbox{or}\ \ \\
	D_{n,\nu}(H)=\displaystyle{\frac{\vert B^{n+1}_{\nu}\vert-\vert A^{n+1}\vert}
	{\vert H^{n+1}\vert}}
	\label{3.88}
	\end{array}
\end{equation}
verifies the following properties:
\begin{equation}\displaystyle{\lim_{\Lambda\rightarrow 0}D_n(\Lambda)}=0 \label{3.89}
\end{equation}
\begin{equation}
	 D_{n,min}(\tilde q, \tilde\Lambda)\leq D_n (H(\tilde q, \tilde\Lambda))\leq  D_{n,max}(\tilde q, \tilde\Lambda)
	\label{3.90}
\end{equation}
  Moreover,  there is a positive finite constant $\delta_{\infty}^{\Lambda}$ such that
\begin{equation}	
	\lim_{n\rightarrow \infty}\frac{ D_{n,min}(\tilde q, \tilde\Lambda)}{3\Lambda 
	n(n-1)}= \frac{1}{ \delta_{\infty}^{\Lambda}}
	\label{3.91}	
\end{equation}

\end{proposition}

\subsubsection{Proof of proposition \ref{prop.3.5}}
By application of propositions \ref{prop.3.3} and \ref{prop.3.4} the properties \ref{3.89}, \ref{3.90}, and \ref{3.91} are directly verified
 \subsubsection{Proof of theorem \ref{th.3.2}}  
 \begin{description}
 \item{i)} We have successively:
 \begin{equation}
 \begin{array}{l}
 a_{\nu}(q,\Lambda)  =\left[N_{3}H^4_{(\nu -1)}\right];  \  {and}\ \tilde a_{(\nu)}(\Lambda)= [ a_{(\nu)}]_{q^{2}+m^{2}=0}\\
 \ \ \\
  \rho_{\nu}(q,\Lambda)=-\Lambda
		\left[\displaystyle{\frac{\partial}{\partial q^{2}}[N_{3}^{(3)}H^4_{(\nu -1)}]} \right] \ {and}\ \  \tilde \rho_{(\nu)}( \Lambda)= [ \rho_{(\nu)}]_{q^{2}+m^{2}=0} \\
		\ \ \\
		  \gamma_{\nu}(\Lambda) =
		 \displaystyle{\frac{6\Lambda\prod_{\ell=1,2,3} H^{2}_{\nu-1}(q_{\ell},\Lambda)}{H^4_{(\nu -1)} (q,\Lambda)}} \ \mbox{and}\ \ \tilde \gamma_{(\nu)}= \left[\gamma_{(\nu)}(q, \Lambda) \right]_{q=0}  \\	
\mbox{where the corresponding upper and lower bounds of $\tilde a_{(\nu)},\tilde \rho_{(\nu)}, \tilde \gamma_{(\nu)}$}\\
 \mbox{are trivially obtained.}\\
 \mbox{Then:}\\		 
\forall(q,\Lambda)\in(\mathcal{E}_{(q)}^{4}\times\R^{+*})\\
				H^{2}_{(\nu)}= (q^2+m^2)(1+\delta_{1,\nu}\Delta_F(q^2))\ \  \mbox{with}\ \ \\
 \delta_{1,\nu}(q,\Lambda)\Delta_F(q^2)=\displaystyle{\frac{\tilde \rho_{(\nu -1)}-\Lambda\lbrace\lbrack
		  N_3H^4_{(\nu -1)}\rbrack -\tilde a_{(\nu -1)}H^2_{(\nu -1)}\Delta_F(q^2)\rbrace}{(\tilde \gamma_{(\nu -1)}+\tilde \rho_{(\nu -1)})}}
		 \end{array}  
		 \label{3.92} 
\end{equation}
 \item{ii)}
$\forall(q,\Lambda)\in(\mathcal{E}_{(q)}^{12}\times\R^{+*}) $
\begin{equation}
	 H^4_{\nu} (q\Lambda) = -\delta_{3,(\nu)}(q,\Lambda)
	 \prod_{\ell=1,2,3}H^{2}_{(\nu)}(q_{\ell},\Lambda)\Delta_{F}(q_{\ell})\ ;
	\label{3.93}
\end{equation} 	
		\begin{equation} \mbox{with:}	\begin{array}{l}
			\delta_{3,(\nu)}(q,\Lambda)=\displaystyle{\frac{6\Lambda}{(\tilde \gamma_{(\nu -1)}+\tilde \rho_{(\nu -1)})+D_{3,(\nu-1)}(H)-\Lambda \tilde a_{(\nu-1)} }}\\
			\ \ \\
			\mbox{and} \quad D_{3,(\nu-1)}(H)= \displaystyle{\frac{\vert B^{4}_{(\nu-1)}\vert-\vert A^{4}_{(\nu-1)}\vert}{\vert H^{4}_{(\nu-1)}\vert}}
			\end{array} 
			\label{3.94}
		\end{equation}
 \begin{equation}\begin{array}{l}
 \mbox{ with}\  \delta_{3,(\nu)}(q,\Lambda)\build\sim_{q\rightarrow\infty}^{}\Lambda\\
	\mbox{and}\ \displaystyle{\lim_{\Lambda\to 0} 
 	\frac{\delta_{3,(\nu)}(q,\Lambda)}{\Lambda}}=6,\ 
 	\mbox{for every }q\in\mathcal{E}_{(q)}^{12}
	\end{array}
	\label{3.95}
 \end{equation}
  Moreover, for every finite fixed $\tilde q\in \mathcal{E}_{(q)}^{12}$  and $\forall\Lambda\in\R^{+*}$
\begin{equation}
	\delta_{3,min} \leq \delta_{3, (\nu)}(\tilde q, \Lambda)\leq\delta_{3,max}  
	\label{3.96}
\end{equation}
 \item{iii)}
	\emph{For every $n\geq 5$: $\forall(q,\Lambda)\in(\mathcal{E}_{(q)}^{4n}\times\R^{+*}) $
}
	\begin{equation}			
			H^{{n+1}}_{\nu}(q,\Lambda)=\frac{\delta_{n,(\nu)}(q,\Lambda)C^{{n+1}}_{\nu}(q,\Lambda)}{3\Lambda n(n-1)}
			\label{3.97}
		\end{equation}
\emph{with}
		\begin{equation}
				\delta_{n,(\nu)}(q,\Lambda)=\frac{3\Lambda n(n-1)}
				{(\tilde \gamma_{(\nu -1)}+\tilde \rho_{(\nu -1)})+D_{n,(\nu-1)}(H)-\Lambda\tilde a_{(\nu-1)}}
				\label{3.98}
			\end{equation}	
\emph{Here, in the denominators of eq. \ref{3.95}, we defined the 
function $D_{n,(\nu-1)}(H)$ by:}
\begin{equation}
	D_{n,(\nu-1)}(H)=\frac{\vert B^{n+1}_{(\nu-1)}\vert-\vert A^{n+1}_{(\nu-1)}\vert}
	{\vert H^{{n+1}}_{\nu-1}\vert}
	\label{3.99}
\end{equation}
\emph{Notice that one is allowed to use the absolute values in view of the 
hypothesis $H_{\nu -1}\in \Phi_R$}

\end{description}
Then, the proof   of theorem \ref{th.3.2} is obtained by application of the particular properties of the  global terms $C^{n+1},\    B^{n+1},\   A^{n+1},\  D_{n}$ presented by propositions \ref{prop.3.2}, \ref{prop.3.3}, \ref{prop.3.4}, and \ref{prop.3.5} that we show in Appendix \ref{Ap.6.2}.

\vspace{3mm}

\section{The $\Phi_{4}^{4}$ nontrivial solution}\ \

In this section we present the construction of the unique nontrivial 
solution of the renormalized $\Phi_{4}^{4}$ equations of motion represented 
by the mapping $\mathcal{M^*}$:
	
We define a closed ball  $S_{r}(H_{T0})\ \subset \Phi_R$  the center of which is the "fundamental" tree type sequence $H_{T0}$ ( introduced in section 3). We show the local contractivity of $\mathcal{M}^{*}$ inside this neighbourhood of $H_{T0}$  and consequently 
the existence and uniqueness of a fixed point of the initial mapping $\mathcal{M}$ 
inside $\Phi_R$. For the construction of the solution we propose an 
iteration of the mapping $\mathcal{M}^{*}$ starting from $H_{T0}$.

\subsection{The  closed ball $S_{r0}(H_{T0})\ \subset \Phi_R$ }
\begin{definition}\label{def.4.1}
 \begin{equation} 
 r(0)=\displaystyle{\sup_{\Lambda,n,q}\left[\displaystyle{\frac{\{ \delta_{n,max}-\delta_{n,min}\}}{\delta_{n,max}};\displaystyle{\frac{\vert H^2_{max}- H^2_{min}\vert}{H^2_{max}}} ;\displaystyle{\frac{\vert \frac{\partial^{(0,1)}}{\partial q^{2}}\vert N_3H^{4}_{max}-  N_3H^{4}_{min}\vert}{\hat M_3^{(0,1)}}} }\right]}
 \label{4.100}
\end{equation}
Here the notation $\partial^{(0,1)}$ means either zero  or first order partial derivative
\begin{equation}\mbox{We define}\qquad  S_{r(0)}(H_{T0})=\left\{H\in \Phi_R: \  \Vert H - H_{T0}\Vert \leq r(0) \right\}
\label{4.101}
\end{equation}
\end{definition}

\subsection{The local
contractivity in $S_{r(0)}(H_{T0})\ \subset \Phi_R$  } \ \ 

\begin{axiom}\label{Th.4.1} \

\begin{description}
\item{i)} The subset (closed ball) $S_{r(0)}(H_{T0})\ \subset \Phi_R$ is a complete metric
 subspace of ${\cal B}_R$.
 \item{ii)} There exists a finite positive
constant
 $\Lambda^*(\approx 0. 04)$
 such that  when
 $\Lambda\in]0, \Lambda^*]$ the mapping ${\cal M}^* $ is contractive inside   $S_{r(0)}(H_{T0})\ \subset \Phi_R$ via the $\Phi_4^4 $- iteration  so,
\item{iii)} The unique nontrivial solution of the $\Phi_4^4$
 equations of motion lies in the neighbourhood $S_{r(0)}(H_{T0})$  of the fundamental sequence
$H_{T0}$ and  is constructed as the limit of the $\Phi_4^4 $- iteration.
 \end{description}

\label{th.4.1}
\end{axiom}

\noindent \textbf{Proof of theorem \ref{th.4.1}}

\begin{itemize}
 	\item  [(i)] By definition, the ball $S_{r(0)}(H_{T0})$ is a closed    subset  of the Banach space $\mathcal{B}_{R}$ so it is also a complete subspace.
 	\item  [(ii)] In Appendix \ref{Ap.6.4} we give the proof of the local contractivity of $\mathcal{M}^{*}$  inside the closed 
ball $S_{r(0)}(H^T_0)\ \subset \Phi_R$ via the $\Phi_4^4 $- iteration. 
In other words we show that, at a given order $\nu$ of the $\Phi_4^4 $- iteration and when
 $\Lambda\in ]0, 0.04]$, there exist  two  real positive continuous
 functions of $\Lambda, \ \ K(\Lambda)<1, \ \ k(\Lambda)<1$ 
such that:
\begin{equation}
 \Vert\mathcal{M}^{*}(H^T_0)-H^T_0\Vert\ \leq \ k(\Lambda)\ r(0)\\
 \label{4.102}
\end{equation}
\begin{equation}\begin{array}{l}
    \Vert\mathcal{M}^{*}(H_{(\nu-1)})-\mathcal{M}^{*}(H_{(\nu-2)})
\Vert\ \leq \ K(\Lambda)\ \Vert H_{(\nu-1)}-H_{(\nu-2)}
\Vert\\
\ \ \\
\qquad\qquad\mbox{with} \quad k(\Lambda)+K(\Lambda)) <1
\end{array}
\label{4.103}
\end{equation}

	\item  [(iii)]
This result is a direct consequence of (ii).	
 \end{itemize} 
 
\vfill\eject

\vfill\eject

\section{APPENDICES} 
\subsection{Proof of theorem \ref{th.3.1}}
\begin{Appendix} 
(\textbf{The non triviality of $\Phi_R$})\label{APP.6.1}

  We consider the fundamental  
 sequence $H_{T0}$ (cf. definition \ref{def.2.5}) and verify successively the properties of $\Phi_R$. Precisely:
 \begin{enumerate}
\item{}
\begin{equation}\begin{array}{l}
 \forall(q,\Lambda)\in(\mathcal{E}_{(q)}^{4}\times\R^{+*}) \\
	H_{T0}^{2} = (q^2+m^2)(1+\delta_{10}(q,\Lambda)\Delta_F )\\
	\mbox{with}\\
	\delta_{10}(q,\Lambda)\Delta_F  =\displaystyle{\frac{-\rho_{0}+ \Lambda \delta_{3,min}([N_{3}\tilde] - [N_{3}\tilde ]_{(q^2+m^2) =0})\Delta_F } {1+\rho_{0}+\Lambda |a_0|}}\\
	\mbox{We  verify:}\\
 \ 	\displaystyle{\lim_{(q^2+m^2)= 0}}H_{T0}^{2}(q,\Lambda)  \Delta_F(q)=1\quad \  (\mbox{or} \ \ \displaystyle{\lim_{(q^2+m^2)= 0}}\delta_{10}(q,\Lambda)\Delta_F =0)\\
 \mbox{and in view of the logarithmic asymptotic behaviour of the $\Phi^4_4$  operation}\\
 \mbox{namely:}\ \  [N_{3}\tilde]\Delta_F\sim_{q\rightarrow\infty}^{}log(q^2+m^2)\\
 H_{T0}^{2}(q,\Lambda) \leq (\Vert q\Vert^{2}+m^2)^{(1+\pi^2/18)}\\
 \mbox{and}\\
   H^2_{min}(q,)  \leq H_{T0}^{2}(q,\Lambda), \ \ \mbox {with}\ \ H^2_{min}(q)=q^2+m^2
	\end{array} 
	\label{6.104}
	\end{equation}
	 $\hspace{10cm}\ \blacksquare$ 
	\item{}
	Moreover we verify  that for every  $n=2k+1, k\in\N^*$  and $\forall\ (q,\Lambda)\in(\mathcal{E}_{(q)}^{4n}\times\R^{+*})$ the functions $H_{T0}^{n +1}$ 
	\begin{equation}\begin{array}{l}
    H_{T0}^{4} = -  \delta_{3,min}(\Lambda)\displaystyle{\prod_{l=1,2,3}H^{2}_{T0}(q_l)\Delta_{F}(q_{l})}\\ 
 H^{n+1}_{T0}(q,\Lambda) = \displaystyle{{\delta_{n, min}(\Lambda)
 C^{n+1 }_{T0}(q,\Lambda) \over 3\Lambda
 n (n-1)}};\\
\end{array}
	\label{6.105}
\end{equation}
(with\  $\{\delta_{n,min}\}_{n\geq 3}$\ the splitting sequence 
of definition \ref{def. 2.4}),
 belong to the class 
$\mathcal{A}_{4n}^{(\alpha_{n}\beta_{n})}$
of Weinberg functions with corresponding asymptotic 
indicatrices  given by:
 $ \forall\  S\subset\mathcal{E}_{(q)}^{4n}$ \begin{equation}
	\alpha_{n}(S)=\left\{
	\begin{array}{l}
		-(n-3) \  \mbox{if } \ S\not\subset\mathcal{K}er\ \lambda_{n}  \\
		0 \ \  \mbox{if}\ S\subset\mathcal{K}er\ \lambda_{n}\\ 
	\beta_{n(S)}=\nu_{(n)}=2n \ \ \forall\  S\subset{\mathcal E}_{(q)}^{4n}\\
	\end{array}
	\right\}		
	\label{6.106}
\end{equation}

 \item{}	Trivially the properties \ref{3.50}, \ref{3.51} \ref{3.52} are satisfied by the definition and bounds of the sequence $\{\delta_{n,min}\}_{n\geq 3}$\ the splitting sequence 
of definition \ref{def. 2.4}),
	\begin{equation}\begin{array}{l}
 \mbox{ with}\   \delta_{3,min} (\Lambda)\build\sim_{q\rightarrow\infty}^{}\Lambda\ 
	\ \ \\
	\mbox{and}\ \displaystyle{\lim_{\Lambda\to 0} 
 	\frac{\delta_{3,min}(\Lambda)}{\Lambda}}=6,\ 
 	\end{array}
	\label{6.107}
 \end{equation}
and
\begin{equation}\begin{array}{l}
 \ \  \delta_{n,min}(q,\Lambda)\build\sim_{q\rightarrow\infty}^{}\Lambda\\
\ \ \\
	\mbox{and}\ \displaystyle{\lim_{\Lambda\to 0} 
 	\frac{\delta_{n,min}(\Lambda)}{\Lambda}}=3n(n-1),\ 
	\end{array}
	\label{6.108}
 \end{equation}
 Moreover
\begin{equation}
	\delta_{n,min}(\Lambda) < 
	\delta_{n,max} (\Lambda) 
		\label{6.109}
\end{equation} 
and there is
 $\delta_{\infty}$ \begin{equation}
	\displaystyle{\lim_{n\to \infty}\delta_{n, max}(\Lambda)}<\delta_{\infty}\ \  \ \forall\ \Lambda\in\R^{+*}:
		\label{6.110}
\end{equation}
\item{} 
 \begin{equation}\begin{array}{l}
	 \gamma_0=1\\	
a_0 = -\delta_{3,min}[N_{3}\tilde ]_{(q^2+m^2) =0};\\	
\ \ \\ 
	 \rho_{0}= \Lambda\delta_{3,min}\lbrack \displaystyle{\frac{\partial}{\partial q^{2}}[N_{3}\tilde ]\rbrack_{(q^2+m^2) =0}} \\
	 \mbox{(Reminder:	$\quad\delta_{3,min}= \displaystyle{\frac{6\Lambda}{1+9\Lambda(1+6\Lambda^2)}}$)}\\	
	\end{array}
	\label{6.111}
\end{equation}
So we trivially obtain that $H_{T0}$ also verifies the property 4.of $\Phi_R$
 (cf. equations \ref{3.53}, \ref{3.54}, \ref{3.55}) for the renormalization constants and this allows to conclude.    \ \ \ \  $\blacksquare$. 
 \end{enumerate}
\end{Appendix}

\subsection{Proof of Proposition \ref{prop.3.3} - (The properties of the global terms $B^{n+1}$)}
\begin{Appendix}\label{Ap.6.2}

 
 Let $H_{\nu}\in  \Phi_R$.
 We first easily  establish the following inequality $\forall   \ \ \mbox{fixed }(\tilde q, \tilde\Lambda)\in(\mathcal{E}_{(q)}^{12}\times]0, 0.05])$:
 \begin{equation}\begin{array}{l}
\mbox{For }\ \ \  D_{3,\nu}(H)=\displaystyle{\frac{\vert B^{4}_{\nu}\vert-\vert A^{4}_{\nu}\vert}
	{\vert H^{4}_{\nu}\vert}}>0\\
 D_{3,min}(H)=\displaystyle{\frac{|B^{4}_{min}|-\vert A^{4}_{max}\vert}{|H^{4}_{max}|}}>0
\end{array}
\label{6.112}
 \end{equation}
  Then we show that: $\forall   \ \ \mbox{fixed }(\tilde q, \tilde\Lambda)\in(\mathcal{E}_{(q)}^{4n}\times]0, 0.05])$ and $n\geq 5$ the sequence:
 \begin{equation}\left\{\tilde\delta^{B}_{n}\right\}_{n=2k+1, k\geq 2}= \displaystyle{\frac{|B^{n+1}_{min}|}{ n(n-1)|H^{n+1}_{max}|}}
\label{6.113}
 \end{equation}
 increases with increasing $n$.
 In other words we prove that:
\begin{equation} \displaystyle{\frac{|B^{n+1}_{min}|}{ n(n-1)|H^{n+1}_{max}|}}\geq 
\displaystyle{\frac{|B^{n-1}_{min}|}{ (n-2)(n-3)|H^{n-1}_{max}|}}
\label{6.114}
\end{equation} 
For further  purposes in our proof we shall use the following recurrence hypothesis which is valid in the first step i.e. for $n=5 \ \mbox{and} \  n-2=3$ when $\Lambda\leq 0,05$
 \begin{equation} \forall \bar {n} \leq n-2,\ \ \displaystyle{\frac{ [N_2|H^{\bar {n}+1}_{min}|]}{{  |H^{\bar {n}+1}_{max}|}}}\geq  \displaystyle{\frac{ [N_2|H^{\bar {n}-1}_{min}|]}{  |H^{\bar {n}-1}_{max}|}} 
 \label{6.115}
 \end{equation}

Now by using definitions \ref{def.2.5},  
we require instead of \ref{6.114} the following condition:
\begin{equation} \displaystyle{\frac{\sum_{\varpi_n(J)}
\lbrack N_2|H^{j_{2}+2}_{min}|]
 [N^{(j_1)}_1|H^{j_{1}+1}_{min}|]}{ n(n-1)|H^{n+1}_{max}|}}\geq 
\displaystyle{\frac{\sum_{\varpi_{n-2}(J)}
\lbrack N_2|H^{j_{2}+2}_{min}|
 [N^{(j_1)}_1|H^{j_{1}+1}_{min}|]}{ (n-2)(n-3)|H^{n-1}_{max}|}}
\label{6.116}
\end{equation} 
 
 Notice that  $\forall n\geq 5$, we can  bound the left hand side sum and respectively the right hand side sum  by their dominant contribution as follows: 
 \begin{equation}
 \displaystyle{\sum_{\varpi_n(J)}
\lbrack N_2|H^{j_{2}+2}_{min}|
 N^{(j_1)}_1|H^{j_{1}+1}_{min}|\rbrack}\geq\frac{n(n-1)}{4}[N_2|H^{n+1}_{min}|]H^2_{min}\Delta _F
 \label{6.117}
 \end{equation}
 and respectively:
 \begin{equation}
 \displaystyle{\sum_{\varpi_{n-2}(J)}
\lbrack N_2|H^{j_{2}+2}_{min}|
 N^{(j_1)}_1|H^{j_{1}+1}_{min}|\rbrack}\geq\frac{(n-2)(n-3)}{4}[N_2|H^{n-1}_{min}]|H^2_{min}\Delta _F
\label{6.118}
 \end{equation}
Then, condition \ref{6.116} becomes:
 \begin{equation} \displaystyle{\frac{ [N_2|H^{n+1}_{min}|]}{{  |H^{n+1}_{max}|}}}\geq  \displaystyle{\frac{  [N_2|H^{n-1}_{min}|]}{  |H^{n-1}_{max}|}} 
 \label{6.119}
 \end{equation}
 By application of definition \ref{def.2.4} the previous condition takes  successively the following forms:
 \begin{equation}
\displaystyle{\frac{\delta_{n,min} [N_2|C^{n+1}_{min}|]}{ \delta_{n,max} |C^{n+1}_{max}|}}\geq \displaystyle{\frac{\delta_{(n-2),min} [N_2|C^{n-1}_{min}|]}{ \delta_{(n-2),max} |C^{n-1}_{max}|}}
\label{6.120}
\end{equation}
 or equivalently, by using definitions  \ref{def.2.5} and proposition \ref{prop.3.2} of the tree terms, 
\begin{equation}
\displaystyle{\frac{\delta_{n,min}\tilde {\mathcal T} _{n} [N_2|H^{n-1}_{min}|]}{ \delta_{n,max}\mathcal{T}_{n} |H^{n-1}_{max}|}}\geq \displaystyle{\frac{\delta_{(n-2),min}\tilde {\mathcal T} _{n-2}  [N_2|H^{n-3}_{min}|]}{ \delta_{(n-2),max} \mathcal{T}_{n-2} |H^{n-3}_{max}|}}
\label{6.121}
\end{equation}
where:
\begin{equation}\begin{array}{l}
    \mathcal{T}_{n}=\frac{(n-3)^{2}}{48}+\frac{(n-3)}{3}+1\\
    \ \ \\
    \mbox{and} \quad \tilde {\mathcal T} _{n}=\frac{(n-3)^{2}}{48}
    \end{array}
    \label{6.122}
\end{equation}

Then, by using the recurrence hypothesis  \ref{6.115} of\  $\tilde n=n-2$ and definitions \ref{def.2.4} we obtain the final equivalent form of condition \ref{6.114}:
 \begin{equation}
 \displaystyle{\frac{[1+3\Lambda(n-2)(n-3)][1+ n(n-1)d_0](n-3)^2 [\frac{(n-5)^{2}}{48}+\frac{(n-5)}{3}+1]}{ 1+3\Lambda n(n-1)][1+(n-2)(n-3)d_0] (n-5)^2 [\frac{(n-3)^{2}}{48}+\frac{(n-3)}{3}+1]}} \geq 1
 \label{6.123}
 \end{equation}
  \begin{figure}[h]
\begin{center}
\hspace*{-5mm}
  \includegraphics[width=10cm]{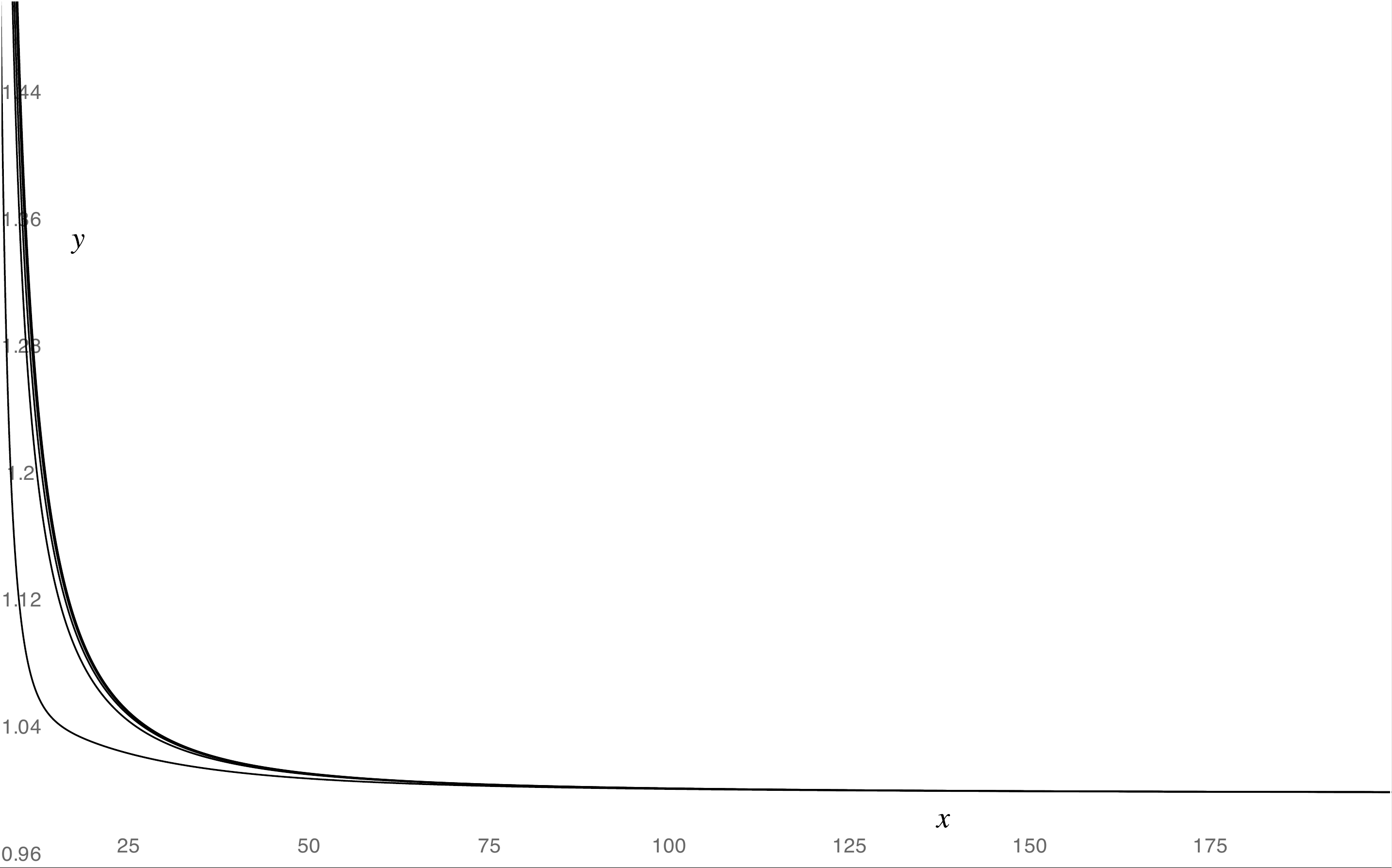}
\end{center} 
 \caption{\small\textrm{For the values of $n$ ($=x$ continuous) in the interval $]7, 200] $ $f_{d0}$ decreases continuously always from bigger values than $1$  up to the limit value of $1$}}
\label{fig.6.6}
\end{figure}
\end{Appendix}

\subsection{Proof of Proposition \ref{prop.3.4} The properties of the global terms $A^{n+1}$}  
\begin{Appendix}\label{Ap.6.3}

 We show that: $\forall   \ \ \mbox{fixed }(\tilde q, \tilde\Lambda)\in(\mathcal{E}_{(q)}^{4n}\times]0, 0.05])$ the sequence
\begin{equation}\left\{\tilde\delta^{A}_{n}\right\}_{n=2k+1, k\geq 3}= \displaystyle{\frac{|A^{n+1}_{max}|}{ n(n-1|H^{n+1}_{max}|}}
\label{6.124}
\end{equation} 
decreases with increasing $n$. In other words we prove that:
\begin{equation} \displaystyle{\frac{|A^{n+1}_{max}|}{ n(n-1)|H^{n+1}_{max}|}}\leq 
\displaystyle{\frac{|A^{n-1}_{max}|}{ (n-2)(n-3)|H^{n-1}_{max}|}}
\label{6.125}
\end{equation} 
As before, by application of definitions \ref{def.2.4},  and proposition \ref{prop.3.2} of the tree terms 
we have:
\begin{equation}\begin{array}{l}
|A^{n+1}_{max}|=  \Lambda\lbrack
N_3|H^{n+3}_{max}|\rbrack\ \ \  ;\  \ \ |A^{n-1}_{max}|=\Lambda\lbrack
N_3|H^{n+1}_{max}|\rbrack\\
\mbox{and}\ \\
\displaystyle{\frac{\Lambda\lbrack
N_3|H^{n+3}_{max}|\rbrack}{ n(n-1)  |H^{n+1}_{max}|}}\leq
 \displaystyle{\frac{\Lambda\delta_{n+2,max} \mathcal{T}_{n+2}\lbrack
N_3|H^{n+1}_{max}|\rbrack}{ n(n-1)\delta_{n,max} \mathcal{T}_{n}|H^{n-1}_{max}|}}
\end{array}
\label{6.126}
\end{equation} 
By comparison with the condition \ref{6.124}  the following function 
$f_{d_1}(n)$ should be   smaller than $1$
\begin{equation}
f_{d_1}(n)=\displaystyle{\frac{\delta_{n+2,max} \mathcal{T}_{n+2}(n-2)(n-3)}{ \delta_{n,max} \mathcal{T}_{n}n(n-1)}}\ \leq 1
\label{6.127}
\end{equation}
\begin{equation}
f_{d_1}(n)=\displaystyle{\frac{(n+1)(n+2)[1+ n(n-1)d_0] [\frac{(n-1)^{2}}{48}+\frac{(n-1)}{3}+1](n-2)(n-3)}{ n(n-1)[1+(n+1)(n+2)d_0] [\frac{(n-3)^{2}}{48}+\frac{(n-3)}{3}+1]n(n-1)}}\ \leq 1
\label{6.128} 
\end{equation}

  \begin{figure}[h]
\begin{center}
\hspace*{-5mm}
 \includegraphics[width=12cm]{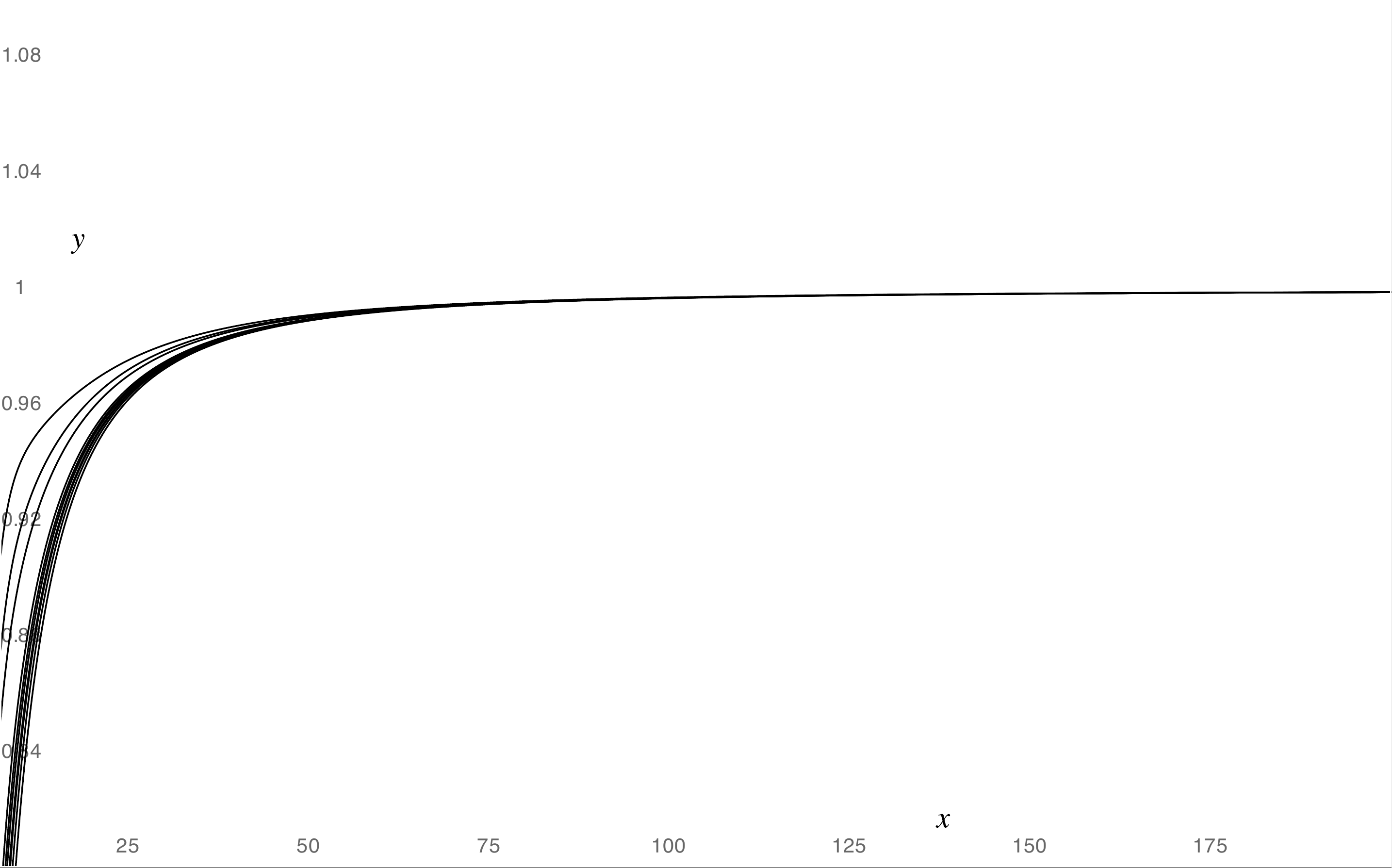}
\end{center} 
 \caption{\small\textrm{For the values of $n$ ($=x$ continuous) in the interval $]7, 200] $ the function $f_{d_0}(n)$ increases continuously  (with positive values always smaller than 1 ) up to the limit value of $1$.}}
\label{fig.7}
\end{figure}
By giving to the numerical  constant $d_0$ different values in the interval [0.02 , 0,45] and after long numerical calculations we can find that the difference between the denominator and numerator is always positive.

For the values of $n$ ($=x$ continuous) in the interval $]7, 200] $ the function $f_{d_0}(n)$ increases continuously  (with positive values always smaller than 1) up to the limit value of $1$.
\begin{remark}
Notice that as far as the $|H^{n+1}_{max}|$'s  with $n\geq 5$  are concerned, the decrease behaviour (with respect the external momenta  (i.e. $\alpha {(S)}\leq -2$)   allows us to take the bounds numerically (at zero external momenta).
\end{remark}

\end{Appendix}
 
\vfill\eject
\subsection{Proof of the local contractivity of the mapping ${\cal M^*}$
or the convergence of the $\Phi^4_4$ iteration inside $Sr(0)$(theorem \ref{Th.4.1})} 

\begin{Appendix}\label{Ap.6.4}
 By 
 the definition \ref{def.2.10}
   \ of the norm $\mathcal{N}$
the  inequalities \ref{4.102} and \ref{4.103} are equivalent to the following:
$\forall\ \ \ \Lambda \in ]0, 0.04]$
\begin{equation}
\begin{array}{l}
\displaystyle{\sup_{(n, q,\Lambda)}}
\left\{\displaystyle{ \frac{\vert H^{n+1}_{\nu}- H^{n+1}_{T0}\vert}{  M_n}; \frac{\vert \frac{\partial^{(0,1)}}{\partial q^{2}}N_{3}(H^{4}_{\nu}-H^{ 4}_{T0})\vert }{\hat M_{(3)}^{(0,1)}};\frac{| N_2(H^{n+1}_{\nu}-H^{ n+1}_{T0})| }{\hat M_{(n,2)}};\frac{|\gamma_{\nu}-\gamma_{0}|}{ {\cal  N}_{\gamma}}}\right\}  \\
\qquad\leq \ k^{(0)}(\Lambda)r_0\quad \mbox{with} \ \ k^{(0)} (\Lambda)<1 \     \end{array}
\label{6.129}
\end{equation}
\begin{equation}
\begin{array}{l} 
\displaystyle{\sup_{(n, q,\Lambda)}}
\left\{\displaystyle{ \frac{\vert H^{n+1}_{\nu}- H^{n+1}_{\nu-1}\vert}{ M_n};\frac{\vert \frac{\partial^{(0,1)}}{\partial q^{2}}N_{3}(H^{4}_{\nu}-H^{ 4}_{\nu-1})\vert }{\hat M_3^{(0,1)}}};\displaystyle{\frac{| N_2(H^{n+1}_{\nu}-H^{ n+1}_{\nu-1})| }{\hat M_{(3,2)}};\frac{|\gamma_{\nu}-\gamma_{\nu-1}|}{{\cal  N}_{\gamma}}}\right\} \\
\qquad \leq \ K^{\nu} (\Lambda)\Vert H_{\nu-1}- H_{\nu-2}\Vert; \quad \mbox{with} \ \ K^{\nu} (\Lambda)<1\\
\ \ \ \\
\qquad\qquad  \mbox{and }
k^{(0)}+K^{\nu}<1
  \end{array}
\label{6.130}
\end{equation}

\begin{enumerate}

\item{} 
\textbf{Proof of \ref{6.129}}

Notice that the condition \ref{6.129} is stronger than the required \ref{4.102} 
(first order image by ${\cal M^*}$ of $H_{T_0}$).

We first obtain the corresponding bounds  for $\nu=1$.
 We start from $n=1, n=3$ and generalize recurrently for every $n\geq 5$ . Then we apply the same procedure for every $\nu\geq 2$.
 \begin{description}
 \item{a)}
 Let $n=1$ 
\begin{equation} 
\displaystyle{ \frac{\vert H^{2}_{1}- H^{2}_{T0}\vert}{M_1}}\leq \displaystyle{\frac{\Lambda\vert N_3(H^{4}_{max}-H^{4}_{min})\vert \Delta_F}{ M_1}}
 \label{6.131}
\end{equation}
then using the definition \ref{def.4.1} of the ball $S_{r(0)}(H_{T0})$, the norm definition \ref{def.2.10} and proposition  \ref{prop.3.2} we finally obtain:
\begin{equation}\begin{array}{l}
\displaystyle{ \frac{\vert H^{2}_{1}- H^{2}_{T0}\vert}{M_1}}\leq k^{(0)}_{1,1} r(0)\\
 \mbox{with:}\ \  k^{(0)}_{1,1}=\displaystyle{\frac{6\Lambda^2(q^2+m^2)^{\pi^{2}/54}(1+6\Lambda^2(q^2+m^2)^{\pi^{2}/54})}{1+6(q^2+m^2)^{\pi^{2}/54}}}
 \end{array}
 \label{6.132}
\end{equation} 
For $n=3$ we have  :
\begin{equation}
	 H^4_{1} (q\Lambda) = -\delta_{3,(1)}(q,\Lambda)
	 \prod_{\ell=1,2,3}H^{2}_{(1)}(q_{\ell},\Lambda)\Delta_{F}(q_{\ell})\ ;
	\label{6.133}
\end{equation} 
so	
\begin{equation}\begin{array}{l}
\displaystyle{ \frac{\vert H^{4}_{1}- H^{4}_{T0}\vert}{M_3}}\leq \ \displaystyle{\frac{\{ \delta_{3,max}-\delta_{3,min}\}}{M_3}} \prod_{\ell=1,2,3}H^{2}_{(max)}(q_{\ell},\Lambda)\Delta_{F}(q_{\ell})+ \\
+3\delta_{3,max}\displaystyle{\frac{\vert H^2_{1}- H^2_{T0}\vert}{{M_3}} \prod_{\ell=1,2}H^{2}_{(max)}(q_{\ell},\Lambda)\Delta_{F}(q_{\ell}) }
\end{array}
\label{6.134}
\end{equation}
And again by the  definition of norms, of $r(0)$ and the previous result for $n=1$ we obtain:
\begin{equation}\begin{array}{l}
\displaystyle{ \frac{\vert H^{4}_{1}- H^{4}_{T0}\vert}{M_3}}\leq \ \Lambda r(0)(1+(18)^2\Lambda ^2 )\quad \mbox{i.e.}\ \  k^{(0)}_{1,3}=\Lambda(1+(18)\Lambda ^2 )\\
\ \qquad \ \mbox{and \ \ \ \ $k^{(0)}_{1,3}<1$\   for \  $\Lambda\leq 0.1$}
\end{array}  
\label{6.135}
\end{equation} 
Now  for every $n\geq 5$ and $\bar n\leq n-2$ we suppose that we have established an analogous inequality, namely:
\begin{equation}\begin{array}{l}
\displaystyle{ \frac{\vert H^{\bar n+1}_{1}- H^{\bar n+1}_{T0}\vert}{M_n}}\leq \   r(0)k^{(0)}_{1,\bar n} \\
\ \mbox{with  \ \ $k^{(0)}_{1,\bar n}<  k^{(0)}_{1,3}$ \ if $\Lambda\leq 0.1$}\ \ 
\mbox{and  we  show that $k^{(0)}_{1,n}< k^{(0)}_{1,3}$} \\
\mbox{without any  supplementary condition on $\Lambda$}
\end{array} 
\label{6.136} 
\end{equation} 

By using definition \ref{def.4.1} of $S_{r(0)}(H_{T0})$, the norm definition \ref{def.2.10} the splitting properties, the bounds \ref{3.70} of  
the tree terms, and the recursion 
we have successively:
\begin{equation}\begin{array}{l}
\displaystyle{ \frac{\vert H^{n+1}_{1}- H^{n+1}_{T0}\vert}{M_n}}\leq \displaystyle{\frac{\{ \delta_{n,max}-\delta_{n,min}\}\vert C^{n+1}_{max}\vert}{3\Lambda n(n-1)M_n}} +\displaystyle{\frac{ \delta_{n,max} \vert C^{n+1}_{1}- C^{n+1}_{T0}\vert}{3\Lambda n(n-1)M_n}}<\\
\ \ \\
< \displaystyle{\frac{r(0)\delta_{n,max}\mathcal{T}_{n}H^2_{(max)} }{n(n-1)\delta_{n,max}M_{1}}} \displaystyle{\{\frac{\vert H^{n-1}_{max}\vert H^2_{(max)}}{M_{n-2}M_1}}+ k^{(0)}_{1,n-2}\frac{H^2_{(max)}}{M_1}  +2k^{(0)}_{1,1}\displaystyle{\frac{\vert H^{n-1}_{max}\vert }{M_{n-2}}\}}<\\
\ \ \\
<\displaystyle{\frac{r(0)(n-3)^2}{48 n(n-1)} } \displaystyle{\{\frac{\vert H^{n-1}_{max}\vert }{M_{n-2}}}+ k^{(0)}_{1,n-2}\frac{H^2_{(max)}}{M_1}  +2k^{(0)}_{1,1}\displaystyle{\frac{\vert H^{n-1}_{max}\vert }{M_{n-2}}\}}
\end{array} 
\label{6.137} 
\end{equation} 
In the last formula we used again the result
of ref. \cite[c]{MM1} about the number $\mathcal{T}_{n}$
 of different partitions inside the tree terms as we did in proposition
  \ref{prop.3.2}.
Now we note that for every $n$ we have:
\begin{equation}
\displaystyle{\frac{\vert H^{n+1}_{max}\vert }{M_{n}}} < \displaystyle{\frac{\vert H^{n-1}_{max}\vert }{M_{n-2}}}
\label{6.138}
\end{equation}
As a matter of fact by application of proposition \ref{prop.3.2}  and in particular the bounds \ref{3.70},  \ref{3.73} and the norm definition \ref{def.2.10}  we can write:
\begin{equation}
\displaystyle{\frac{\vert H^{n+1}_{max}\vert }{M_{n}}}<\displaystyle{\frac{(n-3)^2\vert H^{n-1}_{max}\vert }{48 n(n-1)M_{n-2}}}<\displaystyle{\frac{\vert H^{n-1}_{max}\vert }{48 M_{n-2}}}
\label{6.139}
\end{equation}
It then follows that:
\begin{equation}
\displaystyle{\frac{\vert H^{n+1}_{max}\vert }{M_{n}}}<\displaystyle{\frac{(n-3)^2\vert H^{n-1}_{max}\vert }{48 n(n-1)M_{n-2}}}<\displaystyle{\frac{\vert H^{4}_{max}\vert }{48 M_{3}}} <\Lambda
\label{6.140}
\end{equation}
From these results and the recurrent hypothesis
 \begin{equation} k^{(0)}_{1, n-2}<k^{(0)}_{1, 3}
 \label{6.141}
 \end{equation}
 we have:
\begin{equation}\begin{array}{l}
\displaystyle{ \frac{\vert H^{n+1}_{1}- H^{n+1}_{T0}\vert}{M_n}}<
\displaystyle{\frac{r(0)} {48} } \{\Lambda+ k^{(0)}_{1,3}   +2k^{(0)}_{1,1}\} \\
\mbox{or}\quad \displaystyle{ \frac{\vert H^{n+1}_{1}- H^{n+1}_{T0}\vert}{M_n}}< r(0)  k^{(0)}_{1,n} \ \ \ \mbox{with}\ \ k^{(0)}_{1,n} =\displaystyle{\frac{ k^{(0)}_{1,3} }{16}} 
\end{array} 
\label{6.142} 
\end{equation} 

\begin{figure}[h]
\begin{center}
\hspace*{-5mm}
 \includegraphics[width=12cm]{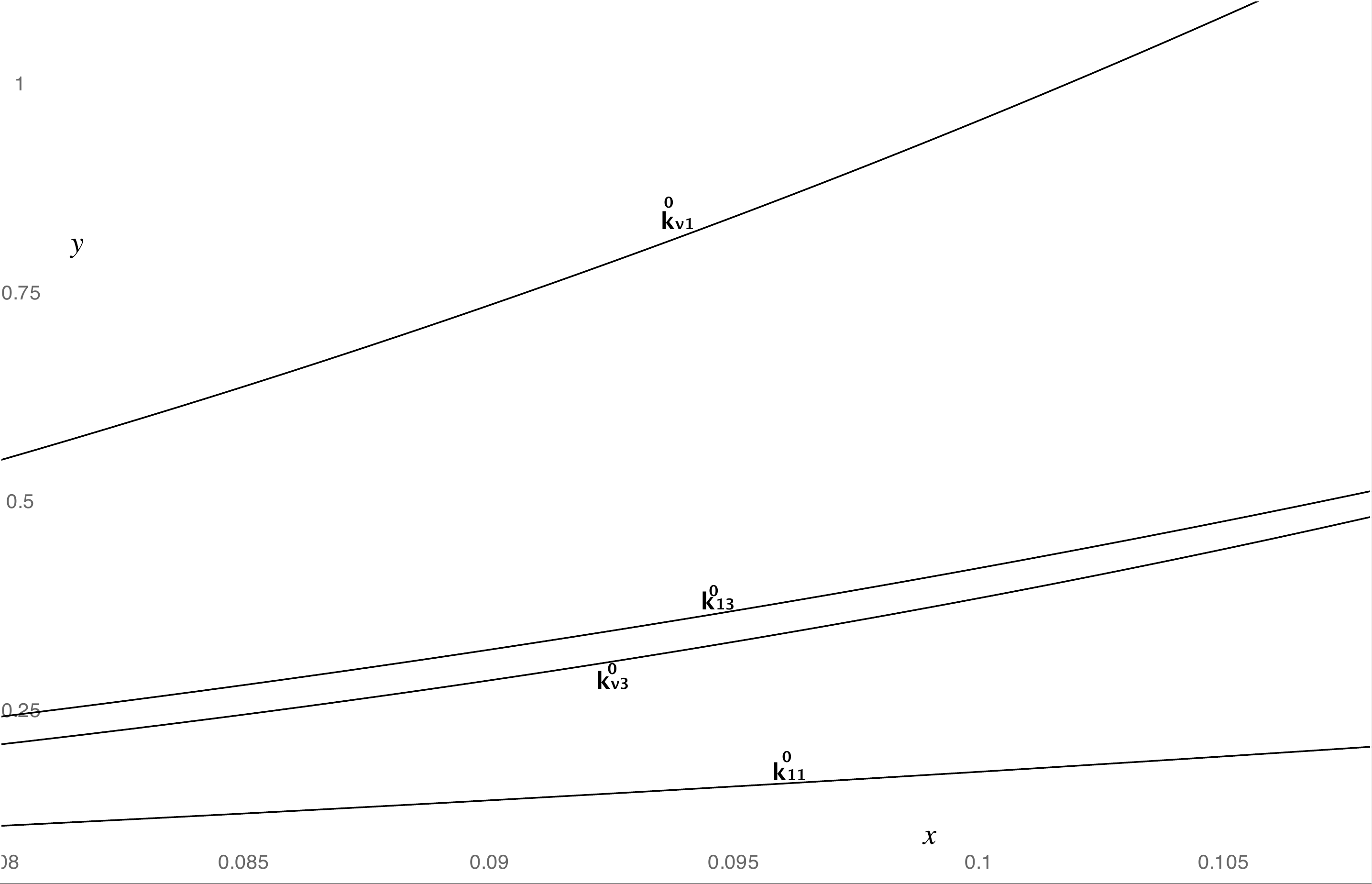}
\end{center} 
 \caption{\small\textrm{The stronger condition to require for the coupling constant  comes from $H^2_{\nu }$ precisely: $\Lambda\leq 0.101$ with $k_{\nu, 1}=0.9925$  while $k_{1, 3}=0.4189$ and corresponding values $k_{\nu, 3}=0.40$ and $k_{1, 1}=0.1846$.}}
\label{fig.8}
\end{figure}

\item{b)} 
In the case of $\nu\geq 2$ we follow an analogous procedure and find similar results . We just notice that for $n=1$ the condition imposed on $\Lambda $ in order that $k^{(0)}_{\nu, 1}<1$ is stronger   than the one of 
every $k^{(0)}_{\nu, n}<1 ,\ \ \mbox{with} \  n\geq 3$ (cf. figure \ref{fig.8}). 

As a matter of fact at every order $\nu\geq 2$ of the  $\Phi_4^4 $-iteration the contributions coming from the values of the renormalization constants $\tilde \gamma,\ \ \tilde \rho, \tilde a$ become nontrivial.

Precisely: 
\begin{equation}\begin{array}{l}\displaystyle{ \frac{\vert H^{2}_{\nu}- H^{2}_{T0}\vert}{M_1}}\leq \displaystyle{\frac{|\tilde \rho_{\nu -1}-\rho_{0}| \tilde \gamma_{0}+\tilde \rho_{0}|+|\tilde \gamma_{\nu -1}- \tilde \gamma_{0}|\tilde \rho_{0}}{|\tilde \gamma_{\nu -1}+\tilde \rho_{\nu -1}|||\tilde \gamma_{0}+\tilde \rho_{0}| | M_1}}\\
\ \ \\
\qquad\qquad+ \displaystyle{\frac{\Lambda\vert N_3H^{4}_{\nu -1}(\tilde \gamma_{0}+\tilde \rho_{0})-N_3H^{4}_{min}(\tilde \gamma_{\nu -1}+\tilde \rho_{\nu -1})\vert \Delta_F}{ |\tilde \gamma_{\nu -1}+\tilde \rho_{\nu -1}|||\tilde \gamma_{0}+\tilde \rho_{0}| | M_1}}\\
\ \ \\
\qquad +\displaystyle{\frac{\Lambda\vert N_3H^{4}_{\nu -1}(\tilde \gamma_{0}+\tilde \rho_{0})-N_3H^{4}_{min}(\tilde \gamma_{\nu -1}+\tilde \rho_{\nu -1})\vert _{(q^2+m^2)=0} H^{2}_{max}\Delta_F}{|\tilde \gamma_{\nu -1}+\tilde \rho_{\nu -1}|||\tilde \gamma_{0}+\tilde \rho_{0}| | M_1}}\\
\ \ \\
\qquad\qquad \ +\Lambda \displaystyle{ \frac{\vert H^{2}_{max}- H^{2}_{min}\vert \vert N_3H^{4}_{max}\vert\Delta_F}{|\tilde \gamma_{\nu -1}+\tilde \rho_{\nu -1}|||\tilde \gamma_{0}+\tilde \rho_{0}| | M_1}}\ 
\end{array}
\label{6.143}
\end{equation}
Then as before by taking into account the  norm definition \ref{def.2.10} and def.\ref{def.4.1} of $S_{r(0)}(H_{T0})$ and $r(0)$ we first have:
\begin{equation}\begin{array}{l}
|\tilde \gamma_{\nu -1}- \tilde \gamma_{0}|\leq r(0) (1+9\Lambda(1+6\Lambda^2))\\
|\tilde \rho_{\nu -1}- \tilde \rho_{0}|\leq \Lambda r(0)  \hat M_3^{(0,1)}\\
|\tilde a_{\nu -1}-  \tilde a_{0}|\leq \Lambda r(0)  \hat M_3^{(0,1)} \end{array}
\label{6.144}
\end{equation}
and finally (after some trivial estimations):
\begin{equation}\begin{array}{l}\displaystyle{ \frac{\vert H^{2}_{\nu}- H^{2}_{T0}\vert}{M_1}}\leq  k_{\nu,1}\ r(0) \ \ \ \ \mbox{with} \  k_{\nu,1}(\Lambda)=48\Lambda^2(1+10\Lambda)\\
\mbox{and} \quad k_{\nu,1}(\Lambda)<1\ \ \mbox{for}: \ \Lambda \leq 0.1 
\end{array}
\label{6.145}
\end{equation} 

Now as before, $\forall n\geq 3$ we estimate the following bounds :
\begin{equation}\begin{array}{l}
\displaystyle{ \frac{\vert H^{4}_{\nu}- H^{4}_{T0}\vert}{M_3}}\leq \ \Lambda r(0)(1+ 144\Lambda ^2 (1+10\Lambda))\\
 \mbox{which means:}\ \  k^{(0)}_{\nu,3}=\Lambda(1+ 144\Lambda ^2 (1+10\Lambda))\\
\ \qquad \ \mbox{and \ \ \ \ $k^{(0)}_{\nu,3}<1$ \ \  for  $\Lambda\leq 0.135$ \qquad  (cf.figure \ref{fig.8})}
\end{array} 
\label{6.146} 
\end{equation} 
and similar results for:
\begin{equation}
  \displaystyle{ \frac{\vert N_3(H^{4}_{\nu}-H^{4}_{T0})\vert}{ \hat M_3^{(0,1)}}} \ \ \ \mbox{and} \displaystyle{ \frac{\vert \frac{\partial}{\partial q^{2}}N_{3}(H^{4}_{\nu}-H^{ 4}_{T0})\vert }{ \hat M_3^{(0,1)}}} 
  \label{6.147}
\end{equation}
Moreover  we find  again recurrently,   using the same arguments as when $\nu=1$  that for all $n\geq 5$:
\begin{equation}
k^{(0)}_{\nu,n}< k^{(0)}_{\nu, 3} < k^{(0)}_{\nu,1} =48\Lambda^2(1+10\Lambda) 
\ \ \ \mbox{\textbf{independent of $\nu$}}
\label{6.148}
\end{equation}

\textbf{Conclusion}: \begin{equation}\begin{array}{l}
\forall \nu\ \geq 2 , \ \Vert  H_{\nu}-H_{T0}\Vert\ \leq \ k^{(0)}(\Lambda)r(0)\quad \mbox{where} \ \\
 k^{(0)}(\Lambda)=48\Lambda^2(1+10\Lambda)<1\ \ \ \forall  \ \Lambda \leq 0.1 
\end{array}
\label{6.149}
\end{equation} 
\end{description}

\item{} 
\textbf{Proof of \ref{6.130}}

 The first step $\nu= 2$ being easily verified, we suppose that for all $\bar \nu\leq \nu-1$  the inequality  \ref{6.130} \  is verified. 
 

\begin{description}
\item{a)}\
Let $n=1$ by  using proposition \ref{prop.3.1} we write:
\begin{equation}\begin{array}{l}\displaystyle{ \frac{\vert H^{2}_{\nu}- H^{2}_{\nu-1}\vert}{  M_1}}\leq \displaystyle{\frac{|\tilde \rho_{\nu -1}-\rho_{\nu -2}| |\tilde \gamma_{\nu -1} +2(\rho_{\nu -1}+\rho_{\nu -2}) |+|\tilde \gamma_{\nu -1}- \tilde \gamma_{\nu -2}|\tilde \rho_{\nu -1}}{|\tilde \gamma_{0}+\tilde \rho_{0}|^2 |   M_1}}\\
\ \ \\ 
\qquad\qquad+ \displaystyle{\frac{\Lambda\vert N_3H^{4}_{\nu -1}(\tilde \gamma_{\nu -2}+\tilde \rho_{\nu -2})-N_3H^{4}_{\nu -2}(\tilde \gamma_{\nu -1}+\tilde \rho_{\nu -1})\vert \Delta_F}{ |(\tilde \gamma_{\nu -1}+\tilde \rho_{\nu -1})(\tilde \gamma_{\nu -2}+\tilde \rho_{\nu -2})|   M_1}}\\
\ \ \\
\qquad +\displaystyle{\frac{\Lambda\vert N_3H^{4}_{\nu -1}(\tilde \gamma_{\nu -2}+\tilde \rho_{\nu -2})-N_3H^{4}_{\nu -2}(\tilde \gamma_{\nu -1}+\tilde \rho_{\nu -1})\vert _{(q^2+m^2)=0} H^{2}_{\nu -1}\Delta_F}{|(\tilde \gamma_{\nu -1}+\tilde \rho_{\nu -1})(\tilde \gamma_{\nu -2}+\tilde \rho_{\nu -2})|   M_1}}\\
\ \ \\
\qquad\qquad \ +\Lambda \displaystyle{ \frac{\vert H^{2}_{\nu-1}- H^{2}_{\nu-2}\vert \vert N_3H^{4}_{\nu-2}\vert_{q^2+m^2=0}\Delta_F}{|(\tilde \gamma_{\nu -1}+\tilde \rho_{\nu -1})(\tilde \gamma_{\nu -2}+\tilde \rho_{\nu -2})|   M_1}}\ 
\end{array} 
\label{6.150}
\end{equation}
Then using the norm definition \ref{def.2.10} and the definitions  of the renormalization constants (cf.proposition \ref{prop.3.1}) we first have:
\begin{equation}\begin{array}{l}
|\tilde \gamma_{\nu -1}- \tilde \gamma_{\nu -2}|\leq \Vert H_{\nu-1}- H_{\nu-2}\Vert {\cal N}_{\gamma}(\Lambda)|_{q^2=0}\\
|\tilde \rho_{\nu -1}- \tilde \rho_{\nu -2}|\leq \Lambda \Vert H_{\nu-1}- H_{\nu-2}\Vert    \hat M_3^{(0,1)}|_{q^2+m^2=0}\
\end{array}
\label{6.151}
\end{equation}
Then, after some elementary estimations the first term of the R.H.S. of \ref{6.131} yields:
\begin{equation}\begin{array}{l}
\displaystyle{\frac{{\cal O}_1}{  M_1}}=\displaystyle{\frac{|\tilde \rho_{\nu -1}-\rho_{\nu -2}| \tilde \gamma_{\nu -2} +|\tilde \gamma_{\nu -1}- \tilde \gamma_{\nu -2}|\tilde \rho_{\nu -2}}{|\tilde \gamma_{0}+\tilde \rho_{0}|^2 |   M_1}}\leq\\
\leq\ K_1^{\nu,1}(\Lambda)\Vert H_{\nu-1}- H_{\nu-2}\Vert \\
\ \ \\
\qquad  \mbox{with}\ \  \ K_1^{\nu,1}(\Lambda)=12\Lambda <1\ \   \mbox{when}\ \Lambda \leq 0.08
\end{array}
\label{6.152}
\end{equation}

We  take the sum of the second and third term of  \ref{6.131} and call it   ${\cal O}_2$.  we obtain:
\begin{equation}\begin{array}{l}
\displaystyle{\frac{{\cal O}_2}{ M_1}}\leq 2\displaystyle{\frac{\Lambda\vert N_3H^{4}_{\nu -1}(\tilde \gamma_{\nu -2}+\tilde \rho_{\nu -2})-N_3H^{4}_{\nu -2}(\tilde \gamma_{\nu -1}+\tilde \rho_{\nu -1})\vert H^{2}_{max}\Delta_F}{(\tilde \gamma_{\nu -1}+\tilde \rho_{\nu -1})  M_1}}\\
\leq  2\Lambda\left\{\displaystyle{\frac{\vert N_3H^{4}_{\nu -1}-N_3H^{4}_{\nu -2}\vert  H^{2}_{max,\nu-1}\Delta_F}{(\tilde \gamma_{\nu -1}+\tilde \rho_{\nu -1})   M_1}}\right\}+\\
+ 2\Lambda H^{2}_{max,\nu-1}\Delta_F\left\{\displaystyle{\frac{\vert N_3H^{4}_{\nu -2,max}\vert\vert\tilde \gamma_{\nu -1}-\tilde \gamma_{\nu -2}\vert+\vert\tilde \rho_{\nu -1}-\tilde \rho_{\nu -2}\vert}{(\tilde \gamma_{\nu -1}+\tilde \rho_{\nu -1})(\tilde \gamma_{\nu -2}+\tilde \rho_{\nu -2})  M_1}}\right\}\\
\mbox{(cf. proposition \ref{prop.3.2} for the definitions of $H^{2}_{max}$ and $H^{4}_{max}$)}\\
\ \ \mbox{and by application again of the norm definition \ref{def.2.10}}\\
\displaystyle{\frac{{\cal O}_2}{ M_1}}\leq \ K_1^{\nu,2}(\Lambda)\Vert H_{\nu-1}- H_{\nu-2}\Vert\\
\mbox{here:} \ \  K_1^{\nu,2}(\Lambda)= 12\Lambda\displaystyle{\frac { H^{2}_{max}\Delta_F\vert N_3H^{4}_{max}\vert \hat M_3}{  M_1}}<12\Lambda\\
\mbox{and again that means  for}\  \ \Lambda \leq  0.08 \ \ \ \   K_1^{\nu,2}<1 
 \end{array}
 \label{6.153}
\end{equation}

The last term of \ref{6.131} that we call  ${\cal O}_3$ yields:
\begin{equation}\begin{array}{l}
\displaystyle{\frac{{\cal O}_3}{\  M_1}}\leq \Lambda \displaystyle{ \frac{\vert H^{2}_{\nu-1}- H^{2}_{\nu-2}\vert \vert N_3H^{4}_{\nu-2}\vert_{q^2+m^2=0}\Delta_F}{|(\tilde \gamma_{\nu -1}+\tilde \rho_{\nu -1})(\tilde \gamma_{\nu -2}+\tilde \rho_{\nu -2})|   M_1}}\leq\\
\ \ \\
\leq \ K_1^{\nu,3}(\Lambda)\Vert H_{\nu-1}- H_{\nu-2}\Vert\quad
\mbox{here:} \ \  K_1^{\nu,3}(\Lambda)= 6\Lambda^2
\end{array}
\label{6.154}
\end{equation}
By using the corresponding bounds from \ref{6.152}, \ref{6.153} \ref{6.154}
we obtain: 
\begin{equation}\begin{array}{l}
\displaystyle{ \frac{\vert H^{2}_{\nu}- H^{2}_{\nu-1}\vert}{\tilde M_1}}< \ K_1^{\nu}(\Lambda) \Vert H_{\nu-1}- H_{\nu-2}\Vert\\
\mbox{here:} \quad \  K_1^{\nu }(\Lambda)= K_1^{\nu,1}+K_1^{\nu,2}+K_1^{\nu,3}=6\Lambda(4+\Lambda)\\  
\ \ \\
\ \mbox{with} \ \ \  K_1^{\nu }<1 \ \ \forall \ \Lambda\leq 0.04
\end{array}
\label{6.155}
\end{equation}
\vfill\eject
\item{b)} Let $n=3$ we write:
\begin{equation}
\displaystyle{ \frac{\vert H^{4}_{\nu}- H^{4}_{\nu-1}\vert}{M_3}}\leq \ A+B
\label{6.156}
\end{equation}
where:
\begin{equation}\begin{array}{l}
A=\displaystyle{\frac{ \delta_{3,\nu}\prod_{\ell=1,2,3}\vert H^{2}_{\nu}(q_{\ell},\Lambda)- H^{2}_{\nu-1}(q_{\ell},\Lambda)\vert\Delta_{F}(q_{\ell})}{M_3}}
\end{array}
\label{6.157}
\end{equation}
and 
\begin{equation}
  B=  \displaystyle{\frac{\vert \delta_{3,\nu}-\delta_{3,\nu-1}\vert\prod_{\ell=1,2,3}H^{2}_{\nu-1}(q_{\ell},\Lambda)\Delta_{F}(q_{\ell})}{M_3}} 
\label{6.158}
\end{equation}
\begin{itemize}
\item[i)] 
By  using again proposition \ref{prop.3.1} the norm definition and the previous result \ref{6.136} of $H^2_{\nu}$ we obtain:
\begin{equation}\begin{array}{l}
A\leq 3\delta_{3,max}\displaystyle{\frac{ K_1^{\nu }M_1\Vert H_{\nu-1}- H_{\nu-2}\Vert}{{M_3}} \prod_{\ell=2,3}H^{2}_{(max)}(q_{\ell},\Lambda)\Delta_{F}(q_{\ell}) }\\
\mbox{or}\ \ \ A\leq K^{\nu}_{3,A}(\Lambda)\Vert H_{\nu-1}- H_{\nu-2}\Vert\\
\ \ \\
\mbox{with}\ \ \  K^{\nu}_{3,A}(\Lambda)=\displaystyle{\frac{3K_1^{\nu }}{\tilde D_{3,min}}}\displaystyle{\prod_{l=2,3}\frac{H^{2}_{(max)}(q_{\ell},\Lambda)}{M_1(q_l)}}\end{array}
\label{6.159}
\end{equation}
\begin{equation}
\mbox{and}\  \quad \tilde D_{3,min}=1+\rho_0+\Lambda|a_0| +0,18\Lambda
\label{6.160}
\end{equation}
\mbox{(cf.definition of $\delta _{3max}$ \ref{def.2.4})}
Now by application of the norm definition (\ref{def.2.10}) of $M_1$ and proposition \ref{prop.3.2} of the definition of $H^{2}_{max}$, we obtain a best estimation of the ratio $\frac{H^{2}_{max}}{M_1}$:
\begin{equation}\begin{array}{l} 
\displaystyle{\frac{H^{2}_{max}}{M_1}}\leq 1-\displaystyle{\frac{6(q^2 +m^2)^{\pi^2/54}(1-\Lambda^2)}{1+6(q^2 +m^2)^{\pi^2/54}}}\\
\mbox{and for sufficiently large $q^2$:}\quad \ \displaystyle{\frac{H^{2}_{max}}{M_1}}\sim 6\Lambda^2
\end{array}
\label{6.161}
\end{equation}
and so:
\begin{equation}
K^{\nu}_{3,A}(\Lambda)< 108\Lambda^4K_1^{\nu }<0.7 K_1^{\nu }\ \ (\forall \Lambda<0.08)
\label{6.162}
\end{equation}
\item[ii)]
As far as   the  term $B$ of the r.h.s. of \ref{6.137} is concerned we use the same arguments as before and we obtain:
\begin{equation}
\displaystyle{\frac{\vert \delta_{3,\nu}-\delta_{3,\nu-1}\vert\prod_{\ell=1,2,3}H^{2}_{\nu-1}(q_{\ell},\Lambda)\Delta_{F}(q_{\ell})}{M_3}}  \leq (B.1) +(B.2)+(B.3)\\
\label{6.163}
\end{equation}
with: 
\begin{enumerate}
\item[ii.1)]
\begin{equation}\begin{array}{l}
(B.1)\leq \\
\leq \displaystyle{\prod_{\ell=1,2,3}\frac{ H^{2}_{\nu-1}(q_{\ell} )}{M_1(q_{\ell}) }\frac{|\tilde \gamma_{\nu -1}- \tilde \gamma_{\nu -2}| + |\tilde \rho_{\nu -1}-\rho_{\nu -2}|  +|\tilde a_{\nu -1}- \tilde a_{\nu -2}|}{ \tilde D_{3,\nu-1}\tilde D_{3,\nu-2}}}\\
\ \ \\
\mbox{Here we used the definitions of the mapping ${\cal M}^*$ (cf. \ref{3.59}  )}\\
\tilde D_{3,\nu-1}=\tilde\gamma_{\nu -1}+\tilde \rho_{\nu -1}+D_{3,\nu-1}-\Lambda \tilde a_{\nu-1}\\ 
\mbox{(and the analogous expression for $\tilde D_{3,\nu-2}$), so}\\
\ \ \\
(B.1)\leq K^{\nu}_{3,B.1}(\Lambda)\Vert H_{\nu-1}- H_{\nu-2}\Vert\\
\mbox{with }\quad   K^{\nu}_{3,B.1}=\displaystyle{\prod_{l=1,2,3}\frac{H^{2}_{(max)}(q_{\ell})}{M_1(q_l)}}\displaystyle{\frac{\gamma_{max} +2\Lambda  \hat M_3^{(0,1)}}{(1+\rho_0+\Lambda|a_0| +0,18\Lambda)^2}}\\
\mbox{or}\quad  K^{\nu}_{3,B.1} \leq (6\Lambda)^3 (1+9\Lambda)\quad \mbox{for sufficiently large $q^2$.}\\
\mbox{and for small $q^2$ } \quad K^{\nu}_{3,B.1}(\Lambda)<1,\ \ \  \forall \Lambda\leq 0.05 
\end{array}
\label{6.164}
\end{equation}
\item[ii.2)] \begin{equation}\begin{array}{l} (B.2)\leq\\ 
\leq 6\Lambda \prod_{\ell=1,2,3} H^{2}_{\nu-1}(q_{\ell} )\Delta_{F}(q_{\ell})
\displaystyle{\frac{|H^4_{\nu-1}-H^4_{\nu-2}|B^4_{\nu-2}-A^4_{\nu-2}|}{M_3|H^4_{\nu-1}||H^4_{\nu-2}|\tilde D_{3,\nu-1}\tilde D_{3,\nu-2}}}
\end{array}
\label{6.165}
\end{equation}
 \begin{equation}\begin{array}{l}
\mbox{But:}\quad |H^4_{\nu-1}|=6\Lambda\displaystyle{\frac{\prod_{\ell=1,2,3} H^{2}_{\nu-1}(q_{\ell} )\Delta_{F}(q_{\ell})}{\tilde D_{3,\nu-2}}} \\ 
\mbox{and}\ \  \  \displaystyle{\frac{|B^4_{\nu-2}-A^4_{\nu-2}|}{|H^4_{\nu-2}|}}=  D_{3,\nu-2}
\end{array}
\label{6.166}
\end{equation}
\begin{equation}\begin{array}{l}
\mbox{so}\  (B.2) \leq K^{\nu}_{3,B.2}(\Lambda)\Vert H_{\nu-1}- H_{\nu-2}\Vert\\ 
\ \ \\
\mbox{where}\ \ \ K^{\nu}_{3,B.2}=\displaystyle{\frac{D_{3,max} }{\tilde D_{3,min}}}=\displaystyle{\frac{9\Lambda(1+6\Lambda^2)}{(1+\rho_0+\Lambda|a_0| +0,18\Lambda)}}\\
\mbox{and}\quad K^{\nu}_{3,B.2}(\Lambda)<1,\ \ \  \forall \Lambda\leq 0.1  \ \ \mbox{(very weak condition)} 
\end{array}
\label{6.167}
\end{equation}
\item[ii.3)] 
\begin{equation}\begin{array}{l}
(B.3)\leq \\
< 6\Lambda \prod_{\ell=1,2,3} H^{2}_{\nu-1}(q_{\ell} )\Delta_{F}(q_{\ell})
\displaystyle{\frac{|B^4_{\nu-1}-B^4_{\nu-2}| }{M_3|H^4_{\nu-1}|\tilde D_{3,\nu-1}\tilde D_{3,\nu-2}}}\\
\mbox{or} \ \\
 (B.3)<\displaystyle{\frac{ 9\Lambda}{M_3\tilde D_{3,\nu-1}} }\left\{\vert N_2(H^{4}_{\nu -1}-H^{4}_{\nu -2})\vert  H^{2}_{max}\Delta_F\right\}\\
 +\displaystyle{\frac{ 9\Lambda}{M_3\tilde D_{3,\nu-1}} }\left\{\vert N_2H^{4}_{max}\vert \vert  H^{2}_{\nu-1}-H^{2}_{\nu-2}\vert\Delta_F\right\}\\
\end{array}
\label{6.168}
\end{equation}
Notice that we have used the sign properties of $B^4$ and $A^4$ together with the definitions of the mapping  (cf. in particular equation \ref{6.146}).
Then, by application of the norm definitions we obtain:
\begin{equation}
\begin{array}{l}
(B.3) < K^{\nu}_{3,B.3}(\Lambda)\Vert H_{\nu-1}- H_{\nu-2}\Vert\\ 
\ \ \\
\mbox{where}\ \ \ K^{\nu}_{3,B.3}=\displaystyle{\frac{18\Lambda H^2_{max} }{\tilde D_{3,min}M_1}}\\
\mbox{and for sufficiently large $q^2$}\quad K^{\nu}_{3,B.3}(\Lambda)<1,\\\forall \Lambda\leq 0.2  \ \ \mbox{(very weak condition)} \\
\mbox{and for small $q^2$ } \quad K^{\nu}_{3,B.3}(\Lambda)<1,\ \ \  \forall \Lambda\leq 0.05 
\end{array}
\label{6.169}
\end{equation}
\end{enumerate}
\end{itemize}
Finally taking into account \ref{6.159}, \ref{6.162}, \ref{6.164}, \ref{6.167} and \ref{6.169} we obtain that 
$$K^{\nu} _3<1 \ \ \ \forall \Lambda\leq 0.05\qquad\qquad\qquad \blacksquare $$
\item{c)} Under weaker conditions on $\Lambda$  and using the analogous procedure, (norm definitions, together with properties in $\Phi_R$ etc...) we find that: 
\begin{itemize} 
\item[c.i)] that there is a positve constant continuous function of $\Lambda$, \ \ 
\begin{equation}\ K_{\gamma}^{\nu}(\Lambda)<1\ \  \mbox{ such that:}
\label{6.170}
\end{equation} 
\begin{equation}
\begin{array}{l} 
\displaystyle{ \frac{|\tilde\gamma_{\nu}-\tilde\gamma_{\nu-1}|}{{\cal  N}_{\gamma}}} \leq \ K_{\gamma}^{\nu} (\Lambda)\Vert H_{\nu-1}- H_{\nu-2}\Vert; \\
 \mbox{with} \ \ K_{\gamma}^{\nu} (\Lambda)=3K_1^{\nu}\gamma_{max}M_1^{-2}|_{q^2=0}+K_3^{\nu}\gamma_{max}^2<1 \\
\ \mbox{(under weaker than the condition $\Lambda\leq 0.05 $)}  \end{array}
\label{6.171}
\end{equation}
\item[c.ii)]
\begin{equation}
\begin{array}{l} 
\displaystyle{ \frac{|\tilde\rho_{\nu}-\tilde\rho_{\nu-1}|}{M_3^{(0,1)}}}\leq \Lambda\displaystyle{ \frac{\vert \frac{\partial}{\partial q^{2}}N_{3}(H^{4}_{\nu}-H^{ 4}_{\nu-1})\vert }{\hat M_3^{(0,1)}}}\leq \ K_{\rho}^{\nu} (\Lambda)\Vert H_{\nu-1}- H_{\nu-2}\Vert; \\
 \mbox{with} \ K_{\rho}^{\nu} =\Lambda K_3^{\nu} \ <\ 1 \ \  
 \mbox{under weaker condition on}\ \Lambda
 \end{array}
\label{6.172}
\end{equation}
\item[c.iii)] \begin{equation}
 \displaystyle{ \frac{|\tilde a_{\nu}-\tilde a_{\nu-1}|}{M_3^{(0,1)}}}\leq \ K_3^{\nu}(\Lambda)\Vert H_{\nu-1}- H_{\nu-2}\Vert  
 \label{6.173}
\end{equation}
\item[c.iv)] By using the basic splitting properties in $\Phi_R$ of $H^4$ and by analogous arguments as above we show that
 \begin{equation}\begin{array}{l}\\
 |N_2H_{\nu}^4|\leq 6\Lambda [N_2^{3}\tilde]\ \displaystyle{\prod_{l=1}^{2}M_{1}(q_l)}\\
\mbox{so that}\ \  \displaystyle{\frac{| N_2(H^{4}_{\nu}-H^{ 4}_{\nu-1})| }{\hat M_{(3,2)}}}\leq \ K_3^{\nu}(\Lambda)\Vert H_{\nu-1}- H_{\nu-2}\Vert  
\end{array}
 \label{6.174}
 \end{equation}
 \end{itemize}
 Finally:
\begin{equation}
\begin{array}{l}  
\displaystyle{\sup_{( q,\Lambda)}}
\left\{\displaystyle{ \frac{\vert \frac{\partial^{(0,1)}}{\partial q^{2}}N_{3}(H^{4}_{\nu}-H^{ 4}_{\nu-1})\vert }{\hat M_3^{(0,1)}}};\displaystyle{\frac{| N_2(H^{4}_{\nu}-H^{ 4}_{\nu-1})| }{\hat M_{(3,2)}};\frac{|\gamma_{\nu}-\gamma_{\nu-1}|}{{\cal  N}_{\gamma}}}\right\} \\
\qquad \qquad \leq \ K_{3,2,\gamma}^{\nu} (\Lambda)\Vert H_{\nu-1}- H_{\nu-2}\Vert; \\
 \mbox{with} \ \  K_{3,2,\gamma}^{\nu} (\Lambda)<1 \ \ \ \mbox{when \  $\Lambda\leq 0.05 $} \qquad \blacksquare $$
  \end{array}
\label{6.175}
\end{equation}

\item{d)} \textbf{Let $n\geq 5$ }\

We suppose that for every $\tilde n\leq n-2$ the first property of  \ref{6.129}
is verified in the following sense: 
\begin{equation}\begin{array}{l}
\forall \Lambda\leq 0.05 \mbox{\ and $\forall$ \ fixed  $\nu$}, \\ \mbox{there is a strictly positive constant (continuous function of $\Lambda$)\ \ $K^{\nu}_{\tilde n}(\Lambda) $ such that:}\\
\displaystyle{\sup_{( q,\Lambda)}}
\left\{\displaystyle{ \frac{\vert H^{\tilde n+1}_{\nu}- H^{\tilde n+1}_{\nu-1}\vert}{ M_{\tilde n}}}\right\} \leq \ K^{\nu}_{\tilde n}(\Lambda) \Vert H_{\nu-1}- H_{\nu-2}\Vert  \quad \mbox{with} \\
   K^{\nu}_{\tilde n}(\Lambda) <K^{\nu}_{\tilde{ n}-2}(\Lambda)\leq\ \dots\leq K^{\nu}_{3}(\Lambda)  
  \end{array}
\label{6.176}
\end{equation}
We show this property for $\tilde n= n$ by using again the  definitions of the norm  (\ref{def.2.10}), of the mapping ${\cal M}^*$ (cf.proposition \ref{prop.3.1}) and the properties in $\Phi_R$ (proposition  \ref{prop.3.2}) . 
\begin{equation}\begin{array}{l}
\displaystyle{ \frac{\vert H^{n+1}_{\nu}- H^{n+1}_{\nu-1}\vert}{M_n}}\leq  A_n+B_n \ \ \mbox{where}\\
A_n=\displaystyle{\frac{ \delta_{n,\nu}\vert C^{n+1}_{\nu} - C^{n+1}_{\nu-1}\vert}{3\Lambda n(n-1)M_n}}\\
B_n=\displaystyle{\frac{\vert \delta_{n,\nu}-\delta_{n,\nu-1}\vert|C^{n+1}_{\nu-1}\vert}{3\Lambda n(n-1)M_n}}  
\end{array}
\label{6.177}
\end{equation} 
\begin{itemize}
\item[i)]
\begin{equation}\begin{array}{l}  
A_n=\displaystyle{\frac{ \delta_{n,\nu}\vert C^{n+1}_{\nu} - C^{n+1}_{\nu-1}\vert}{3\Lambda n(n-1)M_n}}\leq\\
\leq\displaystyle{K^{\nu}_{ { n}-2}\prod_{l= 2,3}\frac{{\cal T}_nH^{2}_{(max)}(q_{\ell},\Lambda)}{n(n-1)M_1(q_l)}}\Vert H_{\nu-1}- H_{\nu-2}\Vert +\\
+\displaystyle{2K^{\nu}_{ 1}\frac{{\cal T}_nH^{n-1}_{(max)}(q_{(n-2)})}{n(n-1)M_{n-2}(q_{(n-2)})}}\Vert H_{\nu-1}- H_{\nu-2}\Vert \\
\mbox{or} \\
A_n\leq K^{\nu}_{n,A} \Vert H_{\nu-1}- H_{\nu-2}\Vert\ \ \  \mbox{with}\\
   K^{\nu}_{n,A}(\Lambda)=\displaystyle{\frac{H^{2}_{(max)}}{24M_1}}\left\{\displaystyle{\frac{K^{\nu}_{ { n}-2}H^{2}_{(max)}}{M_1}}+\displaystyle{\frac{K^{\nu}_{ 1}|H^{n-1}_{(max)}|}{M_{n-2}}}\right\}
\end{array}
\label{6.178}
\end{equation}
We easily verify that under a weaker condition than $\Lambda\leq 0.05$
\begin{equation}
K^{\nu}_{  n,A }\leq K^{\nu}_{  n-2,A}<1 \qquad\ \  \blacksquare
\label{6.179}
\end{equation}
\item[ii)] 
For   the  term $B_n$ of   \ref{6.158}   we use the same arguments as before and we obtain:
\begin{equation} 
B_n= \displaystyle{\frac{\vert \delta_{n,\nu}-\delta_{n,\nu-1}\vert|C^{n+1}_{\nu-1}\vert}{3\Lambda n(n-1)M_n}}\\  \leq |B_{n,1}|+|B_{n,2}|+|B_{n,3}|
\label{6.180}
\end{equation}
with: \begin{enumerate}
\item[ii.1)]\begin{equation}\begin{array}{l}
|B_{n,1}|\leq \\ 
\leq \displaystyle{\vert C^{n+1}_{\nu-1}\vert\frac{|\tilde \gamma_{\nu -1}- \tilde \gamma_{\nu -2}| + |\tilde \rho_{\nu -1}-\rho_{\nu -2}|  +|\tilde a_{\nu -1}- \tilde a_{\nu -2}|}{ M_n\tilde D_{n,\nu-1}\tilde D_{n,\nu-2}}}
\end{array}
\label{6.181}
\end{equation}
By the definitions of the norm  (\ref{def.2.10}) of the mapping ${\cal M}^*$ (cf.prop. \ref{prop.3.1})
and the properties in $\Phi_R$ (proposition  \ref{prop.3.2}) we obtain :  
\begin{equation}
\begin{array}{l}
\tilde D_{n,\nu-1}=\tilde\gamma_{\nu -1}+\tilde \rho_{\nu -1}+D_{n,\nu-1}-\Lambda |\tilde a_{\nu-1}|\\ 
\mbox{(and the analogous expression for $\tilde D_{n,\nu-2}$ \ so)}\\
|B_{n,1}|\leq \displaystyle{\vert H^{n+1}_{\nu-1}\vert\Vert H_{\nu-1}- H_{\nu-2}\Vert\frac{\gamma_{max} +2\Lambda  \hat M_n^{(0,1)}}{ M_n\tilde D_{n,\nu-1}}}\\
\mbox{or}\quad
|B_{n,1}|\leq K^{\nu}_{n,B.1} \Vert H_{\nu-1}- H_{\nu-2}\Vert\quad  
\mbox{with}\\ 
 K^{\nu}_{n,B.1}= \leq\displaystyle{\prod_{l= 2,3}\frac{H^{2}_{(max)}(q_{\ell})(n-3)^2|H^{n-1}_{max}|(\gamma_{max} +2\Lambda  \hat M_n^{(0,1)})}{M_1(q_l)24n(n-1)M_{n-2}\tilde D_{3,min}}} \end{array}
\label{6.182}
\end{equation}
Now using the evident bound:
\begin{equation}\
\displaystyle{ \frac{ |H^{n-1}_{max}|}{  M_{n-2 }}}\leq\displaystyle{\prod_{l= 2,3}\frac{H^{2}_{(max)}(q_{\ell}) |H^{n-3}_{max}|}{M_1(q_l)24 M_{n-4}}}
\label{6.183}
\end{equation}
and the analogous definition \ref{6.163} of\ $  K^{\nu}_{n-2,B.1}$  we have:
\begin{equation}\begin{array}{l}
 K^{\nu}_{n,B.1}  \leq\displaystyle{\prod_{l= 2,3}\frac{H^{2}_{(max)}(q_{\ell}) (n-3)^2 (n-2(n-3)}{M_1(q_l)24n(n-1)(n-5)^2}}K^{\nu}_{n-2,B.1}\\
 \Leftrightarrow\  K^{\nu}_{n,B.1} < K^{\nu}_{n-2,B.1} <1\qquad \qquad\qquad\qquad \qquad\qquad\blacksquare
 \end{array}
 \label{6.184}
\end{equation}
\item[ii.2)] By analogy to $n=3$ 
 \begin{equation}\begin{array}{l}   |B_{n,2}|\leq\\ 
\leq \displaystyle{\frac{\vert C^{n+1}_{\nu-1}\vert}{ M_n\tilde D_{n,\nu-1}\tilde D_{n,\nu-2}}}
\displaystyle{\frac{|H^{n+1}_{\nu-1}-H^{n+1}_{\nu-2}|B^{n+1}_{\nu-2}-A^{n+1}_{\nu-2}|}{|H^{n+1}_{\nu-1}||H^{n+1}_{\nu-2}|}}\\
\ \ \\
\mbox{or} \ \quad
|B_{n,2}|\leq K^{\nu}_{n, B.2} \Vert H_{\nu-1}- H_{\nu-2}\Vert\quad  \mbox{with}  \\ K^{\nu}_{n,B.2}=\displaystyle{\frac{D_{n,max} }{\tilde D_{n,min}}}\sim  \displaystyle{\frac{D_{n-2,max} }{\tilde D_{n-2,min}}}<1
\  \qquad\forall \Lambda\leq 0.05 \qquad\qquad \blacksquare
\end{array}
\label{6.185}
\end{equation}

\item[ii.3)] Then  by an analogous as above procedure we have:
\begin{equation}\begin{array}{l}
|B_n,3|\leq \\
< \displaystyle{\frac{\vert C^{n+1}_{\nu-1}\vert}{ M_n\tilde D_{n,\nu-1}\tilde D_{n,\nu-2}}}
\displaystyle{\frac{|B^{n+1}_{\nu-1}-B^{n+1}_{\nu-2}| }{|H^{n+1}_{\nu-1}|}}\\
\mbox{or} \ \\ 
 |B_n,3|<\displaystyle{\frac{ 3\Lambda n(n-1)}{2M_n\tilde D_{n,\nu-1}} }\left\{\vert N_2(H^{n+1}_{\nu -1}-H^{n+1}_{\nu -2})\vert  H^{2}_{max}\Delta_F\right\}\\
 +\displaystyle{\frac{  3\Lambda n(n-1)}{2M_n\tilde D_{n,\nu-1}} }\left\{\vert N_2H^{n+1}_{max}\vert \vert  H^{2}_{\nu-1}-H^{2}_{\nu-2}\vert\Delta_F\right\}\\
\mbox{or} \\
 |B_n,3|<K^{\nu}_{n, B.3} \Vert H_{\nu-1}- H_{\nu-2}\Vert\quad  \mbox{with}  \\ K^{\nu}_{n,B.3}= \delta_\infty [N_2^n\tilde]_{q=0} \displaystyle{\frac{H^2_{max}}{2M_1}} (1+\displaystyle{\frac{{\cal T}_n}{n(n-1)}})\\
\mbox{and verify} \ \ K^{\nu}_{n, B.3}\sim K^{\nu-2}_{n,B.3}<1, \ \ \ \forall\  \Lambda\leq 0.05 \ \ \qquad\qquad \blacksquare
 \end{array}
\label{6.186} 
\end{equation}
\end{enumerate}
\end{itemize}
\end{description}

Finally by addition of \ref{6.177}, \ref{6.178}, \ref{6.179}, \ref{6.184}, \ref{6.185} and \ref{6.186} we obtain
the proof of the recursion \ref{6.176} and by using also the result \ref{6.149} the proof of the contractivity criterium \ref{6.130} follows.
\ \ $\hspace {5cm}\ \blacksquare$
\end{enumerate}
\end{Appendix}
\end{document}